\documentclass[numbers,sort&compress,10pt,fleqn]{elsarticle}
\usepackage{geometry}
\usepackage{amsmath,amsthm,amssymb,amsbsy,latexsym,graphicx}
\usepackage{rotating}
\usepackage{float}
\usepackage{amsmath}
\usepackage{subfigure}
\usepackage{caption}
\usepackage{multicol}
\usepackage{upgreek}
\usepackage{bm}
\usepackage{mathdots}
\usepackage{relsize}
\usepackage{mathdots}
\usepackage{newfloat}
\usepackage{etoolbox}
\usepackage[bottom]{footmisc}% places footnotes at page bottom
\usepackage[bbgreekl]{mathbbol}
\newgeometry{vmargin={20mm}, hmargin={20mm,20mm}}

\usepackage{color}

\usepackage{natbib}
\usepackage{booktabs}
\usepackage{indentfirst}
\begin{document}
\begin{frontmatter}

\title{A three-dimensional constitutive model for the martensitic transformation in polycrystalline shape memory alloys under large deformation}

\author[ad1]{L. Xu}
\ead{sdf007xulei@tamu.edu}

\author[ad2]{T.~Baxevanis}
%\ead{theocharis@tamu.edu}

\author[ad1,ad3]{D. C.~Lagoudas}
%\ead{lagoudas@tamu.edu}

\cortext[bla]{Corresponding author}

\address[ad1]{Department of Aerospace Engineering, Texas A\&M University, 
College Station, TX 77843, USA}

\address[ad2]{Department of Mechanical Engineering, University of Houston, 
	Houston, TX 77204, USA}

\address[ad3]{Department of Materials Science \& Engineering, Texas A\&M University, 
	College Station, TX 77843, USA}

\begin{abstract}
This work presents a three-dimensional constitutive model for the martensitic transformation in polycrystalline Shape Memory Alloys (SMAs) under large deformation. By utilizing the logarithmic strain and rate, the model is able to account for large strains and rotations that SMA-based components may undertake, but also resolves the artificial stress errors caused by the non-integrable objective rates that are widely used in current commercial finite element software. The model is developed through classical thermodynamic laws combined with the standard Coleman-Noll procedure. The scalar martensitic volume fraction and the second-order transformation strain tensor are chosen as the internal state variables to capture the material response exhibited by polycrystalline SMAs. A detailed implementation procedure of the proposed model is described through a user-defined material subroutine. Numerical experiments considering SMA components including a bar, a beam, a torque tube and a solid flexible structure under stress/thermally-induced phase transformations are investigated via the proposed model, and the results under cyclic loading are compared against the predictions provided by the Abaqus nonlinear solver. The development framework of the proposed model and its implementation procedure can be extended to incorporate other nonlinear phenomena exhibited by SMAs, such as transformation-induced plasticity, viscoplasticity, and damage under large deformation.
\end{abstract}

 
\end{frontmatter}

%\begin{multicols}{2}

\section{INTRODUCTION} \label{sec:intro} 

SMAs belong to a specialized subgroup of multifunctional materials known as active materials, and are capable of recovering their pre-defined geometry when subjected to a thermal stimulus above certain temperatures. This unique property of SMAs is achieved through a solid-to-solid state diffusionless phase transformation between the {high-symmetry}, high-temperature austenitic phase and the {low-symmetry}, low-temperature martensitic phase \cite{lagoudas2008}.  Since the discovery of shape memory effect, SMAs have been extensively investigated as sensors and actuators towards building smart systems integrated with adaptive and morphing features \cite{peraza2014origami,hartl2007aerospace}. Recently, aerospace researchers have considered to use the SMA-based actuators to reconfigure the shape of a supersonic aircraft to meet the noise and efficiency requirements in response to the real-time changing ambient environment, which has the great potential to realize a commercially viable overland civil supersonic flight in the near future \cite{lazzara2019,xu2019}.   

A substantial number of constitutive theories for SMAs have been proposed so far with the majority of them based on the infinitesimal strain theory for small deformation analysis. { Thorough reviews can be found from \cite{raniecki1992,boyd1996,birman1997,raniecki1998,patoor1996,levitas1998,levitas2002,patoor2006,zaki2007,saint2009,hackl2008,chemisky2011,sedlak2012,Zineb2016}.} In general, constitutive models for SMAs can be approximately categorized into three different types: phase-field theory based models, crystal-plasticity theory based models, and classical $J_2$-flow theory based models. The phase-field models, in which order parameters are utilized to differentiate austenitic and martensitic phases, can track microstructure evolution, such as phases front movement, during the phase transformation process \cite{levitas1998,levitas2002,chen2002,steinbach1999,steinbach2006,mamivand2013,zhong2014}. Therefore, phase-field theory based models are well suited to investigate the dynamic nucleation and phase morphology growth for martensitic phase transformation in SMAs. The enormous computational time, however, needed to solve the phase kinetic partial differential equations hinders its popularity to the extent that macroscopic structural response is concerned. In regard to crystal-plasticity theory based models \cite{tham2001,Wang2008,yu2013}, with the consideration of the effect of material microstructure ({e.g.}, crystal orientation, texture, {etc}.), these models are able to capture the anisotropy {in} material response ({e.g.}, tension-compression asymmetry) exhibited by textured polycrystalline SMAs. Similar to phase-field methods, the complex implementation procedure of these models to incorporate the microstructure information makes them computational costly. Following the legacy of $J_2$-flow theory, phenomenological SMA models have attracted attention in the engineering community where repetitive designs and optimization procedures on SMA components are needed to find target shapes. By introducing a set of internal state variables (such as volume fraction and transformation strain tensor), $J_2$ theory-based SMA models are able to simulate the macroscopic response of an SMA component in an efficient way. The simplicity of this model type and its well-established implementation procedure have allowed it to be widely used among real engineering applications \cite{auricchio1997,lagoudas2012,brinson1993finite,brinson1993one,lexcellent1996general,lexcellent2013shape,zaki2007,reese2008}.

Constitutive models based on the infinitesimal strain theory are able to predict SMA response accurately under small deformation situations. However, it has been reported that SMAs can reversibly deform to a relatively large strain regime up to 8\% \cite{shaw2000,jani2014}. Also for specific boundary value problem such as fracture in SMAs, the strain levels close to the crack tip are well beyond 10\% \cite{bh2016,bh2019} within the finite strain regime. In addition to such relatively large strain, SMA-based actuators (e.g., spring, beam, torque tube) may also undergo large rotation during their deployment. For example, an SMA beam component is used as a bending actuator in order to realize a morphing engine shape, where the SMA experienced large bending induced rotation \cite{hartl2007aerospace}. Another example is the SMA tube component utilized as a torsional actuator to repeatedly rotate a deployable and retractable solar panel \cite{wheeler2015}, where the SMA tube is subjected to cyclic large rotation. Combining the aforementioned two facts that SMAs may undergo large strain and rotation, it is necessary to develop a constitutive model based on a finite deformation framework to provide an accurate prediction for the response of SMAs. 

Two kinematic assumptions are often employed in the finite deformation theory, {i.e.}, the multiplicative decomposition of the deformation gradient and the additive decomposition of the strain rate tensor. { In the multiplicative approach, the deformation gradient is usually decomposed into an elastic part multiplied with an inelastic part. Finite strain SMA models based on the multiplicative decomposition can be obtained from literature \cite{ziolkowski2007, reese2008,evangelista2010,wang2017sms,wang2017ijes,stupkiewicz2013,damanpack2017}, among which some advanced capabilities are considered. For examples, Wang and coworkers \cite{wang2017ijes} presented a finite strain SMA model with the fully thermomechanically coupled feature, the consideration of coexistence of different martensitic variants, and accounting for temperature effect on the hysteresis size. In the work of Stupkiewicz and Petryk \cite{stupkiewicz2013}, they proposed a finite strain SMA model to capture the tension-compression asymmetry phenomenon. Damanpack and coworkers \cite{damanpack2017} also developed an SMA model that considers anisotropic behaviors and reorientation in SMAs at finite deformation.
However, it is known that finite strain model based on additive decomposition significantly reduces the complexities of model structure compared to multiplicative models, which in return facilitates the computational efficiency of the finite strain model as a 3-D design tool.} Therefore, they are widely used in current available finite element softwares ({e.g.}, Abaqus, ANSYS). However, to satisfy the principle of objectivity, additive models are required to use an objective rate in its rate form hypoelastic constitutive relation. A number of objective rates (such as Zaremba-Jaumann-Noll rate, Green-Naghdi-Dienes rate, and Truesdell rate) have been proposed to meet this goal. However, those objective rates are not essentially ``objective" because of their failure to integrate the rate form hypoelastic relation to yield a free energy based hyperelastic stress-strain relation \cite{xiao2006}. As a result, spurious phenomena ({e.g.}, shear stress oscillation, artificial stress residuals, {etc}.) are often observed  in the predicted response even for simple elastic materials through these objective rates. More details regarding this issue are addressed in \ref{sec:appendix}.

It was not until the logarithmic rate was proposed in the literature \cite{xiao1997,xiao1997hypo,bruhns1999self,bruhns2001large,bruhns2001self,meyers2003elastic,meyers2006choice} that the previously mentioned self-inconsistency issue related to non-integrable objective rates was resolved. As was proved in the work \cite{xiao1997}, the logarithmic rate of Eulerian logarithmic strain $\mathbf{h}$ is exactly identical to the strain rate tensor $\mathbf{D}$, and the logarithmic strain is the only one among all other strain measures enjoying this important property. Therefore, the finite strain models using logarithmic strain and rate are not only able to capture large strain and large rotation but also are capable of resolving the aforementioned spurious phenomena. This new development in finite deformation theory not only provides solution to classical finite elastoplastic problems for conventional metallic materials \cite{zhu2014,zhu2016}, but also sheds light on the finite strain model development for active materials such as SMAs. {Few SMA models using additive approach can be found from \cite{muller2006,teeriaho2013,xiao2014explicit,yu2015,xu2017,Xu2017trip,xu2018}, but some of the very important SMA phase transformation characteristics have not been addressed among them, such as the smooth transition during the phase transformation, the stress dependent transformation strain to account for the coexistence of oriented/self-accommodated martensitic variants, and a stress dependent critical driving force to consider the effect of applied stress levels on the size of hysteresis loop. To this end, this work presents a finite strain SMA model formulation based on the additive decomposition using the logarithmic strain and rate. As a continuous development from the infinitesimal SMA model \cite{lagoudas2012}, the proposed model has a simple model structure and considers three very important characteristics for SMA response as its infinitesimal counterpart does. These developments combined result in an improved computational efficiency and robustness for the proposed finite strain model to predict the SMA response at large deformation, without introducing additional intermediate state variables, such as Mandel stress, that are utilized in the multiplicative models. It is noted that the primary focus of this work is mainly on the formulation of a finite strain SMA model, rather than the development of a constitutive model that can capture the full complexities of the SMA thermomechanical deformation. Thus, tension-compression, latent-heat effects, reorientation between orientated and self-accommodated martensitic variants, cyclic evolution features (transformation-induced plasticity, two-way shape memory effect at stress-free conditions) are not included here for simplicity. Moreover, this work carefully examines the artificial stress errors caused by using other non-integrable objective rates in current commercial finite element packages. The capability of the proposed model to eliminate such stress errors shows significant importance for the analysis of SMA-based actuators, e.g., SMA beam and SMA torque tube subjected to cyclic large deformation.}

This paper is organized as follows. Section \ref{Preliminary} presents the kinematic preliminaries. Section \ref{sec:framework} concentrates on the model development based on the logarithmic strain and logarithmic rate. The derivation of the consistent tangent stiffness matrix and the consistent thermal matrix are also provided. In section \ref{Implementation}, the detailed implementation procedure for the proposed model is described by using a user-defined material subroutine (UMAT) through the finite element software Abaqus. Numerical examples are studied to demonstrate the capability of the proposed model in Section \ref{Result}. Conclusions are presented in Section \ref{Conc}. A detailed calibration procedure for the material parameters used in this model is also provided in the \ref{sec:calibration}.

%%%%%%%%%%%%%%%%%%%%%%%%%%%%%%%%%%%%%%%%%%%%%%%%%%%%%%%%%%%
%%%%%%%%%%%%%%%%%%%%%%%%%%%%%%%%%%%%%%%%%%%%%%%%%%%%%%%%%%%
%%%%%%%%%%%%%%%%%%%%%%%%%%%%%%%%%%%%%%%%%%%%%%%%%%%%%%%%%%%
\section{Preliminaries} \label{Preliminary}
\subsection{Kinematics}
Let material point $ \mathcal{P} $ from body $ \mathcal{B} $ {be} defined by a position vector $ \mathbf{X} $ in the reference (undeformed) configuration at time $ t_{0} $, and let vector $ \mathbf{x} $ represent the position vector of that material point in the current (deformed) configuration at time $ t $. Therefore ,the deformation process of point $ \mathcal{P} $ between the reference configuration and the current configuration   can be defined through the well-known deformation gradient tensor $\mathbf{F}(\mathbf{x},t)$:
\begin{equation}\label{Deformation}
\mathbf{F}(\mathbf{x},t) =\frac{\partial \mathbf{x}}{ \partial \mathbf{X}}  
\end{equation}
and the velocity field $\mathbf{v}$ can be defined as,
\begin{equation}\label{Velocity}
\mathbf{v} = \dfrac{d \mathbf x}{d t} = \dot{\mathbf x}
\end{equation}
based on the velocity field $\mathbf{v}$, the velocity gradient $\mathbf{L}$ can be derived as,
\begin{equation}\label{eq:V_gradient}
\mathbf{L}= \frac{\partial \mathbf{v}}{ \partial \mathbf{x}}= \mathbf{\dot{F}}\mathbf{F} ^{-1}  
\end{equation}
the following polar decomposition equation for deformation gradient $\mathbf F$ is well known,
\begin{equation}\label{eq:PolarDecom}
\mathbf{F  = RU = VR}
\end{equation}
{where $\mathbf{R}$ is the rotation tensor, $\mathbf{U}$ and $\mathbf{V}$ are the right (or Lagrangian) and the left (or Eulerian) stretch tensors, respectively, by which the right Cauchy-Green tensor $\mathbf{C}$ and the left Cauchy-Green tensor $\mathbf{B}$ can be obtained, as follows,}
\begin{equation}\label{RCG}
\mathbf{C} = \mathbf{F^{T}F} =\mathbf{U}^2 
\end{equation}
\begin{equation}\label{LCG}
\mathbf{B} = \mathbf{FF^{T}} =\mathbf{V}^2 
\end{equation}
where $\mathbf{I} $ is the second order identity tensor. The logarithmic strain (also called Hencky or true strain)  of Lagrangian type $ \mathbf{H} $ and Eulerian type $ \mathbf{h} $ can thus be defined as,
\begin{equation}\label{L_LogStrain}
\mathbf{H} = \frac{1}{2} \ln(\mathbf{ {C}}) =\mathbf{\ln ({U})}
\end{equation}
\begin{equation}\label{E_LogStrain}
\mathbf{h} = \frac{1}{2} \ln(\mathbf{ {B}}) =\mathbf{\ln ({V})}
\end{equation}

It is also well known that the velocity gradient $\mathbf{L}$ can be additively decomposed into a symmetric part, the strain rate tensor $\mathbf{D}$, and an anti-symmetric part, the spin tensor $\mathbf{W}$,
\begin{equation}\label{S_P_tensor}
\mathbf{L = D + W}, \qquad
\mathbf{D} =\dfrac{1}{2} \mathbf{(L+L^{T})}, \vspace{5pt}  \qquad 
\mathbf{W} =\dfrac{1}{2} \mathbf{(L-L^{T})}  
\end{equation}

\subsection{Logarithmic strain, logarithmic rate and logarithmic spin}\label{subsec:objective_rate}
As was mentioned in section \ref{sec:intro}, two widely accepted kinematic assumptions, {i.e.}, the multiplicative decomposition of deformation gradient $\mathbf{F}$ and the additive decomposition of the strain rate tensor $\mathbf{D}$, are usually considered in finite deformation theory. The multiplicative models use a hyperelastic constitutive relation while  a rate form hypoelastic constitutive equation is usually adopted for additive models. The rate form hypoelastic constitutive theory using objective rates has been criticized for its non-integrability because it can not well define an essential elastic material behavior  \cite{simo2006}, this includes many well known objective rates such as Zaremba-Jaumann rate, Green-Naghdi rate, Truesdell rate, {etc}.\cite{xiao2006}.

The aforementioned problems about objective rates were solved in the work by Xiao et al.\cite{xiao1997,xiao1997hypo,xiao2006}, Bruhns et al.\cite{bruhns1999self,bruhns2001large,bruhns2001self} and Meyers et al.\cite{meyers2003elastic,meyers2006choice}, where they proved that the logarithmic rate of the Eulerian logarithmic strain $\mathbf{h}$ is identical with the strain rate tensor $\mathbf{D}$, by which a hypoelastic model can be exactly integrated to a hyperelastic finite strain model \cite{xiao1997}. This unique relationship between logarithmic strain $\mathbf{h}$ and the strain rate tensor $\mathbf{D}$ is expressed as follows,
\begin{equation}\label{eq:Log_strain_rate}
\mathring{\mathbf{h}}^{log} = \dot{\mathbf{h}}+\mathbf{h} \mathbf{ \Omega}^{log}-\mathbf{ \Omega}^{log}\mathbf{h}= \mathbf{D}
\end{equation}
where $ \mathbf{ \Omega}^{log} $ is called the logarithmic spin introduced by \cite{xiao1997} defined as,
\begin{equation}\label{eq:Log_spin}
\mathbf{\Omega}^{log} = \mathbf{W}+ \sum_{i \neq j}^{n}  \big(\frac{1+(\lambda_{i}/\lambda_{j})}{1-(\lambda_{i}/\lambda_{j})}+\frac{2}{\ln (\lambda_{i}/\lambda_{j})}\big) \mathbf{b}_i \mathbf{D} \mathbf{b}_j
\end{equation}
in which $\lambda_{i,j} (i,j=1,2,3) $ are the eigenvalues of left Cauchy-Green tensor $ \mathbf{B} $ and $ \mathbf{b}_{i}, \mathbf{b}_{j} $ are the corresponding subordinate eigenprojections. As along as the logarithmic spin tensor $\mathbf{\Omega}^{log}$ is defined, the second order rotation tensor $\mathbf{R}^{log}$, associated with $\mathbf{\Omega}^{log}$, can be determined through the following differential equation (\ref{eq:R_def}). In general cases, the initial condition is assumed as $\mathbf{R}^{log}|_{t=0}= \mathbf I$.
\begin{equation}\label{eq:R_def}
{\mathbf{\Omega}^{log}}=\dot{\mathbf{R}}^{log}(\mathbf{R}^{log})^T
\end{equation} 
follow the corotational integration definition from \cite{khan1995continuum}, and assume the initial conditions $\mathbf{h}|_{t=0}= \mathbf 0$, equation (\ref{eq:Log_strain_rate}) yields the total logarithmic strain $\mathbf{h}$ after the logarithmic corotational integration, 
\begin{equation}\label{eq:h-D}
\mathbf{h} = \int_{\text{corot.}}\mathbf{D} ~\text{d}t =(\mathbf{R}^{log})^T \bigg (\int_{0}^{t} \mathbf{R}^{log}\mathbf{D}^{e} (\mathbf{R}^{log})^T\text{d}t' \bigg) \mathbf{R}^{log}  
\end{equation}
%

%%%%%%%%%%%%%%%%%%%%%%%%%%%%%%%%%%%%%%%%%%%%%%%%%%%%%%
\subsection{Additive decomposition of logarithmic strain }\label{AdditiveStrain}
The kinematic assumption starts with the additive decomposition of the strain rate tensor $\mathbf D$ into an elastic part $\mathbf{D}^{e}$ plus a transformation part $\mathbf{D}^{tr}$,
\begin{equation}\label{eq:add_D}
\mathbf{D}=\mathbf{D}^{e}+\mathbf{D}^{tr}
\end{equation}

The elastic strain rate part $\mathbf{D}^{e}$ and the transformation strain rate part  $\mathbf{D}^{tr}$ in equation (\ref{eq:add_D}) can be rewritten as $\mathring{\mathbf{h}}^{e\_log}$ and $\mathring{\mathbf{h}}^{tr\_log}$ by virtue of the relation in equation (\ref{eq:Log_strain_rate}) respectively,
\begin{equation}\label{eq:add_h_rate1}
\mathring{\mathbf{h}}^{e\_log}=\mathbf{D}^{e};~~\mathring{\mathbf{h}}^{tr\_log}=\mathbf{D}^{tr}
\end{equation}

By combining equations (\ref{eq:Log_strain_rate}), (\ref{eq:add_D}) and (\ref{eq:add_h_rate1}), the following equation can be obtained,
\begin{equation}\label{eq:add_h_rate2}
\mathring{\mathbf{h}}^{log}=\mathring{\mathbf{h}}^{e\_log}+\mathring{\mathbf{h}}^{tr\_log}
\end{equation}

Similar to the results obtained from equation (\ref{eq:h-D}), equation (\ref{eq:add_h_rate2}) can yield the following relation after applying the logarithmic corotational integration,
\begin{subequations}
	\begin{align}
	\mathbf{h}^{e}  &= \int_{\text{corot.}}\mathbf{D}^{e} ~\text{d}t =(\mathbf{R}^{log})^T \bigg(\int_0^t \mathbf{R}^{log}\mathbf{D}^{e} (\mathbf{R}^{log})^T\text{d}t' \bigg ) \mathbf{R}^{log}  \\
	\mathbf{h}^{tr} &= \int_{\text{corot.}}\mathbf{D}^{tr} ~\text{d}t =(\mathbf{R}^{log})^T \bigg (\int_0^t \mathbf{R}^{log}\mathbf{D}^{tr} (\mathbf{R}^{log})^T\text{d}t' \bigg ) \mathbf{R}^{log}
	\end{align}
\end{subequations}\label{eq:h_coro}

Based on the additive decomposition on the strain rate tensor, combing equations (\ref{eq:add_D}), (\ref{eq:add_h_rate2}) and (\ref{eq:h_coro}), the following additive decomposition on the total logarithmic strain $\mathbf{h}$ can be achieved, {i.e.}, the total logarithmic strain $\mathbf{h}$ can be additively split into an elastic strain like part $\mathbf{h}^{e}$ plus a transformation strain like part $\mathbf{h}^{tr}$. 
\begin{equation}\label{eq:add_h}
\mathbf{h}=\mathbf{h}^{e}+\mathbf{h}^{tr}
\end{equation}

%%%%%%%%%%%%%%%%%%%%%%%%%%%%%%%%%%%%%%%%%%%%%%%%%%%%%%%%%%%
%%%%%%%%%%%%%%%%%%%%%%%%%%%%%%%%%%%%%%%%%%%%%%%%%%%%%%%%%%%
%%%%%%%%%%%%%%%%%%%%%%%%%%%%%%%%%%%%%%%%%%%%%%%%%%%%%%%%%%%
\section{Model Formulation} \label{sec:framework}
\subsection{Thermodynamic framework}
The Gibbs free energy potential $ G $ is defined to be a continuous function dependent on Kirchhoff stress tensor $ \bm{\uptau}$ \footnote{The relationship between Kirchhoff stress $\bm\uptau $ and Cauchy stress $\bm\sigma$ is $\bm{\uptau}= J\bm\sigma $, where $J$ is the determinant of the deformation gradient $\mathbf{F}$, {i.e.}, $J=\text{det}|\mathbf{F}|$. {Assuming phase transformation to be volume preserving, $J$ is approximately equivalent to 1, so $\bm{\uptau} \approx \bm\sigma $}.	Kirchhoff stress $\bm\uptau $ and Eulerian logarithmic strain $\mathbf{h}$, called an energetic conjugate pair \cite{xiao1998objective}, are usually paired up in the formation of free energy potentials.}, Eulerian logarithmic strain $ \mathbf{h} $, temperature $ T $ and a set of internal state variables $ \Upsilon $.
\begin{equation}\label{eq:GIBBS}
G(\bm{\uptau},\mathbf{h},T,\bm{\Upsilon}) = u - \dfrac{1}{\rho_{0}} \bm{\uptau} : \mathbf{h} - sT
\end{equation}
where $\rho_{0}$ is the density in the reference configuration, $ s $ and $ u $ are the specific entropy and internal energy respectively. 
From the $ 2^{nd} $ law of thermodynamics, the dissipation energy $ \mathcal{D} $ can be written in the form of Clausius-Duhem inequality, 
\begin{equation}\label{eq:Dissipation}
\mathcal{D} = \mathbf{\bm{\uptau}:D} - \rho_0 (\dot{u} - T\dot{s}) \geqslant 0
\end{equation}

The logarithmic rate of the Gibbs free energy is taken in equation (\ref{eq:GIBBS}). Note that a scalar subjected to an objective rate equals to its conventional time rate, the following equation is derived. An circle hat denotes the logarithmic rate in the following text for brevity.
\begin{equation}\label{eq:GIBBS_LOG}
\mathring{G}^{log}= \dot{u} - \dfrac{1}{\rho_{0}} \mathring{\bm{\uptau}}^{log} : \mathbf{h} - \dfrac{1}{\rho_{0}} \bm{\uptau} : \mathring{\mathbf{h}}^{log} -s\dot{T}-\dot{s}T
\end{equation}
the following equation is obtained after rearrangement of equation (\ref{eq:GIBBS_LOG}), 
\begin{equation}\label{GIBBS_LOG_sb.}
\dot{u} -\dot{s}T = \dot{G} + \dfrac{1}{\rho_{0}} \mathring{\bm{\uptau}}:\mathbf{h} 
+ \dfrac{1}{\rho_{0}} \bm{\uptau} : \mathring{\mathbf{h}} + s\dot{T}
\end{equation}

Substitute equation (\ref{GIBBS_LOG_sb.}) into Clausius-Duhem inequality (\ref{eq:Dissipation}), the dissipation energy is rearranged as the following,
\begin{equation}\label{eq:Dissipation_f}
\mathcal{D} = -\rho_{0}\dot{G}-\rho_{0}s\dot{T}-\mathring{\bm{\uptau}}:\mathbf{h} \geqslant 0
\end{equation}

Recall that the Gibbs free energy $G(\bm{\uptau},T,\bm{\Upsilon}) $ is a continuous function, chain rule differentiation can be applied on the Gibbs free energy with respect to its independent state variables (i.e., Kirchhoff stress $ \bm{\uptau} $, temperature $ T $ and internal state variables $\bm{\Upsilon}$), which gives,
\begin{equation}\label{eq:Chain_Rule}
\mathring{G} = \frac{\partial G}{\partial \bm{\uptau}}:\mathring{\bm{\uptau}}
+\frac{\partial G}{\partial T}\dot{T}+
\frac{\partial G}{\partial \bm{\Upsilon}}:\mathring{\bm{\Upsilon}}
\end{equation}

Substitute equation (\ref{eq:Chain_Rule}) into equation (\ref{eq:Dissipation_f}), the following expression for the dissipation energy $ \mathcal{D} $ is acquired,
\begin{equation}\label{eq:Dissipation_Cons}
\mathcal{D} = -(\rho_{0} \frac{\partial G}{\partial \bm{\uptau}} + \mathbf{h} ) :\mathring{\bm{\uptau}}
-(\rho_{0} \frac{\partial G}{\partial T} + s ) : \dot{T}
-\rho_{0}  \frac{\partial G}{\partial \bm{\Upsilon}} :\mathring{\bm{\Upsilon}} \geqslant 0
\end{equation}

Following the standard Coleman-Noll procedure, all admissible values for $ \mathring{\bm{\uptau}} $, $\dot{T} $ and $ \mathring{\bm{\Upsilon}} $  have to comply with the dissipation inequality (\ref{eq:Dissipation_Cons}) regardless of thermodynamic paths, thereby the constitutive relationships between stress and strain, entropy and temperature can be inferred as,
\begin{equation}\label{eq:h_Cons}
\mathbf{h} =- \rho_{0}\frac{\partial G}{\partial \bm{\uptau}}
\end{equation}
\begin{equation}\label{eq:entropy_Cons}
s =- \rho_{0}\frac{\partial G}{\partial T}
\end{equation}

Substitute equations (\ref{eq:h_Cons}) and (\ref{eq:entropy_Cons}) into equation (\ref{eq:Dissipation_Cons}), the following reduced form of the dissipation inequality is acquired,
\begin{equation}\label{eq:Dissipation_State_V}
-\rho_{0}  \frac{\partial G}{\partial \bm{\Upsilon}} :\mathring{\bm{\Upsilon}} \geqslant 0
\end{equation}

\subsection{Constitutive modeling for SMAs}
\subsubsection{Thermodynamic potential}
The formulation of the proposed model is based on the thermodynamic framework presented in section \ref{sec:framework} and the early SMA model developed by Lagoudas and coworkers \cite{boyd1996,lagoudas2012} for small deformation analysis.  The model is able to predict the pseudoelastic (isothermal) and actuation (isobaric) response under large deformation including both large strain and large rotation. A quadratic Gibbs free energy potential $G$ is introduced in equation (\ref{eq:GIBBS_explicit}), in which Kirchhoff stress tensor $\bm{\uptau}$ and temperature $T$ are chosen as the independent state variables. The martensitic volume fraction $ \xi $ and the second order transformation strain tensor $\mathbf{h}^{tr}$ are chosen as internal state variables $ \mathbf{\Upsilon}=\{ \xi,\mathbf{h}^{tr}\} $ to capture the material response exhibited by polycrystalline SMAs. The Gibbs free energy potential $G$ is employed as follows,
\begin{equation}\label{eq:GIBBS_explicit}
\begin{aligned}
G=  -\dfrac{1}{2 \rho_{0}} \bm{\uptau} : \mathcal{S}\bm{\uptau} - \dfrac{1}{\rho_{0}}  \bm{\uptau} :[~\bm{\alpha}(T-T_0)+\mathbf{h}^{tr}]  +c \Big[(T-T_0)-T\ln (\dfrac{T}{T_0}) \Big] -s_0(T-T_0)+u_0+ \dfrac{1}{\rho_{0}}f(\xi)
\end{aligned}
\end{equation}
where $\mathcal{S}$ is the effective compliance tensor calculated by equation (\ref{eq:S_mix}), $\mathcal{S}^A$ is the compliance tensor for austenitic phase while $\mathcal{S}^M$ is for martensitic phase, and $\Delta\mathcal{S}$ is the phase difference for the compliance tensor. The effective stiffness tensor $\mathcal{C}$ can be gained by taking the inverse of the effective compliance tensor, {i.e.}, $\mathcal{C}=\mathcal{S}^{-1}$. $ \bm{\alpha}$ is the second order thermoelastic expansion tensor, $ c $ is the effective specific heat, $ s_0$ and $ u_0 $ are the effective specific entropy and effective specific internal energy at the reference state. All the aforementioned effective variables are determined from equation (\ref{eq:alpha}) to (\ref{eq:u0}). $ T $ represents the temperature at current state while $ T_0 $ is the temperature at reference state.
\begin{equation}\label{eq:S_mix}
\mathcal{S}(\xi)=\mathcal{S}^A + \xi(\mathcal{S}^M-\mathcal{S}^A)=\mathcal{S}^A + \xi\Delta\mathcal{S}
\end{equation}
\begin{equation}\label{eq:alpha}
\bm{\alpha}(\xi)=\bm{\alpha}^A + \xi(\bm{\alpha}^M-\bm{\alpha}^A)=\bm{\alpha}^A + \xi\Delta\bm{\alpha}
\end{equation}
\begin{equation}\label{eq:c}
{c}(\xi)~= c^A ~+ \xi(c^M-c^A)~~=c^A + \xi\Delta c
\end{equation}
\begin{equation}\label{eq:s0}
{s}_0(\xi)= {s}_0^A + \xi({s}_0^M-{s}_0^A)~~={s}_0^A + \xi\Delta {s}_0
\end{equation}
\begin{equation}\label{eq:u0}
{u}_0(\xi)= {u}_0^A + \xi({u}_0^M-{u}_0^A)~~={u}_0^A + \xi\Delta {u}_0
\end{equation}

{
A smooth hardening function $f(\xi)$ is proposed in equation (\ref{eq:Smooth_hardeing}) to account for the hardening effects in polycrystalline SMAs, such as the plastic strain accumulation after the training procedure, imperfections located at the grain boundary, and nano-precipitates hardening effects, etc.\cite{lagoudas2008}, where three additional intermediate material parameters $a_1,a_2,a_3$ and  four curve fitting parameters $n_1,n_2,n_3,n_4$ are introduced to better treat the smooth transition behaviors at the initiation and completion of phase transformation. }
\begin{equation}\label{eq:Smooth_hardeing}
\begin{aligned}
f(\xi)=   \begin{cases} \dfrac{1}{2} a_1\Big(  \xi  + \frac{\xi^{n_1+1}} {n_1+1}+ \frac{(1-\xi)^{n_2+1}} {n_2+1} \Big)+a_3\xi ~, \; \dot{\xi}>0, \vspace{5pt} \\ 
\dfrac{1}{2} a_2\Big(  \xi  + \frac{\xi^{n_3+1}} {n_3+1}+ \frac{(1-\xi)^{n_4+1}} {n_4+1} \Big)-a_3\xi ~, \; \dot{\xi}<0 \end{cases}\\
\end{aligned} 
\end{equation}

Following the standard Coleman-Noll procedure described in section \ref{sec:framework}, the explicit form for constitutive relation (\ref{eq:h_Cons}) between stress and strain is derived as follows. Note that the nonlinearity in this constitutive relation is implied by the transformation strain $\mathbf{h}^{tr}$.    
\begin{equation}\label{eq:h_Cons_f}
\mathbf{h} = - \rho_{0}\frac{\partial G}{\partial \bm\uptau}=\mathcal{S}{\bm\uptau}+\bm\alpha(T-T_0)+ \mathbf{h}^{tr}
\end{equation}   
the explicit form for constitutive relation (\ref{eq:entropy_Cons}) between entropy $s$ and temperature $T$ can also be derived as,
\begin{equation}\label{eq:entropy_Cons_f}
s =- \rho_{0}\frac{\partial G}{\partial T}=\dfrac{1}{\rho_{0}} \bm{\uptau}:\bm\alpha+c\ln (\dfrac{T}{T_0}) + s_0
\end{equation}
the reduced form of the dissipation inequality (\ref{eq:Dissipation_State_V}) can be rewritten in terms of the chosen internal state variables $ \bm{\Upsilon}=\{ \xi,\mathbf{h}^{tr}\} $ as,
\begin{equation}\label{eq:Dissipation_strict_V2}
-\rho_{0}  \frac{\partial G}{\partial \mathbf{h}^{tr}} :\mathring{\mathbf{h}}^{tr} -\rho_{0}  \frac{\partial G}{\partial \xi}\dot{\xi} \geqslant 0
\end{equation}
%
%\newline

\subsubsection{Evolution equation of internal state variables}\label{sec:Evolution}
The evolution equation for the internal state variables $ \bm{\Upsilon}=\{ \xi,\mathbf{h}^{tr}\} $ is presented here. It is proposed that the logarithmic rate of the transformation strain $\mathbf{h}^{tr} $ is proportional to the rate change of the martensitic volume fraction $\xi$. This proportional evolution rule is adopted by following the principle of maximum dissipation such that among all the admissible thermodynamic paths, the one dissipating the most energy is chosen during the SMAs phase transformation process \cite{qidwai2000}. The idea of maximum dissipation for inelastic materials is not new, it was also widely employed for plastic deformed materials to derive the associated flow rule \cite{hill1948}. It is worth pointing out that the rate applied on the transformation strain is the logarithmic rate rather than the conventional time rate. The explicit evolution rule is as follows,
\begin{equation}\label{eq:Trans_Evol}
{\mathring{\mathbf h}}^{\textit{tr}}= \bm{\Lambda}  \dot{\xi},  \ \  \bm\Lambda=\begin{cases}\bm{\Lambda}^{\textit{fwd}}, \; \dot{\xi}>0, \vspace{5pt} \\ \bm{\Lambda}^{\textit{rev}}, \; ~\dot{\xi}<0, \end{cases}\\
\end{equation}
where $\bm\Lambda^{\textit{fwd}}$ is called the forward transformation direction tensor and $\bm\Lambda^{\textit{rev}}$ is called the reverse transformation direction tensor. They are defined as,
\begin{equation}\label{eq:direction}
\bm\Lambda^{\textit{fwd}}=
\frac{3}{2} H^{\textit{cur}} 
\frac{\bm{\uptau}^{'}}{\bar{\bm\uptau}},  \ \bm\Lambda^{rev}=
\frac{\mathbf h^{\textit{tr-r}}}{{\xi}^{\textit{r}}}.
\end{equation}
in which, $ \bm{\uptau}^{'} $ is the deviatoric part of Kirchhoff stress tensor calculated by {\small $ \bm{\uptau}^{'} =\bm{\uptau} -{\small \frac{1}{3}}\textrm{tr}(\bm{\uptau})~\mathbf{1} $}, $\mathbf{1} $ is the second order identity tensor. The effective Mises equivalent stress is given by $ \bar{\bm\uptau} ={\small \sqrt{{{\small \frac{3}{2}}\bm\uptau}^{'}:\bm{\uptau}^{'}}}$. $\mathbf h^{\textit{tr-r}}$ and ${\xi}^{r}$ represent the transformation strain value and martensitic volume fraction at the reverse transformation starting point. { It is common among available SMA models that the magnitude of the inelastic recoverable transformation strain is the same for full transformation under any applied stress levels. This is true when the stress levels is high enough to generate maximum oriented martensitic variants. However, if the applied stress level is not sufficiently high, self-accommodated martensitic variants will be generated. This renders the value of transformation strain less than it is in the high stress level case (i.e., the stress dependency of the magnitude of the transformation strain). Therefore, an exponential function $H^{\textit{cur}}$ dependent on current stress levels is introduced to calculate the current transformation strain as shown in equation (\ref{eq:Hcur}), where $H^{\textit{max}}$ is the maximum (or saturated) transformation strain and $\textit{k}_t$ is a curve fitting material parameter.}
\begin{equation}\label{eq:Hcur}
H^{\textit{cur}}(\bm\uptau)= H^{\textit{max}}(1-e^{-\textit{k}_t { \bar{\bm\uptau}}})
\end{equation}

\subsubsection{Transformation function}\label{Trans_Func}
The objective in this part is to define a proper transformation criterion to determine the occurrence of the phase transformation. Recall the reduced form for dissipation energy is given by inequility (\ref{eq:Dissipation_strict_V2}) and the relation between $\mathbf{h}^{tr}$ and $\xi$ is defined through evolution equation (\ref{eq:Trans_Evol}). Substituting evolution equation (\ref{eq:Trans_Evol}) into reduced form dissipation inequality (\ref{eq:Dissipation_strict_V2}), the following equation is obtained,
\begin{equation}\label{eq:Dissipation_xi}
(\bm\uptau:\bm\Lambda-\rho_{0}  \frac{\partial G}{\partial \xi})\dot{\xi}=\pi\dot{\xi}\geqslant 0 
\end{equation}
the above equation implies that all the dissipation energy is directly a result of the change in the martensitic volume fraction. Based upon this, a scalar variable $\pi$,  called the thermodynamic driving force conjugated to the martensitic volume fraction $\xi$, can thus be defined. Substitution of Gibbs free energy potential $G$ in equation (\ref{eq:GIBBS_explicit}) into equation (\ref{eq:Dissipation_xi}) yields the explicit expression for $ \pi $ as follows,
\begin{equation}\label{eq:Driving_Force}
\begin{aligned}
\pi(\bm\uptau,T,\xi)=\bm\uptau:\bm\Lambda+
\dfrac{1}{2}\bm\uptau:{\Delta}\mathcal{S}\bm\uptau+\bm\uptau:{\Delta}\bm{\alpha}(T-T_0)-\rho_0\Delta c  
\big[ T-T_0 -T\ln(\dfrac{T}{T_0}) \big ] + \rho_0\Delta s_0 T - \rho_0\Delta u_0 - \frac{\partial f(\xi)}{\partial \xi}
\end{aligned}
\end{equation}
where $\Delta \mathcal{S},\Delta \mathbf{\alpha}, \Delta c, \Delta s_0$, and $ \Delta u_0 $ are the phase differences on compliance tensor, thermal expansion tensor, specific heat, reference entropy and reference internal energy, respectively. It can be observed that the thermodynamic driving force $\pi$ is a function of Kirchhoff stress $\bm{\uptau}$, temperature $T$ and martenstic volume fraction $\xi$. This indicates that the phase transformation process can be activated by two independent sources, namely either the stress or temperature, which correlates quite well with the experimentally observed stress-induced and thermally-induced phase transformations in SMAs. To proceed to the goal of defining a transformation criteria, it is assumed that whenever the thermodynamic driving force $\pi$ reaches a critical value $Y $ ($ -Y $), the forward (reverse) phase transformation takes place. Therefore a transformation function $\Phi$ can be defined as the transformation criteria to determine the transformation occurrence as follows,
\begin{equation}\label{eq:Transfor_Fun}
\normalfont{\Phi}=\begin{cases}~~\pi - Y, \; \dot{\xi}>0, \vspace{5pt} \\ -\pi - Y, \; \dot{\xi}<0, \end{cases}\\
\end{equation}

{In the infinitesimal strain theory based SMA model \cite{lagoudas2012}, a reference critical value $Y_0$ and an additional parameter $D$ were introduced into $Y$, through which the thermodynamical critical value $Y$ becomes a function dependent on applied stress levels, see in equation (\ref{Critical_Y}). Such treatment let the model consider the effect of applied stress levels on the size of hysteresis loop. This capability is provided through capturing the different slopes $C_A, C_M$ in the effective stress-temperature phase diagram. The explicit derivation is provided from equation (\ref{eq:Reduced_5_Parameters}) to equation (\ref{eq:Reduced_D}) at the model calibration part in \ref{sec:calibration}.}
\begin{equation}\label{Critical_Y}
Y(\bm{\uptau}) = \begin{cases}Y_0 + D\bm\uptau:\bm\Lambda^{\textit{fwd}}, \; \dot{\xi}>0, \vspace{5pt} \\ Y_0 + D\bm\uptau:\bm\Lambda^{\textit{rev}}, \; \dot{\xi}<0, \end{cases}\\
\end{equation}

As a consequence of the application of the principle of maximum dissipation \cite{qidwai2000}, the so-called Kuhn-Tucker constraints are placed on the proposed model, which are stated as follows for the forward and reverse cases respectively,  
\begin{equation}\label{eq:Kuhn-Tucker}
\begin{aligned}
\dot{\xi} \geqslant 0; \quad \Phi(\bm\uptau,T,\xi)= ~~\pi - Y \leqslant 0;  \quad  \Phi\dot{\xi}=0;~~~\bf{(A\Rightarrow M)}\\
\dot{\xi} \leqslant 0; \quad \Phi(\bm\uptau,T,\xi)= -\pi - Y \leqslant 0; \quad   \Phi\dot{\xi}=0;~~~ \bf{(M\Rightarrow A)}
\end{aligned}
\end{equation}

\subsection{Consistent tangent stiffness and thermal matrix}\label{sec:Jacobian}
In this section, a detailed derivation of the consistent tangent stiffness matrix and the thermal matrix is provided to complete the proposed model. For most typical displacement-based finite element softwares, such as Abaqus, the consistent tangent matrices are often required to be provided in the UMAT so that the finite element solver can achieve a fast convergence for the global equilibrium equations. Normally, consistent tangent matrices can be expressed in the rate form shown in equation (\ref{eq:Jacobian}), where $\mathcal{L}$ is called the consistent tangent stiffness matrix and $\Theta$ is the consistent thermal matrix.
\begin{equation}\label{eq:Jacobian}
\mathring{\bm{\uptau}}  =  {\mathcal{L}}\mathring{\mathbf h} 
+ \Theta \dot T
\end{equation}
applying the logarithmic rate on constitutive equation (\ref{eq:h_Cons_f}) yields, 
\begin{equation}\label{eq:rate_cons}
\mathring{\bm{\uptau}}  = {\mathcal{C}}~[\mathring {\mathbf h}  -  \bm\alpha\dot T - ({\Delta\mathcal{S}}\bm{\uptau} +  \Delta{\bm\alpha}(T-T_0) + \mathbf\Lambda )\dot \xi~]
\end{equation}
taking chain rule differentiation on the transformation function equation (\ref{eq:Transfor_Fun}) gives,
\begin{equation}\label{eq:rate_TransFun}
\dot{\Phi}  = \partial_{\bm{\uptau}}\Phi:\mathring{\bm\uptau} + \partial_{T}\Phi\dot{T} + \partial_{\xi}\Phi\dot{\xi} = 0  
\end{equation}
substituting equation (\ref{eq:rate_cons}) back into equation (\ref{eq:rate_TransFun}) to eliminate $\mathring{\bm{\uptau}}$ and solving it for $\dot \xi$, the following expression for $\dot \xi$ can be obtained,
{
\begin{equation}\label{eq:rate_xi}
\dot{\xi}  = -\dfrac{\partial_{\bm\uptau}\Phi: \mathcal{C} \mathring {\mathbf h} +(\partial_{T}\Phi- \partial_{\bm\uptau}\Phi: \mathcal{C} \bm \alpha)\dot T}{\partial_{\xi}\Phi- \partial_{\bm\uptau}\Phi: \mathcal{C} (\Delta\bm S\bm\uptau+\bm\Lambda+\Delta{\bm\alpha}(T-T_0))} 
\end{equation}
substituting equation(\ref{eq:rate_xi}) back into the rate form constitutive equation(\ref{eq:rate_cons}) to eliminate $\dot \xi$, considering the phase difference of the thermal expansion coefficients can be ignored for martensite and austenite phase, the final explicit expression corresponding to equation (\ref{eq:Jacobian}) can be obtained as follows, }
\begin{equation}\label{eq:Jacobian_exp}
\mathring{\bm{\uptau}}  = \Big[ \mathcal{C}+\dfrac{[\mathcal{C}({\Delta\mathcal{S}}\bm{\uptau}+ \bm\Lambda )] \otimes  [\mathcal{C} \partial_{\bm\uptau}\Phi]}{\partial_{\xi}\Phi- \partial_{\bm\uptau}\Phi: \mathcal{C}(\Delta\mathcal{S}\bm\uptau+\bm\Lambda)} \Big] \mathring{\mathbf{h}} 
+ \Big[ 
- \mathcal{C}\bm\alpha +
\dfrac{ \mathcal{C}({\Delta\mathcal{S}}\bm{\uptau}+ \bm\Lambda)  (\partial_{T}\Phi - \partial_{\bm\uptau}\Phi: \mathcal{C}\bm\alpha )}
{\partial_{\xi}\Phi- \partial_{\bm\uptau}\Phi: \mathcal{C}(\Delta\mathcal{S}\bm\uptau+\bm\Lambda)} 
\Big] \dot T 
\end{equation}
in which the consistent tangent stiffness matrix ${\mathcal{L}}$ is, 
\begin{equation}
{\mathcal{L}}=\mathcal{C}+\dfrac{[\mathcal{C}({\Delta\mathcal{S}}\bm{\uptau}+ \bm\Lambda ] \otimes  [\mathcal{C} \partial_{ \bm\uptau}\Phi]}{\partial_{\xi}\Phi- \partial_{\bm\uptau}\Phi: \mathcal{C}(\Delta\mathcal{S}\bm\uptau+\bm\Lambda)} 
\end{equation}
and the consistent thermal matrix $\Theta$ is, 
\begin{equation}
\Theta = 
- \mathcal{C}\bm\alpha + 
\dfrac{ \mathcal{C}({\Delta\mathcal{S}}\bm{\uptau}+ \bm\Lambda )  (\partial_{T}\Phi - \partial_{\bm\uptau}\Phi: \mathcal{C}\bm\alpha )}
{\partial_{\xi}\Phi- \partial_{\bm\uptau}\Phi: \mathcal{C}(\Delta\mathcal{S}\bm\uptau+\bm\Lambda)} 
\end{equation}
In order to fully determine the explicit values for $\mathcal{L}$ and ${\Theta}$ during the implementation section for the proposed model, the explicit expressions of the following terms $\partial_{\bm\uptau}\Phi$, $\partial_{\xi}\Phi$, $\partial_{T}\Phi$ used in above equations are derived in \ref{sec:cons_append}.

\begin{figure}[h]
	\centering
	\includegraphics[width=0.9\textwidth]{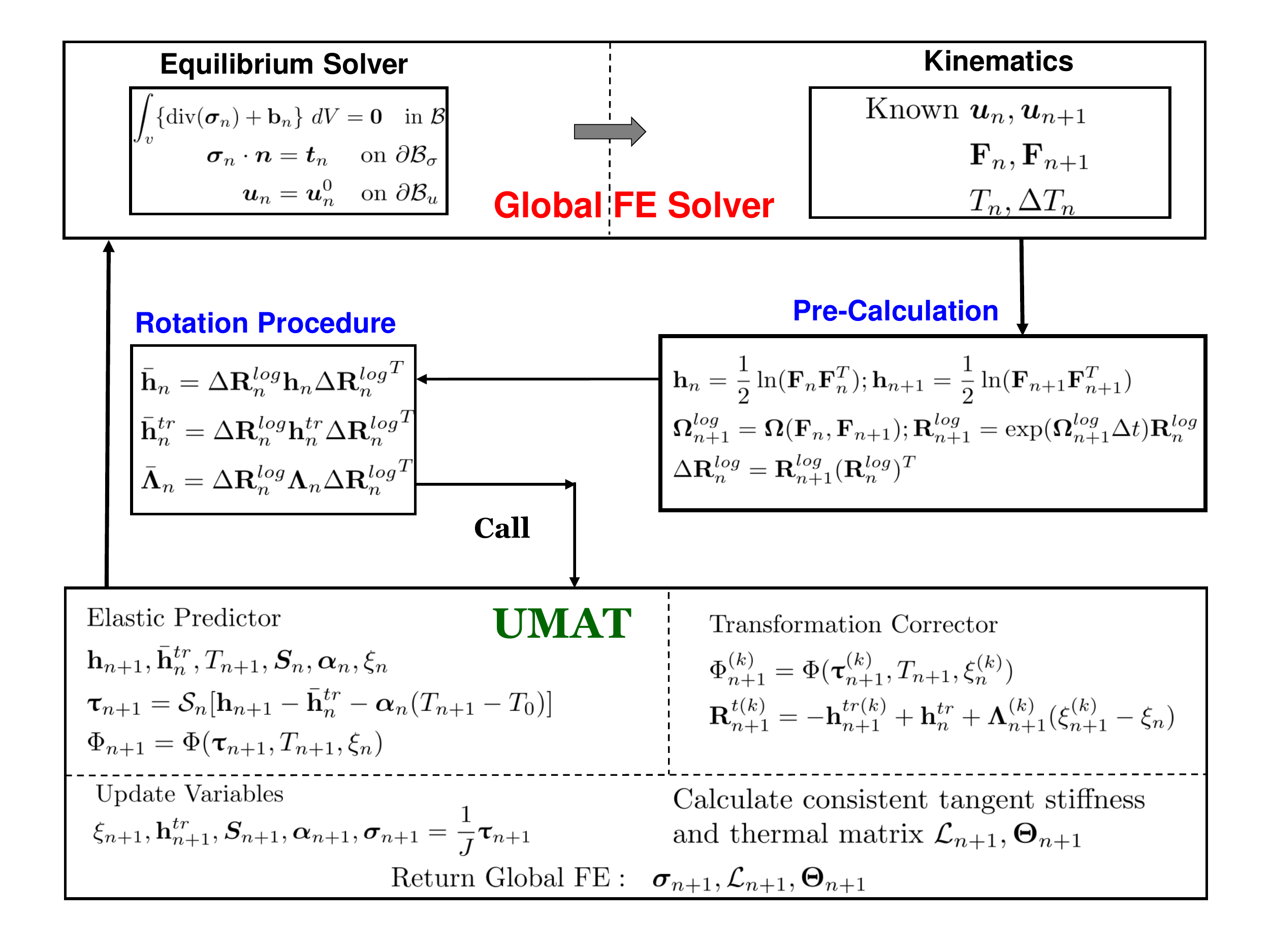}\vspace{-0.2cm}
	\caption{ {Flowchart for the used variables in the proposed model and the UMAT integration with the global FE solver.}}
	\label{fig:Flowchart}
\end{figure}

%%%%%%%%%%%%%%%%%%%%%%%%%%%%%%%%%%%%%%%%%%%%%%%%%%%%%%%%%%%%%%%%%%%%%%%%%
%%%%%%%%%%%%%%%%%%%%%%%%%%%%%%%%%%%%%%%%%%%%%%%%%%%%%%%%%%%%%%%%%%%%%%%%%
%%%%%%%%%%%%%%%%%%%%%%%%%%%%%%%%%%%%%%%%%%%%%%%%%%%%%%%%%%%%%%%%%%%%%%%%%

\section{Numerical implementation}\label{Implementation}
This section focuses on the implementation of the proposed model within finite element (FE) solvers to solve boundary value problems (BVPs). The implementation flowchart is shown in figure \ref{fig:Flowchart}. While typically stress and strain information are provided from FE solver, the initial input information used in this model are only the temperatures $T_n,~\Delta T_n$ and deformation gradients at current step $\mathbf{F}_n$ and next step $\mathbf{F}_{n+1}$. The reason for using only these information is that other tensorial variables have been rotated by the finite element (FE) solver before they are used as inputs, in which the rotation tensor is calculated based on the other non-integrable objective rates. This consequently leads to the artificial stress errors described in section \ref{sec:intro}. During the implementation for the proposed model, a pre-calculation and a rotation procedure are employed before calling the main UMAT subroutine. In the pre-calculation procedure, the logarithmic strain at current step $\mathbf{h}_n$ and next step $\mathbf{h}_{n+1}$ are calculated based on $\mathbf{F}_n$ and $\mathbf{F}_{n+1}$. The incremental rotation tensor $\Delta\mathbf{R}_n^{log}$ based on the logarithmic rate can be calculated by using the exponential map scheme \cite{simo2006,miehe1996,zhu2014}. In the rotation procedure, tensorial variables including $\mathbf{h}_n$, $\mathbf{h}^{tr}_n$ and $\mathbf{\Lambda}_n$ are rotated from step $n$ configuration to the new configuration at step $n+1$ by using the obtained $\Delta\mathbf{R}_n^{log}$, thus, the so-called principle of objectivity is preserved.

The rest of implementation procedure consists of two steps, the first step is called the thermoelastic predictor and the second step is called the transformation corrector. During the initialization of thermoelastic step, the total strain $\mathbf h_{n+1}$ and current temperature $T_{n+1} = T_{n} + \Delta T_{n}$ are provided. The initial internal state variables $\mathbf{\Upsilon}_{n+1}^{(0)}$ are assumed to be the same as $\mathbf{\Upsilon}_{n}$ for the initial Kuhn-Tucker consistency checking, {i.e.}, $\Phi^{(0)}_{n+1} \leqslant 0$. If the initial value of $\Phi^{(0)}_{n+1}$ satisfies the consistency checking, the n+1 step is detected as a thermoelastic response and the UMAT returns to global FE solver for next increment. In case of consistency condition violated, the second step called the transformation corrector is activated to find the updated internal state variables $\mathbf{\Upsilon}_{n+1}^{(k)}$ in order to regain the Kuhn-Tucker consistency. A detailed summary for the implementation procedure is listed in the table \ref{table:implementation}.

\subsection{Thermoelastic Prediction}
Take the $(n+1)^{th}$ step as an example to go through the thermoelastic prediction process. The total strain tensor $\mathbf h_{n+1}$ and the temperature $T_{n+1}$ are provided from Pre-Calculation procedure, and the initial internal state variables $\mathbf{\Upsilon}_{n+1}^{(0)}$ are assumed the same as $\mathbf{\Upsilon}_{n}$,
\begin{equation}\label{eq:implement_SDV}
\mathbf{h}^{tr(0)}_{n+1}=\mathbf{h}^{tr}_{n}; ~~~\xi^{(0)}_{n+1}=\xi_{n}
\end{equation}

Based on equation (\ref{eq:implement_SDV}), the initial guess for Kirchhoff stress $\bm\uptau^{(0)}_{n+1}$ can be calculated through the constitutive equation (\ref{eq:implement_stress}). Here the integer in the upper parenthesis represents that how many iterations have been done during the transformation correction procedure, and integer zero means that this step is just an initial guess in the thermoelastic procedure. The initial calculation for stress $\bm\uptau^{(0)}_{n+1}$ can be obtained,
\begin{equation}\label{eq:implement_stress}
\begin{aligned}
\bm\uptau^{(0)}_{n+1}= \mathcal{C}_{n}\Big[  \mathbf h_{n+1}  - \mathbf{h}^{tr(0)}_{n+1}  - \bm\alpha^{(0)}_{n+1}(T_{n+1}-T_{0})    \Big]  
\end{aligned}
\end{equation}

After the calculation of $\bm\uptau^{(0)}_{n+1}$, the value of transformation function  $\Phi^{(0)}_{n+1}$ can be evaluated based on equations (\ref{eq:Driving_Force}) and (\ref{eq:Transfor_Fun}) for the initial Kuhn-Tucker consistency checking,
\begin{equation}\label{eq:implement_phi}
\begin{aligned}
\Phi^{(0)}_{n+1}= \Phi(\bm\uptau^{(0)}_{n+1},T_{n+1},\mathbf{\Upsilon}^{(0)}_{n+1}) 
\end{aligned}
\end{equation}

If the calculated value of transformation function $\Phi^{(0)}_{n+1}$ remains under the transformation surface (i.e., $\Phi^{(0)}_{n+1}\leqslant\text{'tol'}$, 'tol' is usually set to be $10^{-6}$), step n+1 is detected as a thermoelastic response. Therefore the values of current state variables $\bm\uptau^{(0)}_{n+1}$ and $\mathbf{\Upsilon}^{(0)}_{n+1}$ are accepted as correct and the UMAT proceeds to the global FE solver for the next increment. In case the transformation surface is violated (i.e. $\Phi^{(0)}_{n+1} \geqslant \text{tol}$), the transformation corrector step is activated to find the updated state variables until the consistency equation (\ref{eq:Kuhn-Tucker}) is preserved.   

\subsection{Transformation Correction}
This part addresses the iterative procedures required for the transformation corrector to restore the Kuhn-Tucker consistency. In general, the transformation corrector is nothing but a set of Newton-Raphson iterations on equations (\ref{eq:R1}) and (\ref{eq:R2}) to find the updated internal state variables. Take the $k^{th}$ local iteration for example, the corrector is activated to find a set of $\mathbf{\Upsilon}^{(k)}_{n+1}$ which makes the residual terms $\mathbf R_{n+1}^{tr(k)}$ in equation (\ref{eq:R1}) and transformation function $\Phi^{(k)}_{n+1}$ in equation (\ref{eq:R2}) less than 'tol'. 
\begin{equation}\label{eq:R1}
\mathbf R_{n+1}^{tr(k)} = -{\mathbf{h}}^{tr(k)}_{n+1}+{\mathbf{h}}^{tr}_{n}+\bm\Lambda_{n+1}^{(k)}(\xi_{n+1}^{(k)}-\xi_{n})\\
\end{equation}
\begin{equation}\label{eq:R2}
\Phi^{(k)}_{n+1} ~~= \Phi(\bm{\uptau}^{(k)}_{n+1},T_{n+1},\xi^{(k)}_{n+1})
\end{equation}

This objective is equivalent to the following convergence conditions,
\begin{equation}\label{eq:NR_Criterion}
\begin{aligned}
|\xi_{n+1}^{(k+1)}-\xi_{n+1}^{(k)}|\leqslant \text{tol}~; ~ |{\mathbf{h}}^{tr(k+1)}_{n+1}-{\mathbf{h}}^{tr(k)}_{n+1}|\leqslant\text{tol}~
\end{aligned}
\end{equation}

Use the standard Newton-Raphson procedure \footnote{The explicit expression for the Jacobian matrix during this Newton-Raphson iteration in equation (\ref{eq:Jacob_matrix}) is quite complicated. The symbolic calculation tool in MATLAB is used here to find the Jacobian matrix, and the authors suggest interested readers to utilize this method to perform the tedious calculation.} to solve equations (\ref{eq:R1}) and (\ref{eq:R2}),
\begin{equation}\label{eq:Jacob_matrix}
% ...
\arraycolsep=1.0pt\def\arraystretch{2.3}
\left[\begin{array}{c} \Delta \xi_{n+1}^{(k+1)} \\ {\Delta\mathbf{h}}^{tr(k+1)}_{n+1} \end{array} \right] 
=~-\left[\begin{array}{c} 
\dfrac{\partial \Phi^{(k)}_{n+1}}{\partial \xi} ~~ \dfrac{\partial \Phi^{(k)}_{n+1}}{\partial \mathbf{h}^{tr} } \\ 
\dfrac{\partial \mathbf R_{n+1}^{tr(k)} }{\partial \xi} ~~ \dfrac{\partial \mathbf R_{n+1}^{tr(k)} }{\partial \mathbf{h}^{tr} } \\ 

\end{array} \right]^{-1} 
\left[\begin{array}{c} \Phi^{(k)}_{n+1} \\ \mathbf R_{n+1}^{tr(k)}  \end{array} \right] 
% ...
\end{equation}

The following results on internal state variables at $(k+1)^{th}$ iteration can be obtained,
\begin{equation}\label{eq:Jacob_matrix_rearrange}
% ...
\arraycolsep=1.0pt\def\arraystretch{2.3}
\left[\begin{array}{c} \xi_{n+1}^{(k+1)} \\ {\mathbf{h}}^{tr(k+1)}_{n+1} \end{array} \right] 
=\left[\begin{array}{c} \xi_{n+1}^{(k)} \\ {\mathbf{h}}^{tr(k)}_{n+1}  \end{array} \right] 
+\left[\begin{array}{c} \Delta \xi_{n+1}^{(k+1)} \\ {\Delta\mathbf{h}}^{tr(k+1)}_{n+1} \end{array} \right] 
\end{equation}

Once the converged values of $\{\mathbf{h}^{tr(k+1)}_{n+1},\xi^{(k+1)}_{n+1}\}$ are found, the current transformation corrector step is labeled as finished and the UMAT proceeds to the next increment. Otherwise the Newton-Raphson procedure exits at this step after certain number of iterations and the current finite element increment step stops.\\\\

\begin{minipage}{0.8\textwidth}
	\captionof{table}{The implementation procedure for the proposed finite strain SMA model.} \label{table:implementation}\vspace{-0.2cm}
	\begin{tabular}{lp{0.9\textwidth}}\toprule[0.3mm]
		& 1.\textit{Initialization}
		\vspace{-0.25cm}
		\begin{itemize}
			\itemsep-0.25em 
			\item Conduct pre-calculation and rotation procedures.
			\item $k=0;  \xi^{(0)}_{n+1}=\xi_{n};  \bm{h}^{tr(0)}_{n+1}=\bm{h}^{tr}_{n};$
		\end{itemize}\vspace{-0.5cm}\\	
		& 2.\textit{Thermoelastic Predictor}
		\vspace{-0.25cm}
		\begin{itemize}
			\itemsep-0.25em 
			\item $\bm\uptau^{(0)}_{n+1}= \mathcal{C}^{(0)}_{n+1}[  \bm h_{n+1}  - \bm{h}^{tr(0)}_{n+1}  -\bm\alpha(T_{n+1}-T_{0}) ] $
			\item Calculate $\Phi^{(k)}_{n+1}$.
			\item IF $\Phi^{(0)}_{n+1} \leqslant tol  $, GOTO 4 (thermoelastic response).
			\item IF $\Phi^{(0)}_{n+1} > tol  $, GOTO 3  (transformation happens).
		\end{itemize}\vspace{-0.5cm}\\
		%%%%%%
		%
		& 3.\textit{Transformation Corrector}
		\vspace{-0.23cm}
		\begin{itemize}
			\itemsep-0.12em 
			\item Calculate residual matrix \newline
			$\mathbf R_{n+1}^{tr(k)} = -{\bm{h}}^{tr(k)}_{n+1}+{\bm{h}}^{tr}_{n}+\bm\Lambda_{n+1}^{(k)}(\xi_{n+1}^{(k)}-\xi_{n})$\newline
			$\Phi^{(k)}_{n+1} ~~= \Phi(\bm{\uptau}^{(k)}_{n+1},T_{n+1},\xi^{(k)}_{n+1})$
			\item Perform the Newton-Raphson iterations in equation (\ref{eq:Jacob_matrix}).
			%		\item Calculate   $d\mathbf U_{n+1}^{(k)}$ for $k$ step, (equation (\ref{eq:NR_Criterion})).
			\item Update variables $\xi_{n+1}^{(k+1)},{\bm{h}}^{tr(k+1)}_{n+1},\mathcal{S}^{(k+1)}_{n+1}$\newline
			$\xi_{n+1}^{(k+1)}       \quad =\xi_{n+1}^{(k)}+\Delta\xi_{n+1}^{(k+1)}$\newline
			${\bm{h}}^{tr(k+1)}_{n+1}~={\bm{h}}^{tr(k)}_{n+1}+\Delta{\bm{h}}^{tr(k+1)}_{n+1} $\newline
			$\mathcal{S}^{(k+1)}_{n+1}\quad=\mathcal{S}^{A}+\xi_{n+1}^{(k+1)}\Delta\mathcal{S}$	
			\item IF $\Phi^{(k+1)}_{n+1} \geqslant tol  $, GOTO step 3 for the next local iteration, $k=k+1$.\newline
			ELSE GOTO step 4
		\end{itemize}\vspace{-0.5cm}\\
		%%%%%%
		
		& 4.\textit{Calculate consistent stiffness matrix $\mathcal{L}$ and thermal matrix $\mathbf{\Theta}$}.
		\vspace{-0.2cm}
		\begin{itemize}
			\itemsep-0.3em 
			\item ${\mathcal{L}}=\mathcal{C}+\dfrac{[\mathcal{C}({\Delta\mathcal{S}}\bm{\uptau}+ \bm\Lambda ] \otimes  [\mathcal{C} \partial_{\bm\uptau}\Phi]}{\partial_{\xi}\Phi- \partial_{\bm\uptau}\Phi: \mathcal{C}(\Delta\mathcal{S}\bm\uptau+\bm\Lambda)} $
			\item $\Theta = 
			- \mathcal{C}\bm\alpha + 
			\dfrac{ \mathcal{C}({\Delta\mathcal{S}}\bm{\uptau}+ \bm\Lambda )  (\partial_{T}\Phi - \partial_{\bm\uptau}\Phi: \mathcal{C}\bm\alpha )}
			{\partial_{\xi}\Phi- \partial_{\bm\uptau}\Phi: \mathcal{C}(\Delta\mathcal{S}\bm\uptau+\bm\Lambda)} $	
		\end{itemize}\vspace{-0.3cm}\\
		
		& 5.Exit UMAT and proceed to the global FE solver for the next increment\\	
		\bottomrule[0.3mm]
	\end{tabular}
\end{minipage}\vspace{0.6cm}

%%%%%%%%%%%%%%%%%%%%%%%%%%%%%%%%%%%%%%%%%%%%%%%%%%%%%%%%%%%%%%%%%%%%%%%%%
%%%%%%%%%%%%%%%%%%%%%%%%%%%%%%%%%%%%%%%%%%%%%%%%%%%%%%%%%%%%%%%%%%%%%%%%%
%%%%%%%%%%%%%%%%%%%%%%%%%%%%%%%%%%%%%%%%%%%%%%%%%%%%%%%%%%%%%%%%%%%%%%%%%

\section{Numerical Results}\label{Result}
In this section, the proposed model is used to predict the stress/thermally-induced phase transformations in SMAs subjected to general three-dimensional thermo-mechanical loading. Several numerical examples are presented here to test the capabilities of this model to account for large strains and rotations, and also to resolve the artificial stress errors issue. First, a parametric analysis on a uniaxial SMA bar is studied to show that the proposed model is able to consider the geometry nonlinearity induced by large strains. Second, two BVPs, {i.e.}, an SMA beam and an SMA torque tube subjected to stress-induced phase transformations, are tested as large rotation cases. To show the model is able to resolve the artificial stress errors issue, the cyclic response of the beam and the torque tube are obtained via the proposed model, and the results are compared against the predictions obtained by the Abaqus nonlinear solver\footnote{As the nonlinear solver is activated for implicit analysis (i.e., select NLGEOM on), Abaqus automatically use the logarithmic strain as its strain measure, and the Jaumman rate is the utilized objective rate to account for the large rotation\cite{abaqus2014}.}. Next, an isobaric BVP of an SMA torque tube subjected to varying thermal loading is investigated to predict the thermally-induced phase transformation. In the end, to show the model is able to capture the non-proportional local stress and strain evolutions, a 3-D solid flexible structure undergoing a self-expanding process is studied. The proposed model is anticipated to be further validated against experimental data of NiTi and NiTiHf SMAs under uniaxial and other non-uniform loading conditions. The ultimate objective is to validate the capability of the proposed model to predict the response of SMA-based actuators, such as SMA beams and torque tubes, which are intended to be integrated with the future supersonic transport aircrafts to realize the morphing capabilities to reduce the sonic boom noise.

%%%%%%%%%%%%%%%%%%%%%%%%%%%%%%%%%%%%%%%%%%%%%%%%%%%%%%%%%%%%%%%%%%%%%%%%%%%%%%%%%%%%%%%%%%%%%%%
\subsection{SMA bar under isothermal loading}\label{subsec:bar}
\begin{figure}[t!]%
	\hspace{-0.5cm}
	\subfigure[Load and displacement curve]{%
		\label{fig:LD_curve}
		\includegraphics[width=0.5\textwidth]{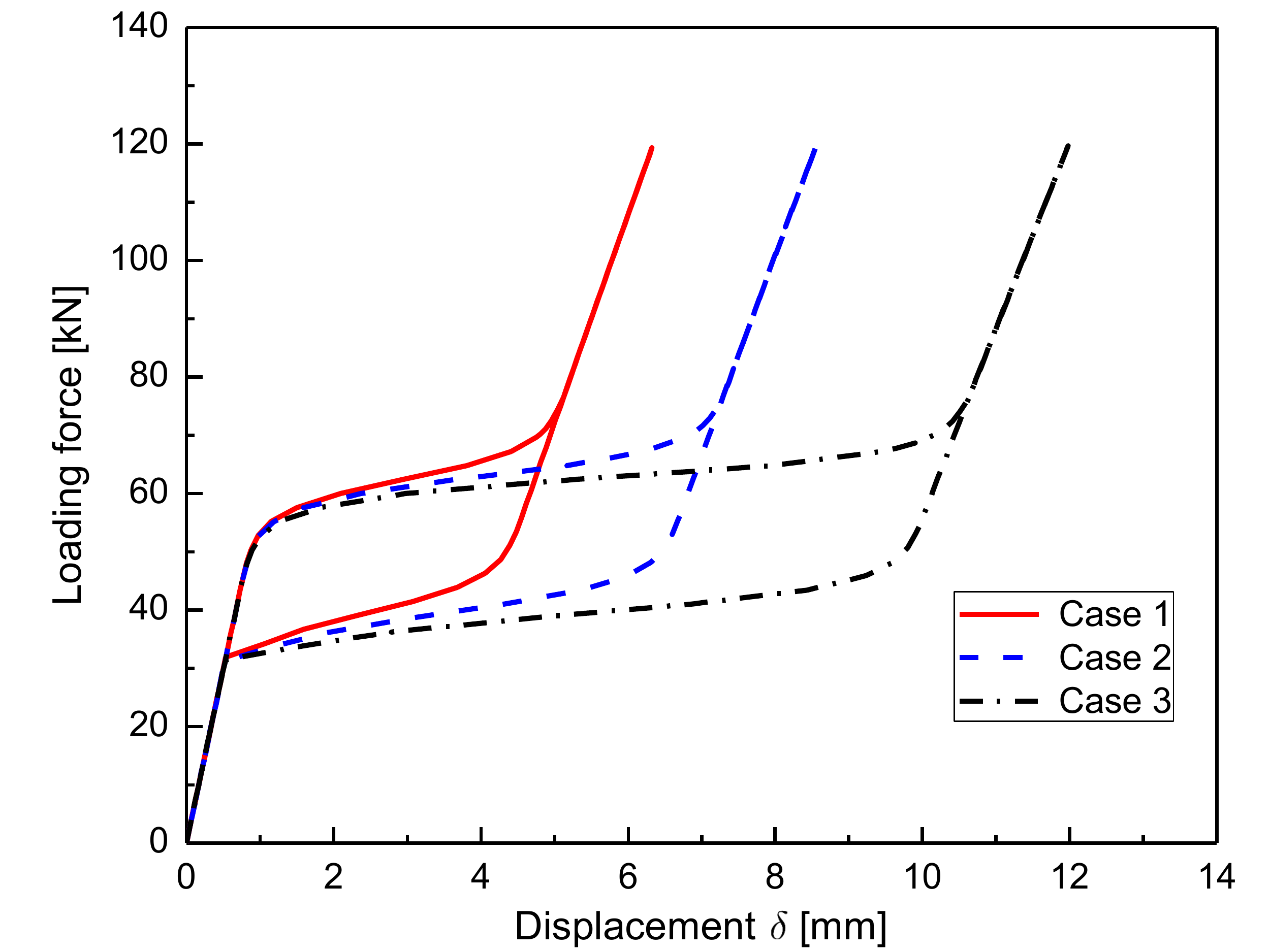}}%
	\subfigure[Stress and strain curve]{%
		\label{fig:SS_curve}%
		\includegraphics[width=0.5\textwidth]{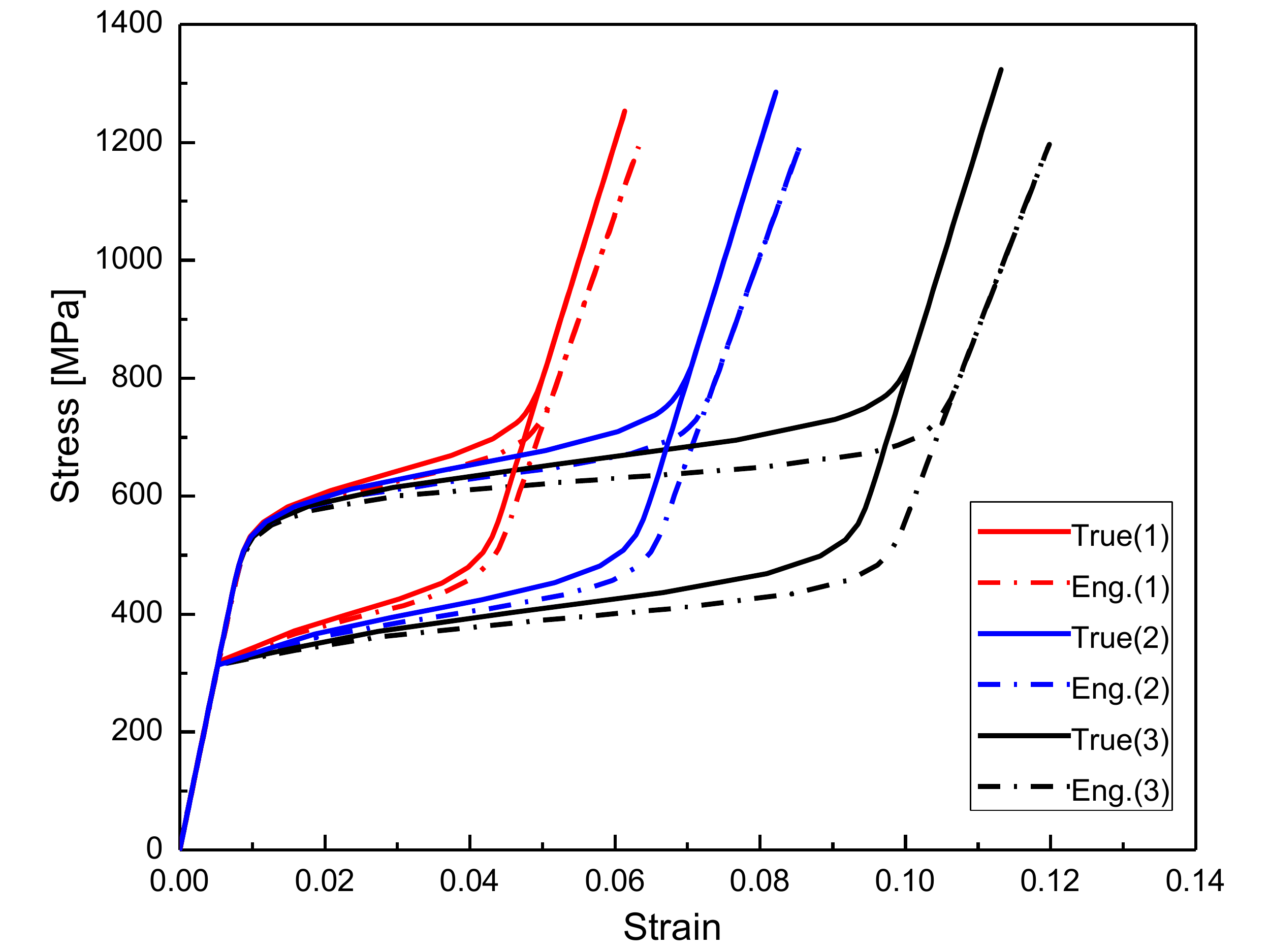}}%
	\caption{Three sets of load-displacement curves are interpreted into the engineering scale stress-strain curve for the calibration of infinitesimal model, and the true stress-strain curve for the calibration of proposed model. (engineering scale is denoted by Eng. and true scale is denoted by True.)}
	\label{fig:Cali_results}
\end{figure}

To test the capability of the proposed model to account for the effects of large strain, an SMA prismatic bar is studied under uniaxial isothermal loading condition. A parametric study is performed with the maximum transformation strain $H^{max}= 3\%, 5\%, 8\%$ to represent three different loading cases. {A group of representative material parameters (two material parameter groups combined)  used in this example are listed in table \ref{tab:MaterialProperty_bar} referenced from \cite{lagoudas2008,lagoudas2012}.} The SMA prismatic bar has a length $L=100$ (mm) and an square cross section with an edge length $ a=10 $ (mm). It is subjected to a proportional force loading up to 120 (kN) then unloading to 0 (kN), the temperature is kept constant at $380$K throughout the process. Generally, the load-displacement curves provided from such uniaxial experiments are interpreted into the engineering scale stress-strain curves to facilitate the model calibration. However, when the materials experience a strain that is no longer considered small, the geometry nonlinearity due to such strain has to be taken into consideration. To demonstrate that the proposed model accounts for this, the calibrated values of elastic modulus ($E_A, E_M$) and maximum transformation strain ($H^{max}$) based on the true stress-strain curve are compared against the values from its infinitesimal counterpart. Three sets of load-displacement curves are generated shown in figure \ref{fig:LD_curve}. They are interpreted into stress-strain curves in two scales, i.e, the true stress (Cauchy stress) versus the true strain (logarithmic strain) curve and the engineering stress (nominal stress) versus engineering strain (infinitesimal strain) curve. By using the calibration procedure described in \ref{sec:calibration}, the calibrated values of $E_A, E_M$ and $H^{max}$ summarized in table \ref{tab:cali_MP}.

{
Table \ref{tab:cali_MP} shows that the values of $E_A$ are identical in both the two scales. However, the values of $E_M$ in engineering scale change from 35.48 GPa to 32.05 GPa, which indicates a material softening. Actually, such material softning is not real. Instead, it is the effect of disregarding the geometric nonlinearity induced by large strain as described previously. In this case, the geometric nonlinearity means that the bar needs to contract its cross section to compensate for its elongation to preserve the volume conservation. Disregarding the change of cross section results in an unreal decreasing on the values of $E_M$. By doing the calibration based on the true stress-strain curve instead of the engineering one, the proposed model is able to exclude the geometry nonlinearity induced by large strain, so that the calibrated values of $E_M$ remain the same in the three loading cases from true scale. Besides, The values of $H^{max}$ are also worth to be noted. Although $H^{max}$ shows different values in the two scales, a relationship exists between the true scale $H^{max}$ and the engineering scale $H^{max}_{\text{eng}}$, {i.e.}, $H^{max}=\ln(1+H^{max}_{\text{eng}})$. Based on the results from this parametric study, it is shown that the infinitesimal strain assumption may no longer be considered as an accurate approximation when the strain regime is beyond 3\%. In order to account for the effects caused by large strain, a finite strain model to consider the geometry nonlinearity is required even in a uniaxial case.}

\begin{table}[t!] 
	\centering
	\caption{ {A set of representative material parameters used for the parametric numerical study \cite{lagoudas2008,lagoudas2012}.}}\vspace{-0.2cm}
	\renewcommand{\arraystretch}{1}
	\begin{tabular}{c|lr|ll} \toprule
		Type                         &Parameter                        & Value                                   &Parameter            & Value  \\                                       \midrule
		&$E_A$                            & 60   [GPa]                              & $C_A$                & 8  [MPa/K]\\
		&$E_M$                            & 40   [GPa]                              & $C_M$               & 6  [MPa/K]\\
		Key material parameters       &$\nu_A=\nu_M$                          & 0.3~~~~~~~~                              & $M_s$                & 333  [K]\\
		12                      &$\alpha_A=\alpha_M$    &    1.0$\times$10$^{-5}$ [K$^{-1}$]                           & $M_f$                &220  [K]\\
		& $ H^\textit{max}$                        & 3\%, 5\%, 8\%          & $A_s$                 &274  [K]\\
		& $k_t$                      &    0.02   & $A_f $                 & 370  [K]\\                                   \midrule
		
		Smooth hardening parameters 		&  $n_1$         &   0.5                &  $n_3$             & 0.5 \\
		4							   &  $n_2$         &   0.5                &  $n_4$        	 & 0.5 \\                                   
		\bottomrule
	\end{tabular}
	\label{tab:MaterialProperty_bar}
\end{table}

\begin{table}[t!]
	\centering
	\caption{Elastic modulus and transformation strain calibrated based on engineering and true scale stress-strain curves.}\vspace{-0.2cm}
	\begin{tabular}{@{}llccc|lcccc@{}}
		\toprule
		&Engineering scale    & Case 1  & Case 2  & Case 3 &~~~True scale    & Case 1  & Case 2  & Case 3 &\\ \midrule
		&~~~~~~$E_A$~[GPa] &60.00 & 60.00 &60.00 &~~~$E_A$~[GPa] &60.00 & 60.00 &60.00 &\\ 
		&~~~~~~$E_M$[GPa] &35.48 & 34.06 & 32.05	&~~~$E_M$[GPa] &40.00  &40.00  &40.00&\\ 
		&~~~~~~$H^{max}_{\text{eng}}$ &$3.04\%$  &$5.13\%$ &$8.33\%$&~~~~~$H^{max}$ &$3.00\%$  &$5.00\%$ &$8.00\%$&\\ 
		\bottomrule
	\end{tabular}\label{tab:cali_MP}
\end{table}

%%%%%%%%%%%%%%%%%%%%%%%%%%%%%%%%%%%%%%%%%%%%%%%%%%%%%%%%%%%%%%%%%%%%%%%%%%%%%%%%%%%%%%%%%%%%%%%
\subsection{SMA beam under isothermal loading}
\begin{figure}[H]
	\centering\hspace*{-1.0cm}
	\includegraphics[width=0.5\textwidth]{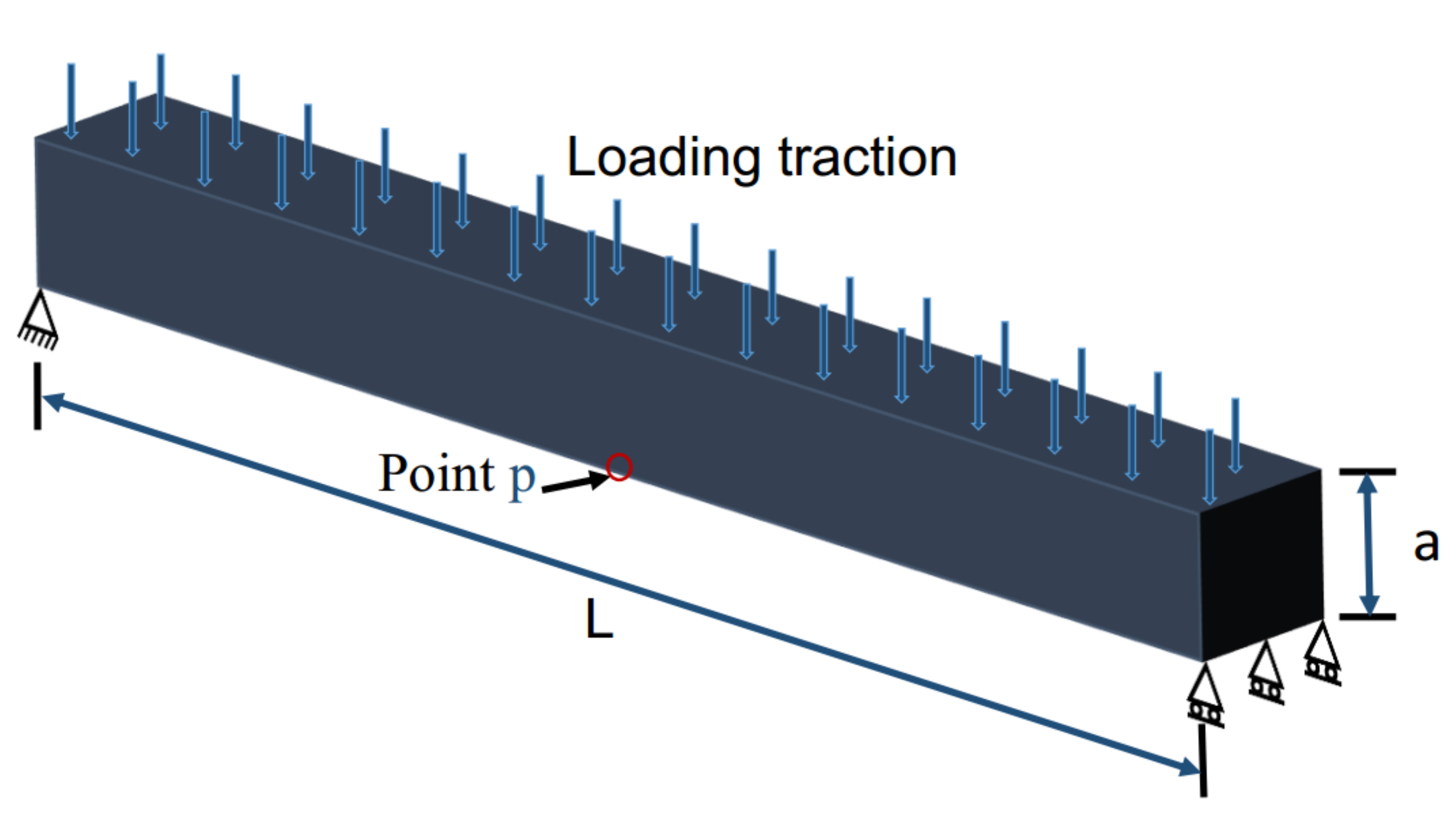}
	\caption{Schematic for the SMA beam subjected to isothermal bending load condition at constant temperature 306 K}
	\label{fig:Beam_Schrmatic}
\end{figure}

The second BVP considered here is an SMA beam subjected to isothermal loading shown in figure \ref{fig:Beam_Schrmatic}. The SMA beam component has been investigated as bending actuators in \cite{hartl2007aerospace} to realize a morphing variable-geometry chevron in order to change the outer engine shell shape to achieve specific aerodynamic characteristics. While only one loading cycle was considered in the previous study, this example examines the cyclic material and structural response. The studied beam has the same geometry as the SMA bar in section \ref{subsec:bar}. Refer to figure \ref{fig:Beam_Schrmatic}, the beam is simply supported with one node being fixed to suppress the rigid body motion, and the upper face is subjected to a traction that ramps up to 24 (MPa) then decreases to 0 (MPa). Temperature is kept constant at $380$K throughout the whole numerical experiment. Material parameters used in this simulation are summarized in table \ref{tab:MaterialProperty_bar2}. The cyclic material and structural response are obtained by the proposed model for a material point p (in figure \ref{fig:Beam_Schrmatic}) located at the middle bottom position, and are compared against the results obtained from the Abaqus nonlinear solver. 
\begin{table}[H] 
	\centering
	\caption{Calibrated values of material parameters for equiatomic NiTi \cite{lagoudas2012}.}\vspace{-0.2cm}
	\renewcommand{\arraystretch}{1}
	\begin{tabular}{c|lr|ll} \toprule
		Type                         &Parameter                        & Value                                   &Parameter            & Value  \\                                       \midrule
		&$E_A$                            & 60   [GPa]                              & $C_A$                & 7.8  [MPa/K]\\
		&$E_M$                            & 60   [GPa]                              & $C_M$               & 7.3  [MPa/K]\\
		Key material parameters       &$\nu_A=\nu_M$                          & 0.3~~~~~~~~                              & $M_s$                & 333  [K]\\
		12                      &$\alpha_A=\alpha_M$    &    1.0$\times$10$^{-5}$ [K$^{-1}$]                           & $M_f$                &220  [K]\\
		& $ H^\textit{max}$                        & 4.7\%          & $A_s$                 &274  [K]\\
		& $k_t$                      &    0.021   & $A_f $                 & 370  [K]\\                                   \midrule
		
		Smooth hardening parameters 		&  $n_1$         &   0.5                &  $n_3$             & 0.5 \\
		4							   &  $n_2$         &   0.5                &  $n_4$        	 & 0.5 \\                                   
		\bottomrule
	\end{tabular}
	\label{tab:MaterialProperty_bar2}
\end{table}

\begin{figure}[h]%
	\hspace{-0.5cm}
	\subfigure[Stress-strain curve predicted by proposed model]{%
		\label{fig:Beam_Lei_SS}
		\includegraphics[width=0.5\textwidth]{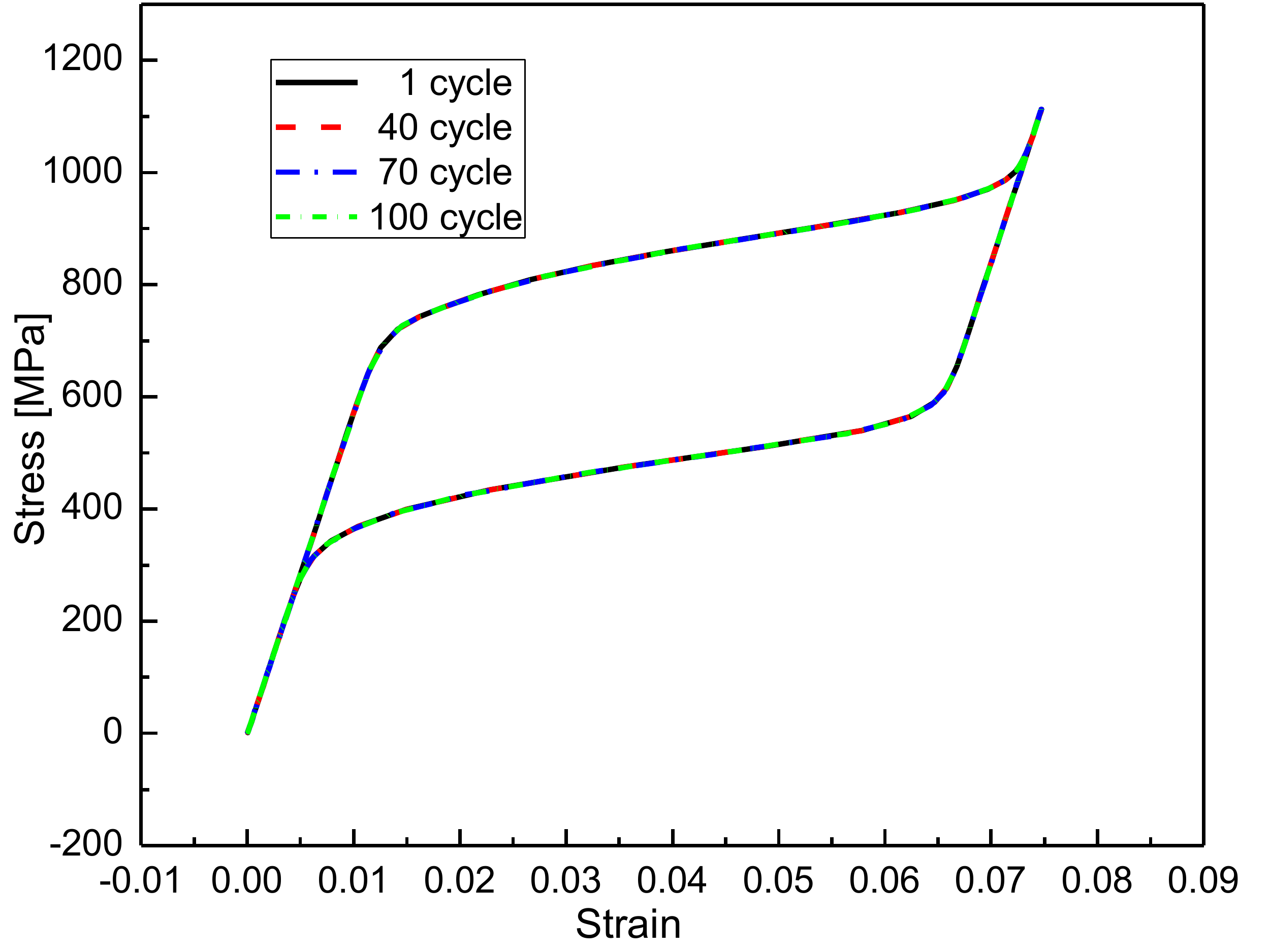}}%
	\subfigure[Stress-strain curve predicted by Abaqus nonlinear solver]{%
		\label{fig:Beam_Ori_SS}%
		\includegraphics[width=0.5\textwidth]{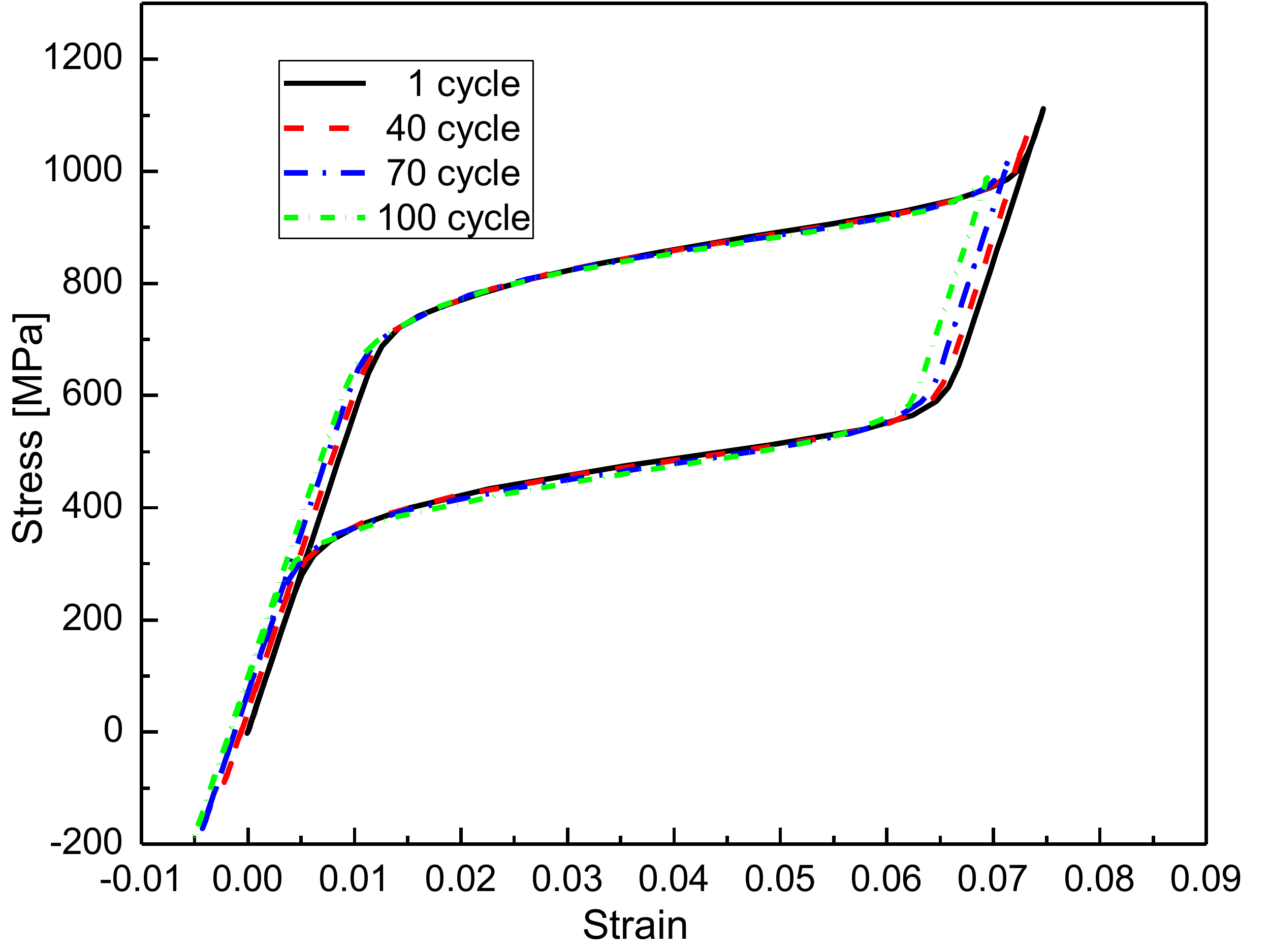}}%
	
	\caption{The cyclic stress-strain response for a bottom surface point under isothermal loading condition.}
	\label{fig:Beam_SS}
\end{figure}

\begin{figure}[H]%
	\hspace{-0.5cm}
	\subfigure[Stress-strain curve predicted by proposed model]{%
		\label{fig:Beam_Lei_SS_top}
		\includegraphics[width=0.5\textwidth]{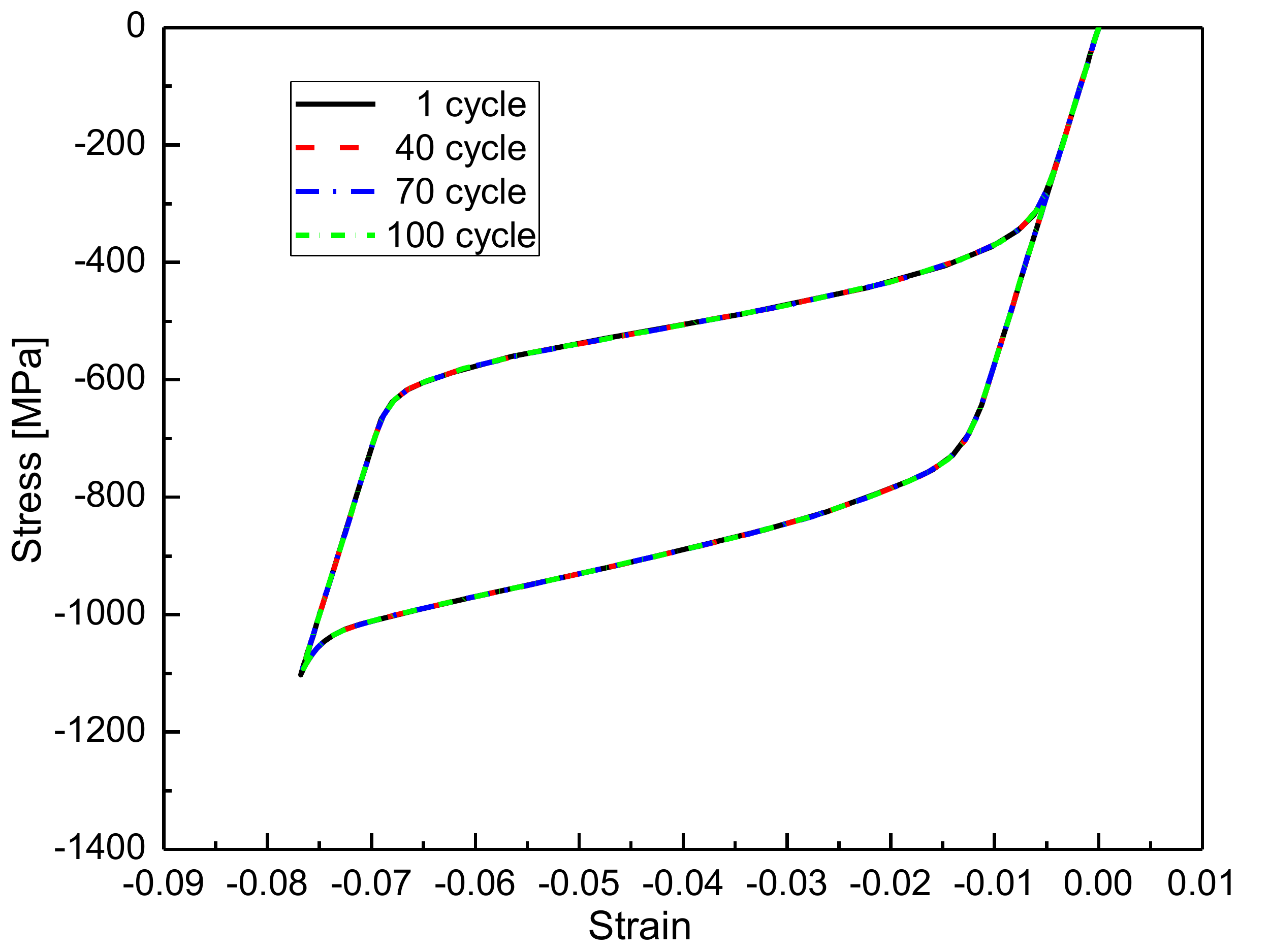}}%
	\subfigure[Stress-strain curve predicted by Abaqus nonlinear solver]{%
		\label{fig:Beam_Ori_SS_top}%
		\includegraphics[width=0.5\textwidth]{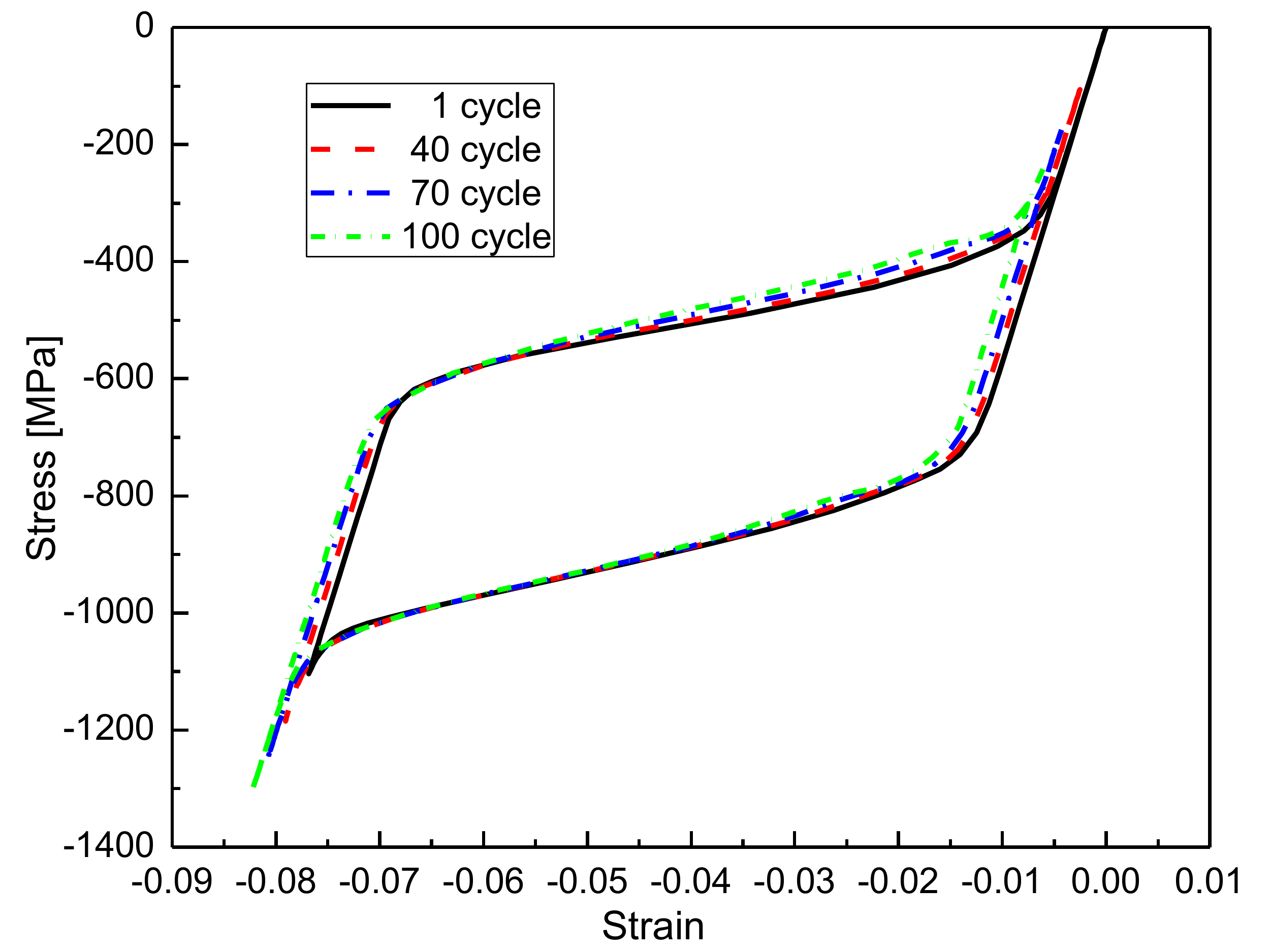}}%
	
	\caption{ {The cyclic stress-strain response for an upper surface point under isothermal loading condition. }}
	\label{fig:Beam_SS_top}
\end{figure}

\begin{figure}[h]%
	\hspace{-0.5cm}
	\subfigure[Load-displacement curve predicted by proposed model]{%
		\label{fig:Beam_Lei_PD}
		\includegraphics[width=0.5\textwidth]{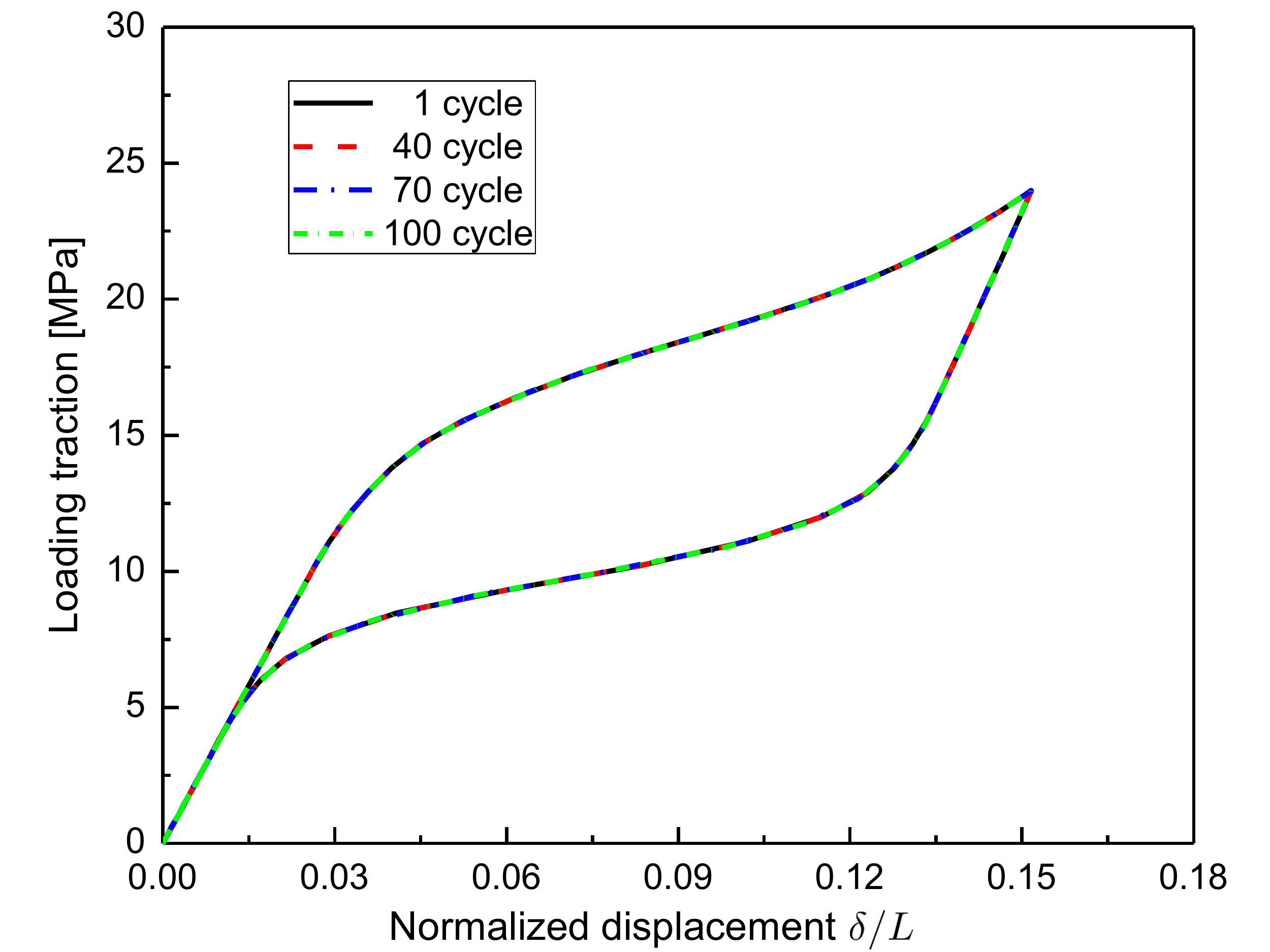}}%
	\subfigure[Load-displacement curve predicted by Abaqus nonlinear solver]{%
		\label{fig:Beam_Ori_PD}%
		\includegraphics[width=0.5\textwidth]{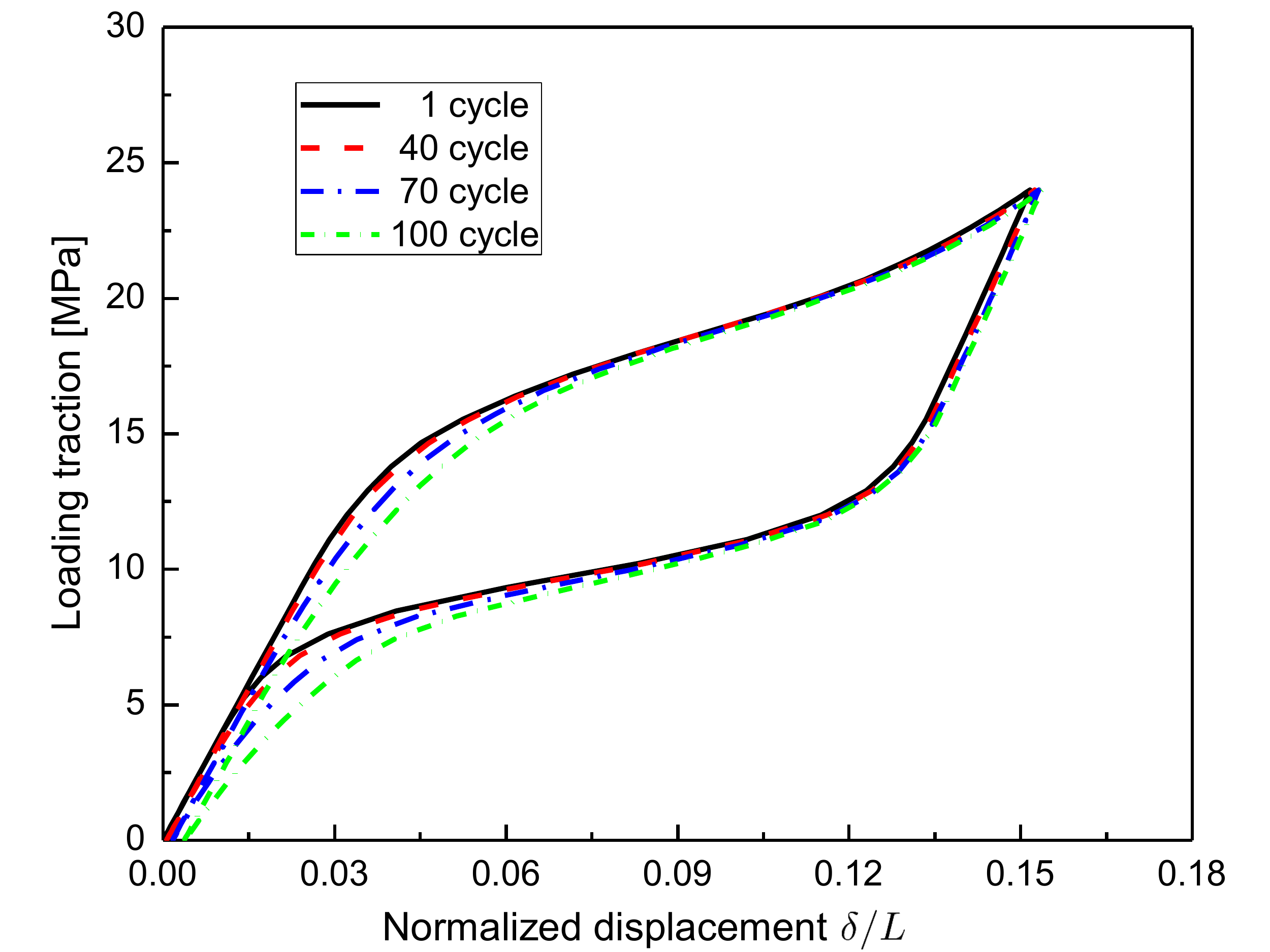}}%
	
	\caption{The cyclic load-displacement response for an SMA beam under isothermal loading condition.}
	\label{fig:Beam_PD}
\end{figure}

{
Figure \ref{fig:Beam_SS} shows the obtained cyclic longitudinal stress-strain curve for material point p under tension. As shown in figure \ref{fig:Beam_Lei_SS}, the proposed model provided a stable material response, while the Abaqus nonlinear solver predicted a shifting, instead of stable, response shown in figure \ref{fig:Beam_Ori_SS}. The observation from these results indicate that the spurious material response is obtained due to the usage of non-integrable objective rates in Abaqus as discussed in the introduction. Although the  initial several loading cycles are almost the same in the results provided by Abaqus nonlinear solver, the accumulation of artificially introduced stress errors, around -2 MPa for each cycle, gradually drifts the material response left downwards throughout the 100 loading cycles. In total, -200 MPa stress residuals together with -0.6\% remnant strains are observed at the end. Such stress errors consist of almost 18\% of the maximum stress levels experienced by material point p. As a comparison, figure \ref{fig:Beam_SS_top} shows 100 stress-strain curves for another material point subjected to compression at the middle of beam upper surface. The result shows an opposite shifting trend in contrast to the results of point p. Again, a stable compressive stress-strain curve are predicted by the proposed model while the Abaqus nonlinear solver predicts a shifting one. In addition, figure \ref{fig:Beam_PD} shows the obtained cyclic load-displacement curves for point p. It can be seen that the proposed model predicted a stable structural response while the Abaqus nonlinear solver predicted a shifting structural response. Based on these results, it is demonstrated that the Abaqus nonlinear solver can no longer produce reliable results for the SMA beam subjected to 100 bending cycles. Therefore, the proposed model with the capability to eliminate the stress errors is required for the SMA beam subjected to cyclic loading.  } 

%Note that 1000 increment steps are specified in this analysis which can be considered small enough to exclude the increment size effect.     

\subsection{SMA tube under isothermal loading}\label{sec:torque_tube_isothermal}
\begin{figure}[H]
	\centering
	%\hspace{-1.5cm}
	\subfigure[Full size torque tube geometry]{
	\includegraphics[width=0.5\textwidth]{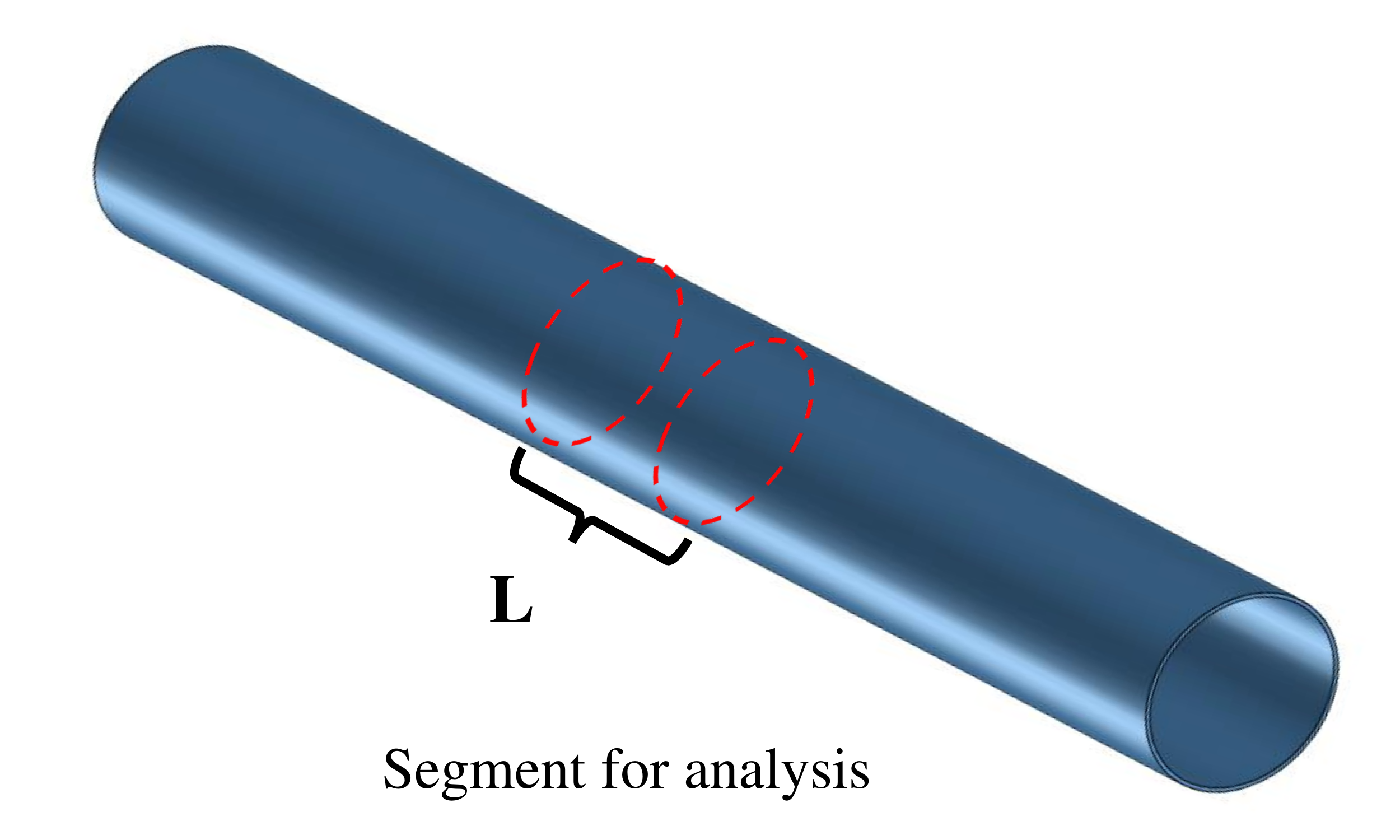}}
	\hspace{1cm}
	\subfigure[Reduced size torque tube geometry]{
	\includegraphics[width=0.35\textwidth]{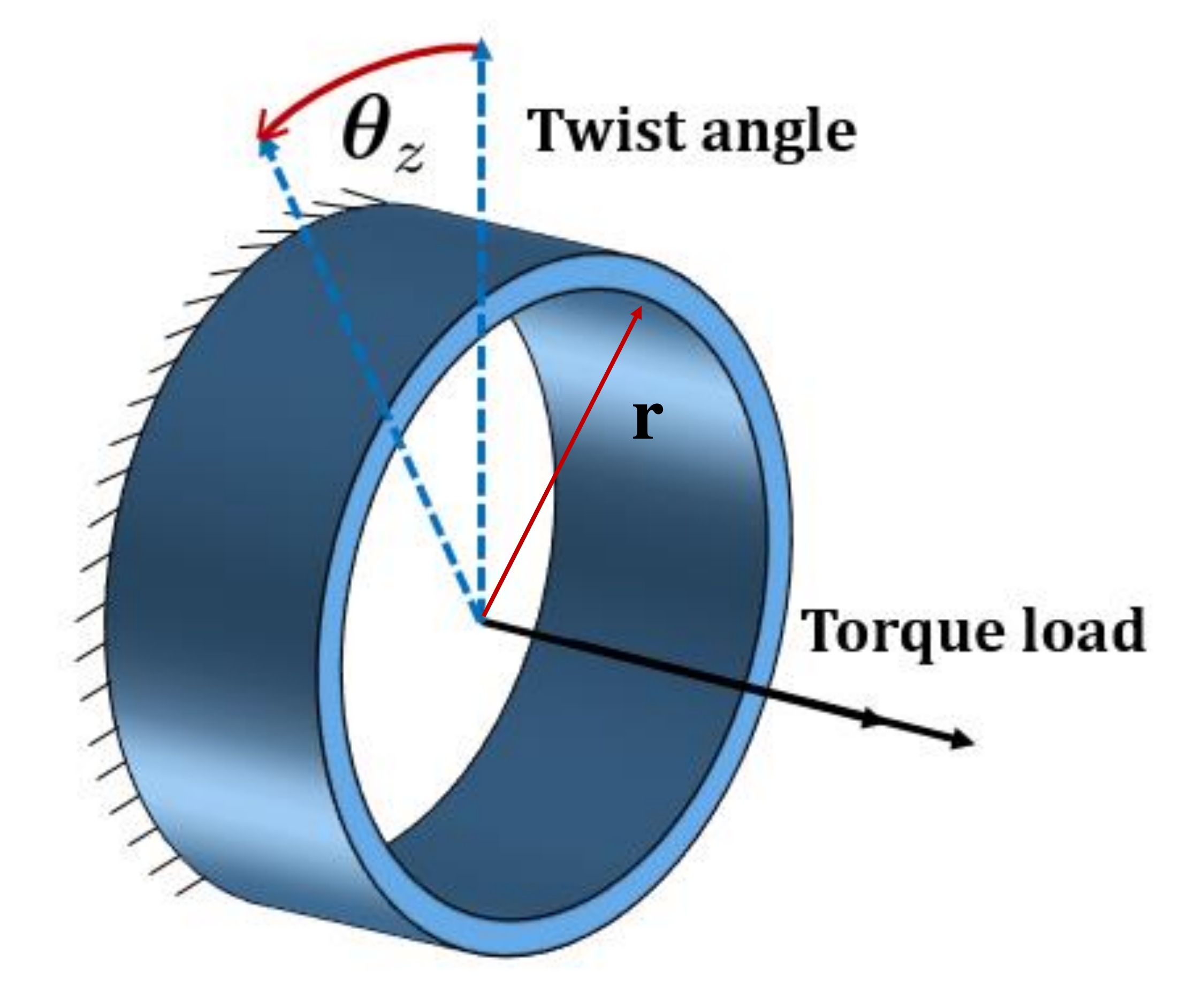}}
	%\vspace{2cm}
	\caption{Schematic for a three-dimensional cylindrical SMA torque tube subjected to isothermal torsion loading.}
	\label{fig:Tube_Schematic}
\end{figure}

In this subsection, the BVP of a three-dimensional SMA torque tube under torsion loading is studied. Refer to figure \ref{fig:Tube_Schematic}(a), the tube has an inner radius $r=3.0$(mm) and thickness $t/r=0.1$. In order to reduce the computational cost, a representative tube segment $L/r =2/3$ is analyzed here. Boundary conditions are depicted in figure \ref{fig:Tube_Schematic}(b), the tube left face is fixed and the right face is subjected to a torsion loading. The torque proportionally increases to 25 (N$\cdot$m) then unloads to 0 (N$\cdot$m), the temperature is kept constant at $380$ K. The torque tube undergoes a fully forward phase transformation from austenitic phase to martensitic phase followed by a reverse phase transformation from martensitic phase to austenitic phase. The material parameters used in this simulation are from table \ref{tab:MaterialProperty_bar2}.

\begin{figure}[H]
	\hspace{0.5cm}
	\subfigure[Proposed model]{
	\includegraphics[width=0.45\textwidth]{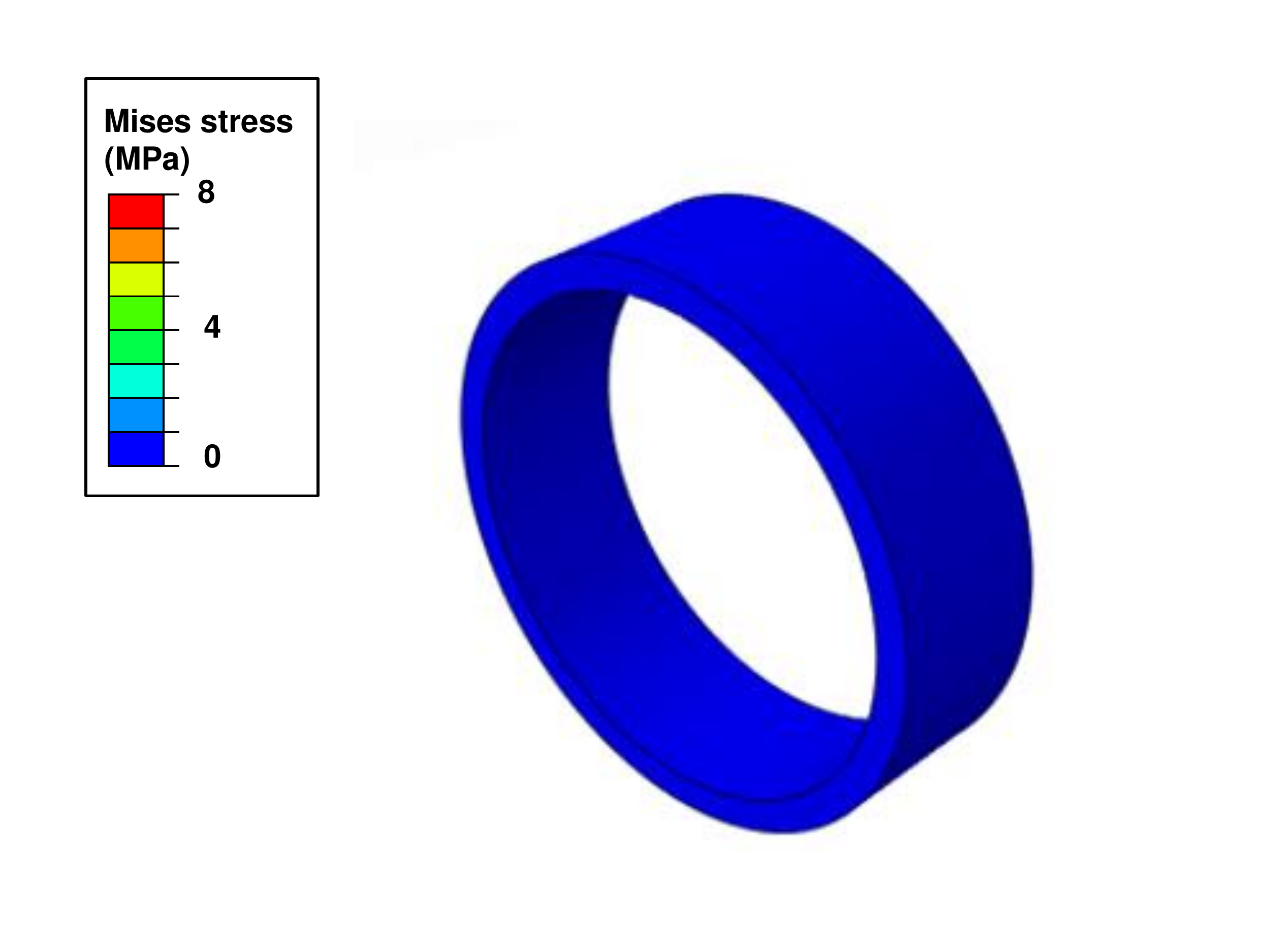}}%
	%\hspace{1.2cm}
	\subfigure[Abaqus nonlinear solver]{%
	%\label{fig:Stent_SDV}%
	\includegraphics[width=0.45\textwidth]{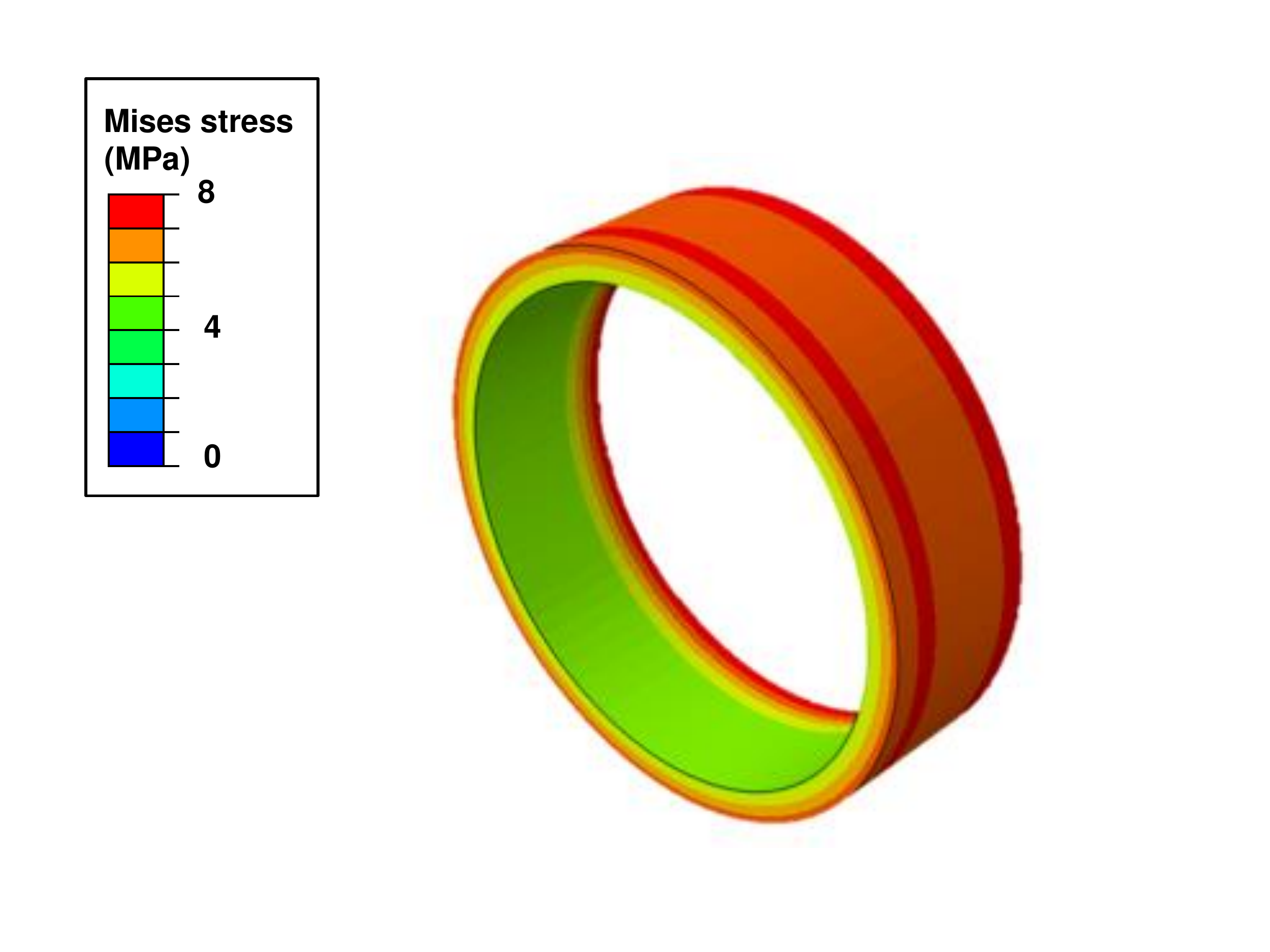}}%
\vspace{-0.2cm}
\caption{Mises stress residuals accumulated after one loading cycle for the torque tube predicted by the proposed model and the Abaqus nonlinear solver.}
\label{fig:Tube_Contour}
\end{figure}

\begin{figure}[H]%
	\hspace{-0.5cm}
	\subfigure[Shear stress-strain curve predicted by proposed model]{%
		\label{fig:Tube_Lei_SS}
		\includegraphics[width=0.5\textwidth]{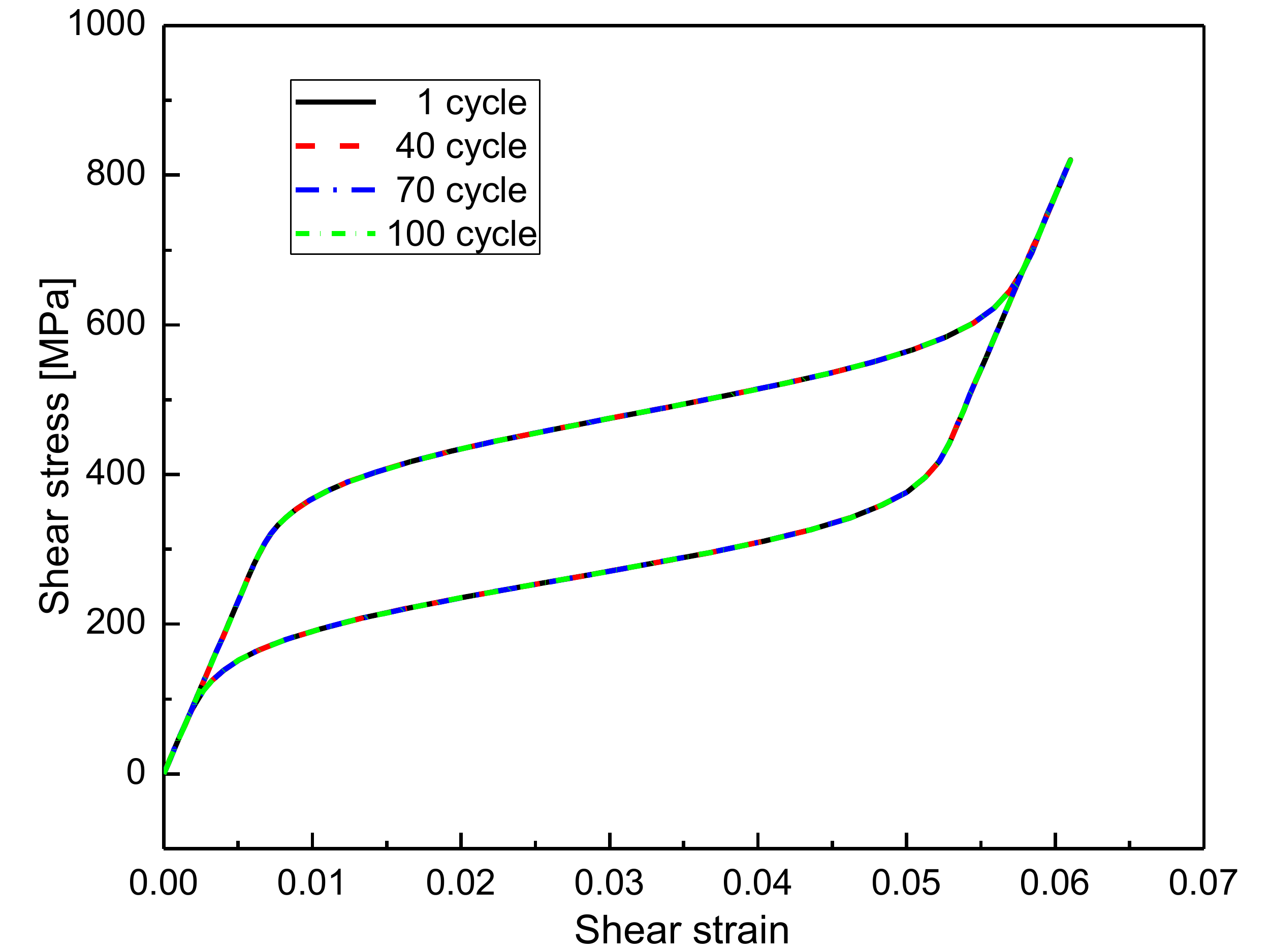}}%
	\subfigure[Shear stress-strain curve predicted by Abaqus nonlinear solver]{%
		\label{fig:Tube_Ori_SS}%
		\includegraphics[width=0.5\textwidth]{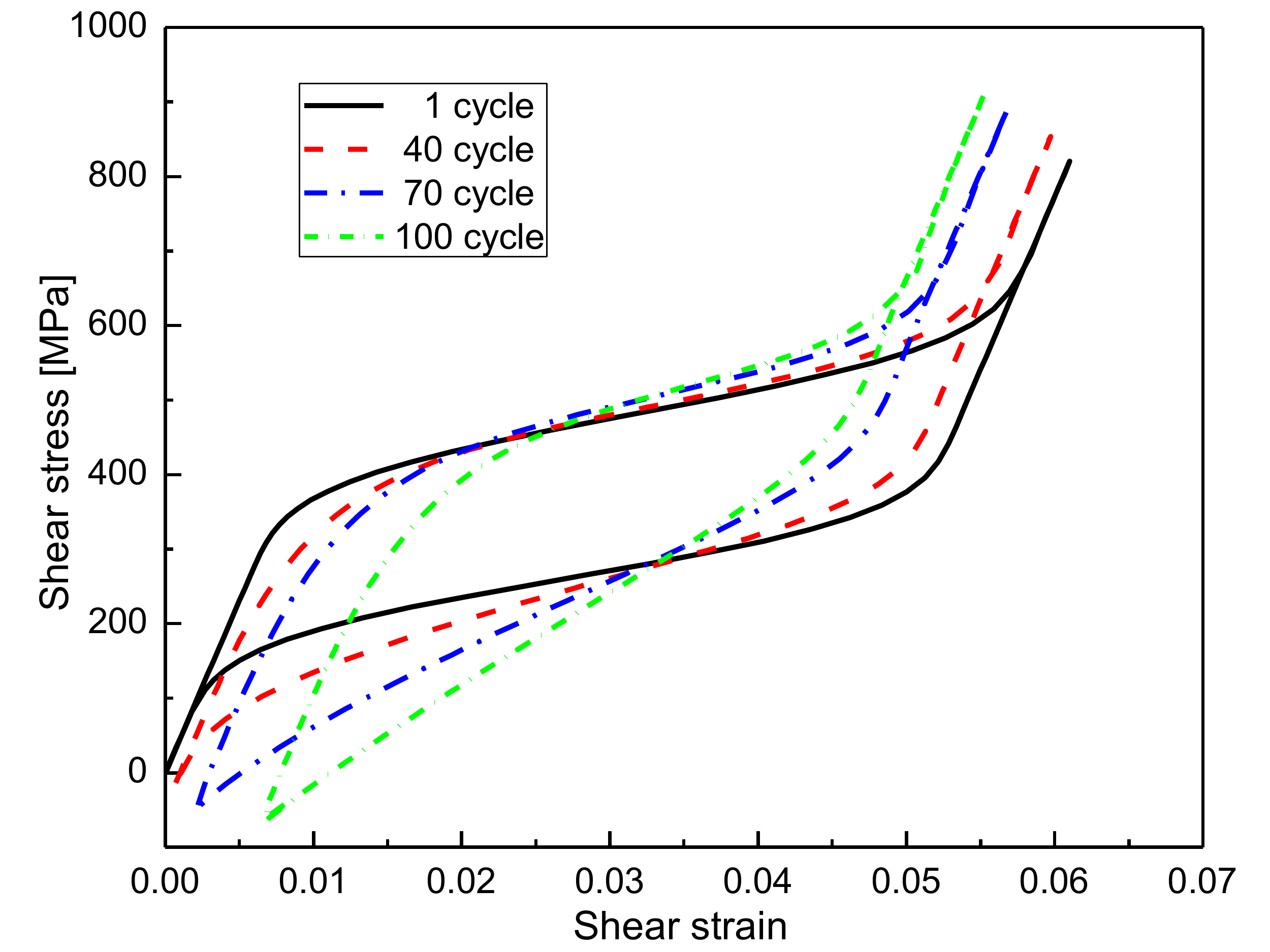}}%
	\caption{The cyclic stress-strain response for an SMA tube under isothermal loading condition.}
	\label{fig:Tube_SS}
\end{figure}

\begin{figure}[H]%
	\hspace{-0.5cm}
	\subfigure[Load-displacement curve predicted by proposed model]{%
		\label{fig:Tube_Lei_PD}
		\includegraphics[width=0.5\textwidth]{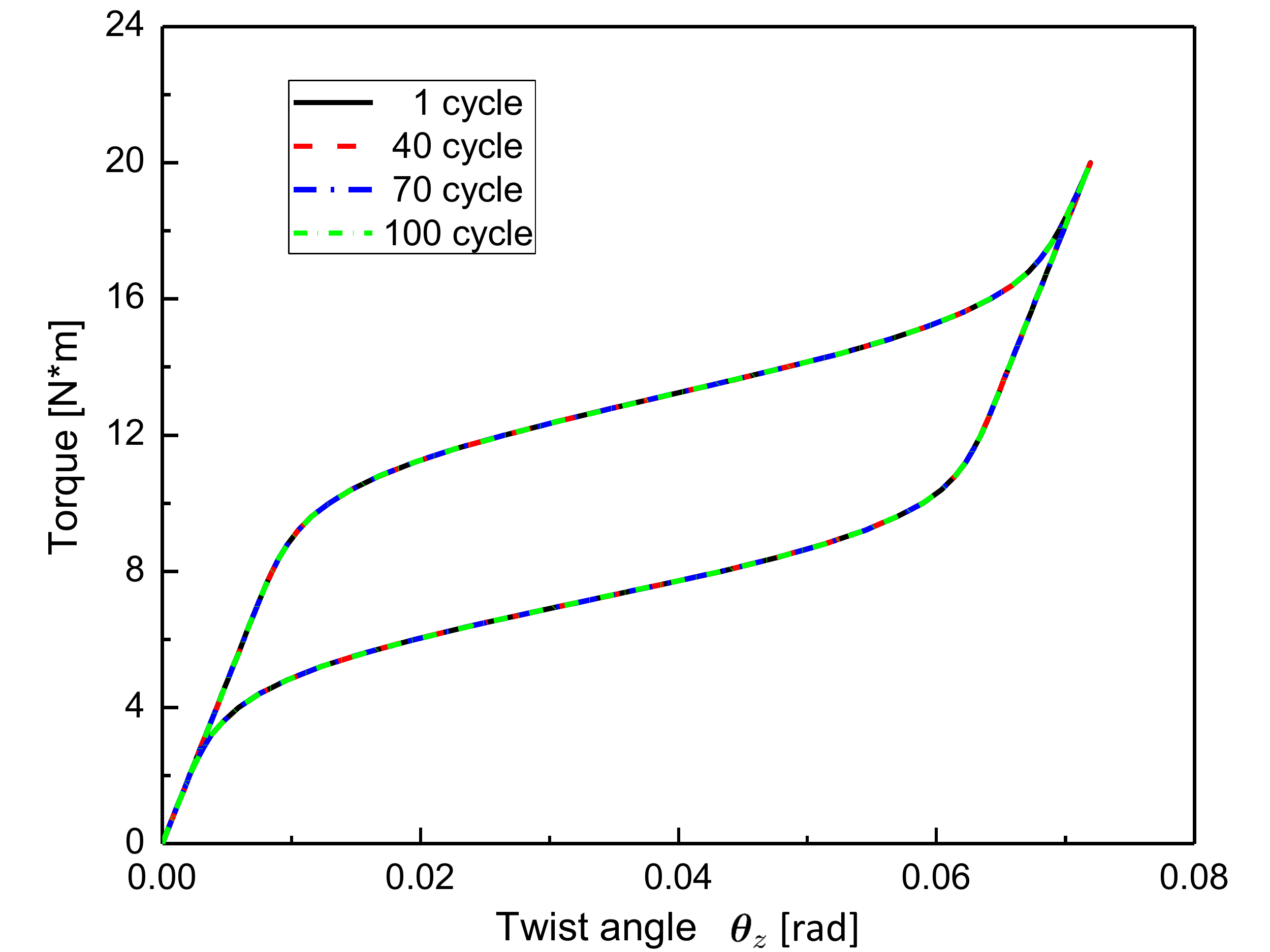}}%
	\subfigure[Load-displacement curve predicted by Abaqus nonlinear solver]{%
		\label{fig:Tube_Ori_PD}%
		\includegraphics[width=0.5\textwidth]{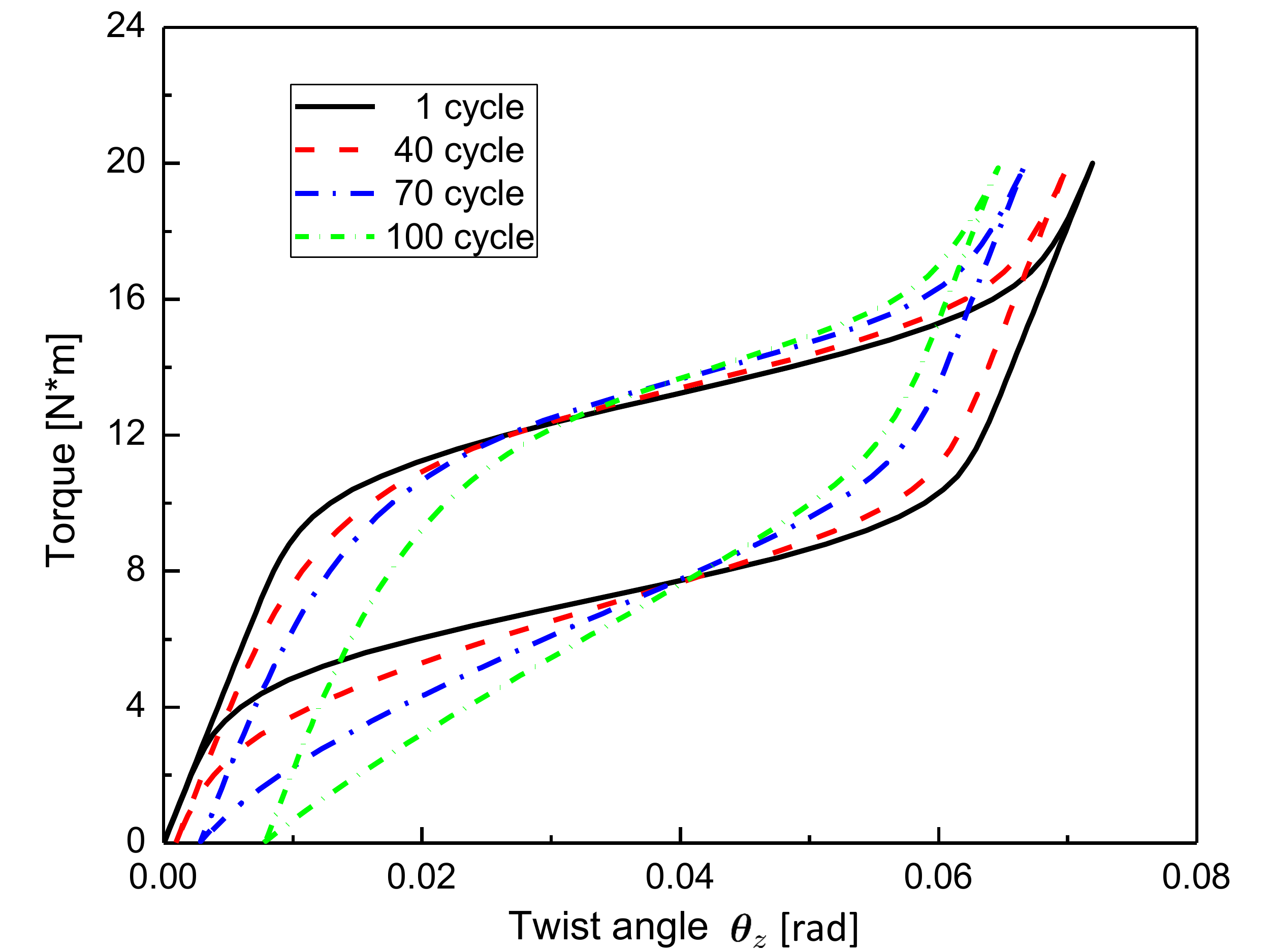}}%
	\caption{The cyclic load-displacement response for an SMA tube under isothermal loading condition.}
	\label{fig:Tube_PD}
\end{figure}

The cyclic shear stress-strain response of a material point from the tube outer surface at the right end is predicted by the proposed model and Abaqus nonlinear solver. As shown in figure \ref{fig:Tube_Ori_SS}, similar to the results observed in the SMA beam case, a shifting response is predicted by the Abaqus nonlinear solver due to the accumulation of shear stress errors. In contrast, a stable response is predicted by the proposed model shown in figure \ref{fig:Tube_Lei_SS}. More specifically, figure \ref{fig:Tube_Contour} shows the magnitude of stress residual accumulated after one loading cycle. The value of Mises stress residual predicted by the proposed model is almost zero compared to a value around 4 MPa predicted by the the Abaqus nonlinear solver. As a result of the accumulation of such shear stress errors, the shear stress levels required to start the forward phase transformation spuriously decreases in the case of Abaqus nonlinear solver shown in figure \ref{fig:Tube_Ori_SS}. Besides, the maximum shear stress levels at the end of forward transformation increases, and the shape of hysteresis loop also changes. The cyclic structural response of the torque tube is also provided in figure \ref{fig:Tube_PD} by plotting the applied torque versus the twist angle $\theta_z$. It can be seen that a stable structural response is predicted by the proposed model shown in figure \ref{fig:Tube_Lei_PD} compared to a shifting structural response predicted by the Abaqus nonlinear solver shown in figure \ref{fig:Tube_Ori_PD}. From the observation on these results, it is seen that the Abaqus nonlinear solver is not able to predict reliable results for the SMA torque tube subjected to 100 shearing cycles any more. Thus, the proposed model that can resolve the shear stress errors is required for the SMA torque tube subjected to cyclic torsion loading. 

%The comparison shows that the proposed model is able to eliminate the stress errors. 

\subsection{SMA tube under isobaric loading}\label{sec:torque_tube_isobaric}
In order to test the capability of the proposed model to predict the thermally-induced phase transformation in SMAs, a three-dimensional cylindrical SMA tube is studied under cyclic isobaric torsional loading, {i.e.}, subjected to a constant torsion load with temperature variation cycles. The SMA torque tubes has been investigated as rotational actuators to realize a morphing wing during the plane take-off and cruise regime \cite{saunders2014,mabe2014,calkins2016,icardi2009}. The design and optimization of such SMA-based morphing structure requires a thorough understanding on the response of SMA torque tubes subjected to cyclic isobaric loading. To that end, the SMA tube component is analyzed under cyclic isobaric loading conditions. The model has the same geometry and material information as the tube simulation in section \ref{sec:torque_tube_isothermal}. The loading condition is as follows, a 3 N$\cdot$m torque load is applied to the tube right end and the temperature varies between $250$ K and $390$ K for 100 cycles. Cyclic shear strain-temperature and $\bm\theta_z$-temperature curves are obtained via the proposed model and the Abaqus nonlinear solver.
%As found in the previous examples, the artificial stress residuals caused by other non-objective rates, \emph{i.e.}, Jaumman rate used in Abaqus nonlinear solver, builds up during the cyclic loading, which in return causes a shifting material and structural response. To demonstrate the effectiveness of the proposed model to eliminate the artificial stress residuals,

\begin{figure}[H]%
	\hspace{-0.5cm}
	\subfigure[Temp.-shear strain curve predicted by proposed model]{%
		\label{fig:Tube_Lei_TS}
		\includegraphics[width=0.5\textwidth]{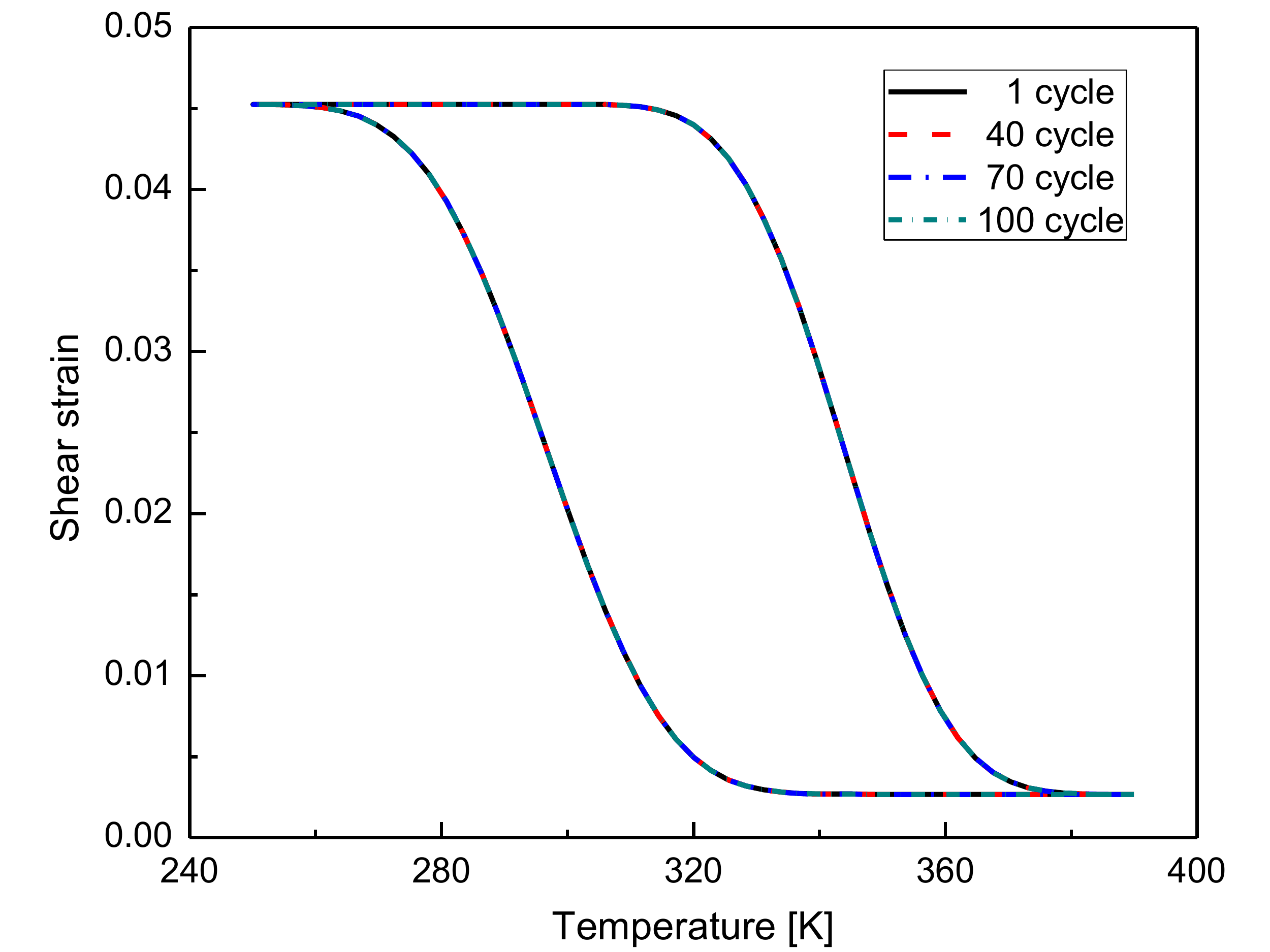}}%
	\subfigure[Temp.-shear strain curve predicted by Abaqus nonlinear solver]{%
		\label{fig:Tube_Ori_TS}%
		\includegraphics[width=0.5\textwidth]{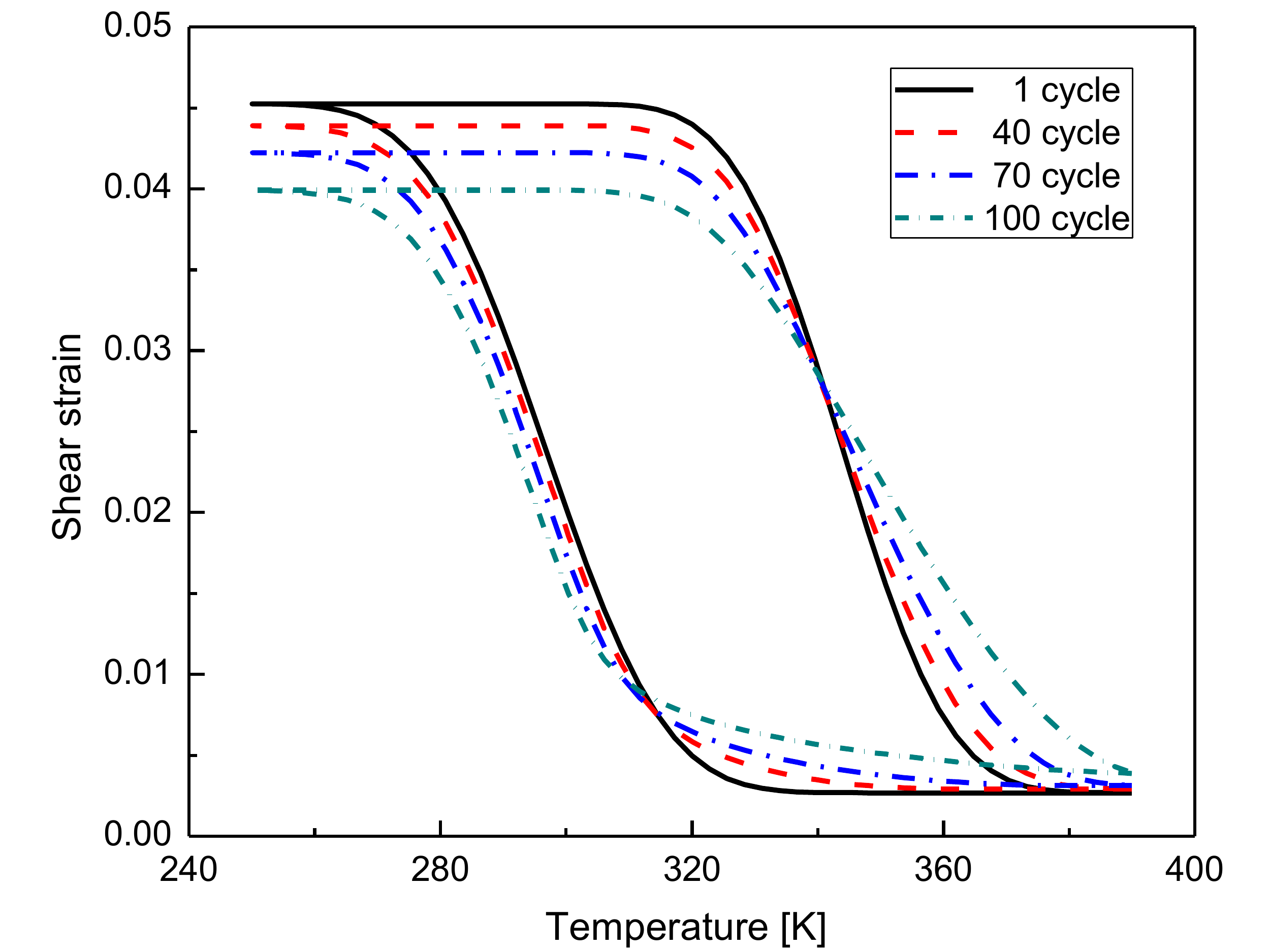}}%
	\caption{The cyclic temperature-shear strain response for an SMA tube under isobaric loading condition.}
	\label{fig:Tube_TS}
\end{figure}

\begin{figure}[H]%
	\hspace{-0.5cm}
	\subfigure[Temperature-$\theta_z$ curve predicted by proposed model]{%
		\label{fig:Tube_Lei_TA}
		\includegraphics[width=0.50\textwidth]{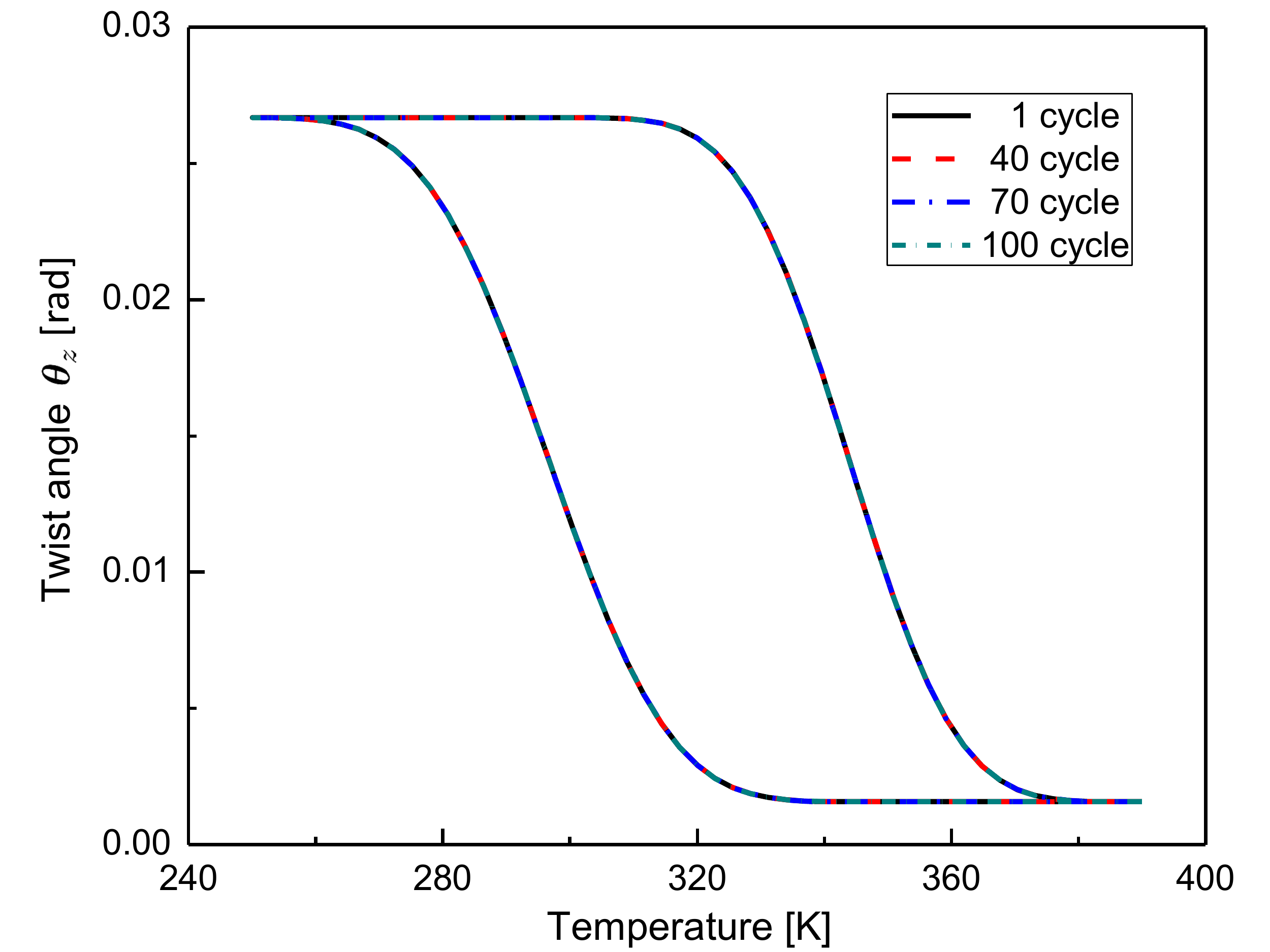}}%
	\subfigure[Temperature-$\theta_z$ curve predicted by Abaqus nonlinear solver]{%
		\label{fig:Tube_Ori_TA}%
		\includegraphics[width=0.50\textwidth]{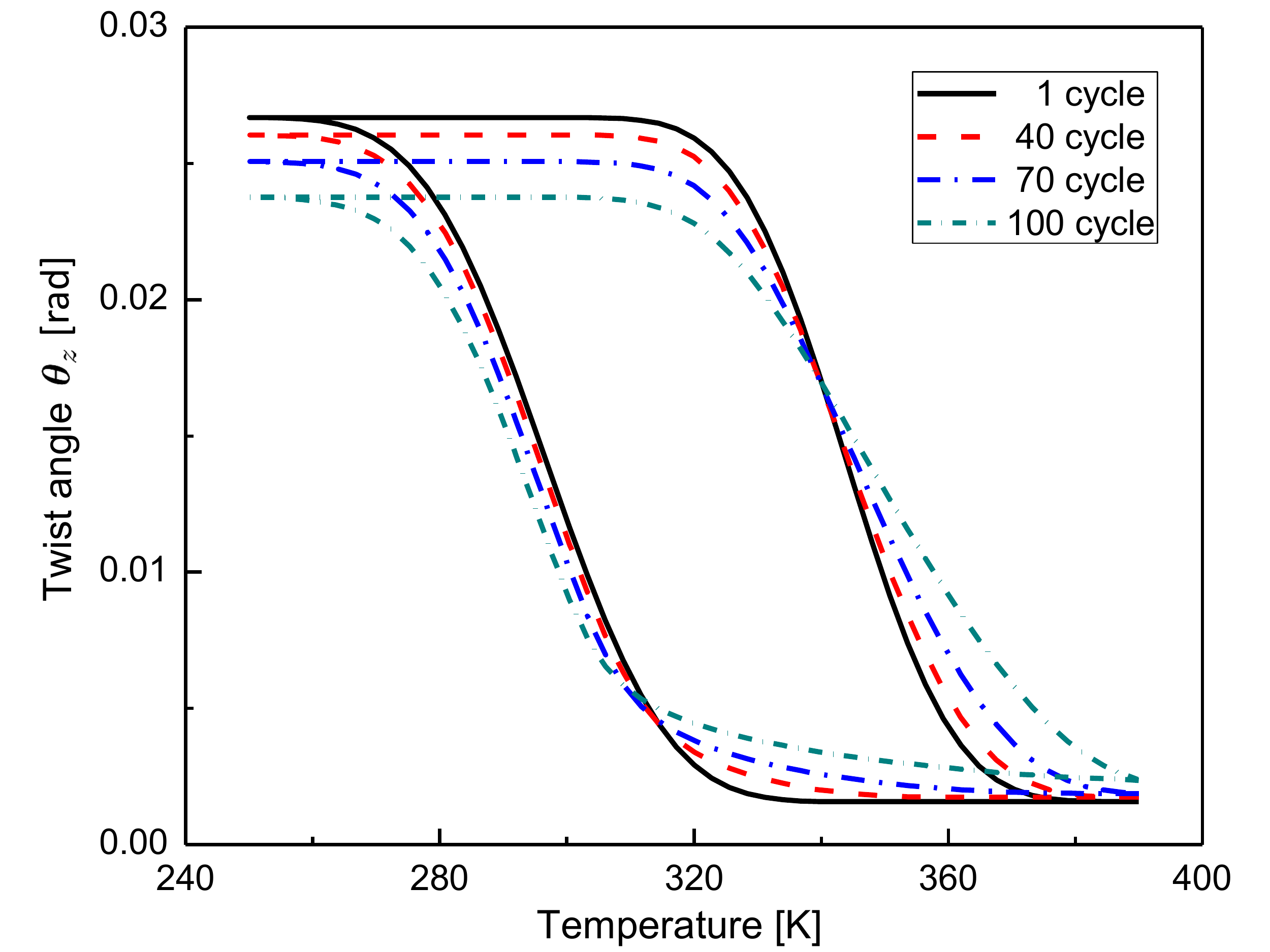}}%
	\caption{The cyclic temperature-$\theta_z$ response for an SMA tube under isobaric loading condition.}
	\label{fig:Tube_TA}
\end{figure}

{As it is shown in figure \ref{fig:Tube_Lei_TS}, a stable cyclic shear strain-temperature response for the tube is predicted by the proposed model. In contrast, a shifting cyclic strain-temperature response is predicted by using the Abaqus nonlinear solver shown in figure \ref{fig:Tube_Ori_TS}. More specifically, it can be seen that the isobaric response drifts downwards at $T=240$ K and is lifted up at $T=390$ K from cycle to cycle due to the stress error accumulation. Similar to the observation on the strain-temperature response, the cyclic $\bm\theta_z$-temperature response is stable in the case of proposed model, and it is a shifting response predicted by the Abaqus nonlinear solver. The comparison on $\bm\theta_z$-temperature response is plotted in figure \ref{fig:Tube_TA}. Based upon the analysis of SMA tube subjected to thermal loading cycles, it is shown that the accumulated stress errors from Abaqus nonlinear solver result in an shifting cyclic isobaric response, and such artificial stress errors can be eliminated by using the proposed model.}    

\begin{table}[H] 
	\centering
	\caption{Calibrated values of material parameters for NiTi (50.8 at.\% Ni)\vspace{-0.3cm} \cite{lagoudas2012}.}
	\renewcommand{\arraystretch}{1}
	\begin{tabular}{c|lr|ll} \toprule
		Type                         &Parameter                        & Value                                   &Parameter\footnotemark             & Value  \\                                       \midrule
		&$E_A$                            & 32.5   [GPa]                              & $C_A$                & 3.5  [MPa/K]\\
		&$E_M$                            & 23.0   [GPa]                              & $C_M$               & 3.5  [MPa/K]\\
		Key material parameters       &$\nu_A=\nu_M$                          & 0.3~~~~~~~~                              & $M_s$                & 264  [K]\\
		12                      &$\alpha_A=\alpha_M$    &    2.2$\times$10$^{-5}$ [K$^{-1}$]                           & $M_f$                &160  [K]\\
		& $ H^\textit{max}$                        & 3.3\%          & $A_s$                 &217  [K]\\
		& $k_t$                      &    N/A   & $A_f $                 & 290  [K]\\                                   \midrule
		
		Smooth hardening parameters 		&  $n_1$         &   0.17                &  $n_3$             & 0.25 \\
		4							   &  $n_2$         &   0.27                &  $n_4$        	 & 0.35 \\                                   
		\bottomrule
	\end{tabular}
	\label{tab:MaterialProperty_bar3}
\end{table}
\footnotetext{The values of transformation temperatures $(M_s, M_f, A_s, A_f)$ are referenced from \cite{Johnson2015} in order to realize the self-expanding process within human body environment, the rest of values of the material parameters are taken from \cite{lagoudas2012} as they are not provided from \cite{Johnson2015}.}

\begin{figure}[h]%
	\centering
	%\hspace{1cm}
	\subfigure[The expanded and crimped state for the flexible structure]{%
		\label{fig:Stent_shape}%
		\includegraphics[width=0.45\textwidth]{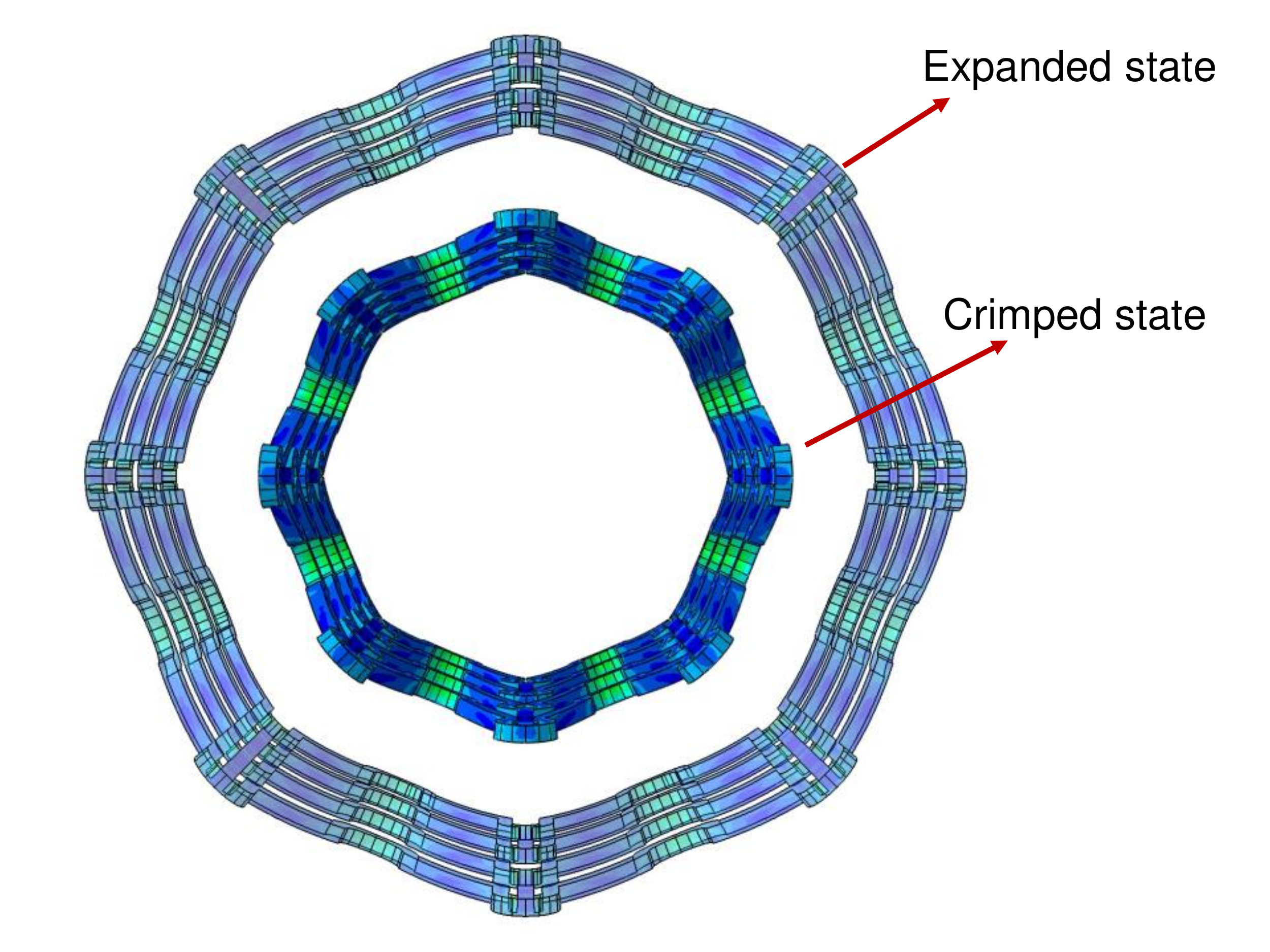}}%
	\subfigure[ {Loading path of the flexible structure}]{%
		\label{fig:stent_path}%
		\includegraphics[width=0.55\textwidth]{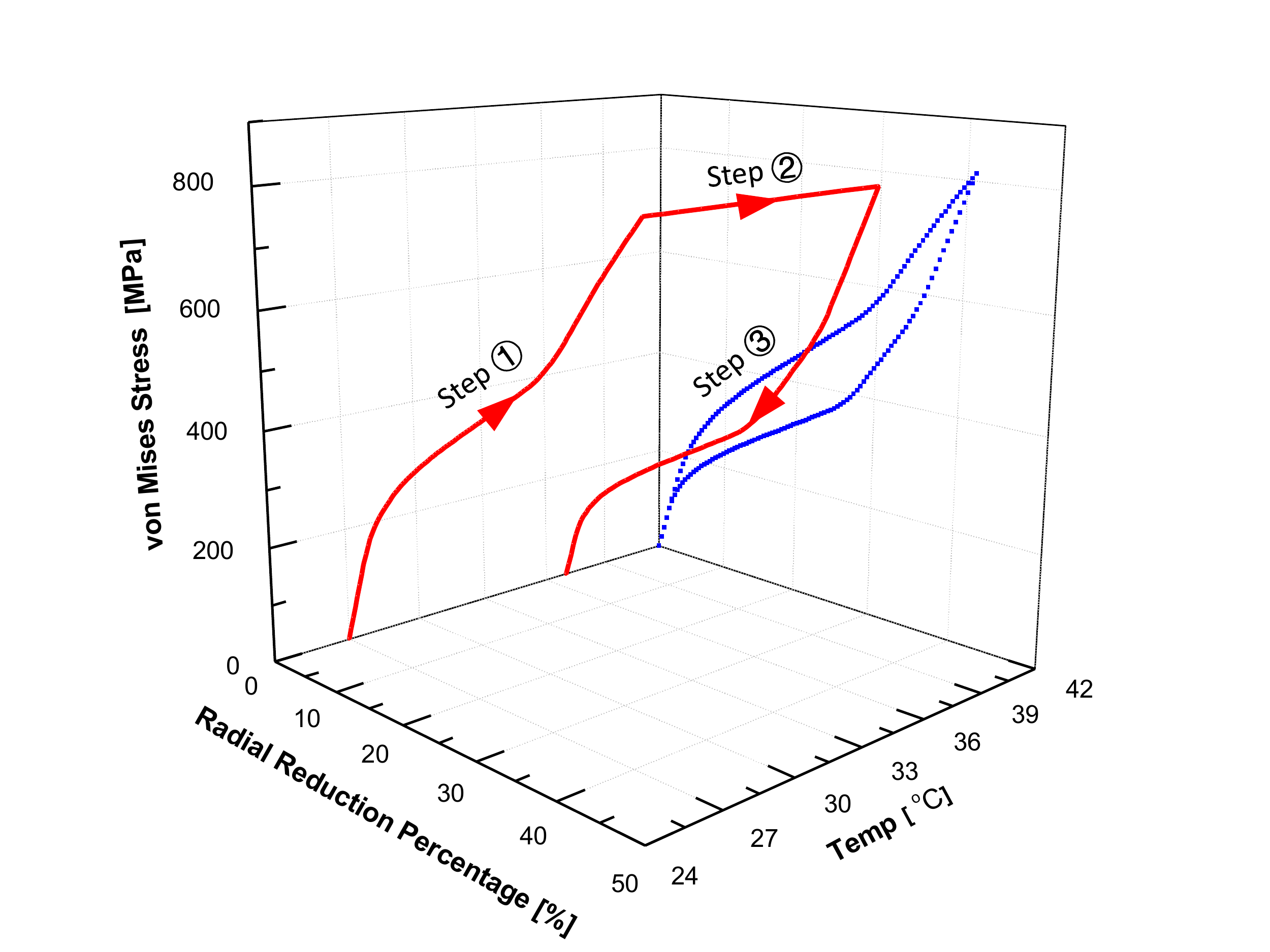}}%
	
	\centering 
	\hspace*{-1cm}
	\subfigure[von Mises stress]{%
		\label{fig:Stent_stress}%
		\includegraphics[width=0.45\textwidth]{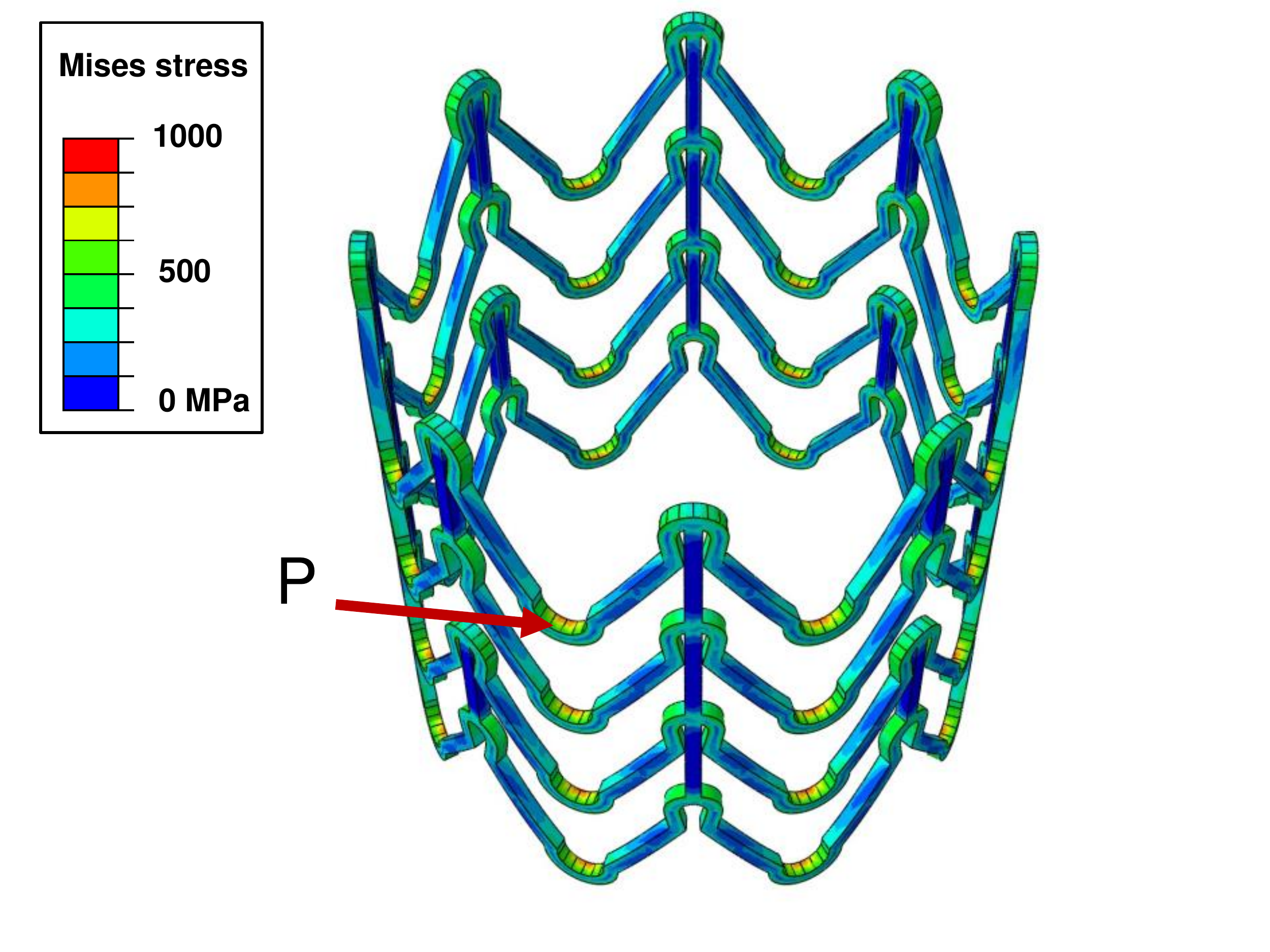}}%
	\hspace*{1.2cm}
	\subfigure[Martensitic volume fraction]{%
		\label{fig:Stent_SDV}%
		\includegraphics[width=0.45\textwidth]{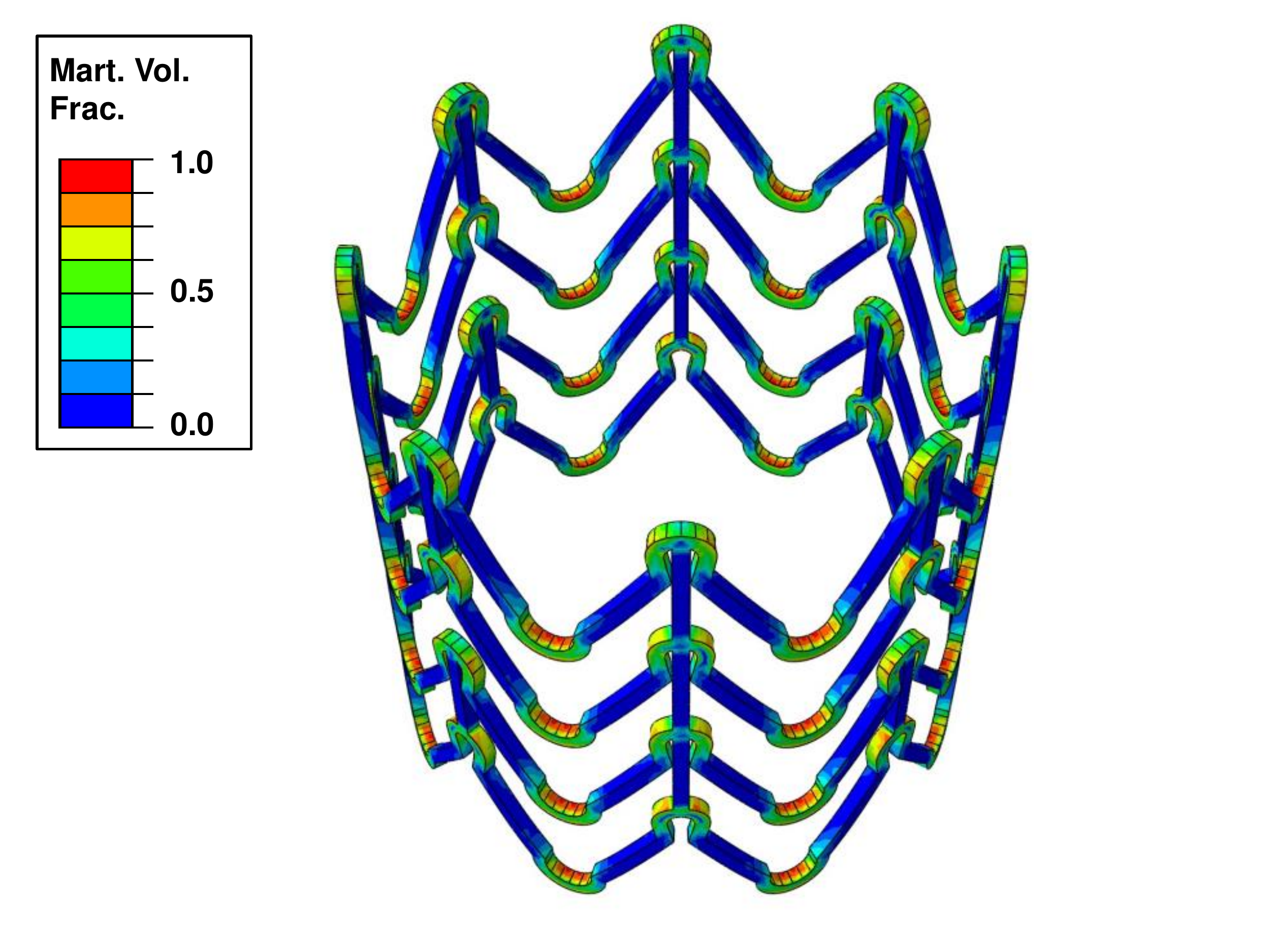}}%
	\caption{ {The expanded and crimped shapes, loading path, von Mises stress distribution and martensitic volume fraction contour of the SMA flexible structure during the self-expanding analysis.}}
\end{figure}

\subsection{3-D analysis of a flexible SMA structure}\label{sec:stent}
{In this section, a 3-D solid SMA flexible structure is studied to demonstrate that the proposed model can be used as a 3-D structural design tool. This SMA flexible structure can be used as a stent to provide a less invasive method for administering support to diseased arteries, veins or other vessels in the human body. They are crimped into a smaller shape outside the body then inserted into the diseased artery. After being delivered into the desired position, the flexible structure expands automatically by using the phase transformation of SMAs \cite{jay2018Stent}. The loading path of this self-expanding structure can be described with three steps, i.e., 1) The flexure is firstly crimped outside the body by external constraints and attached to a constraint container device called catheter or cannula. 2) The flexure is inserted into the body while the temperature increases from the room temperature to the body temperature. 3) The structure recovers its original shape when the constraint is removed \cite{stoeckel2004,kapoor2017,liu2008}. Although there are analysis for similar type of flexible structure by other researchers \cite{arghavani2011, hartl2010}, among which only a small unit cell of the structure is analyzed. Here a full scale 3-D SMA flexible structure designed for repairing aortic dissection \cite{jay2018Stent,jay2018Dacron} is studied to provide the global response of the structure during the self-expanding process. The structure is 32 (mm) long, 25.4 (mm) in outer diameter with 4 struts and 0.5 (mm) thick in radial direction. Due to the curvature of the structure strut, the SMA flexure experiences large rotation and stress concentrations around the hinge part, which in return results in a complex local non-proportional stress and strain evolutions at the hinge. The loading path of this analysis is indicated by the red curve in figure.\ref{fig:stent_path}. The material parameters used in this simulation are from table \ref{tab:MaterialProperty_bar3}.}

{
The expanded and crimped shapes of the SMA flexure during the analysis are shown in figure \ref{fig:Stent_shape}. As illustrated in figure \ref{fig:Stent_stress}, stress concentration due to the strut curved part is observed. The von Mises contour indicates that a local non-proportional stress field is evolved at the hinge location during the crimping process. As shown in figure \ref{fig:Stent_SDV}, while the straight strut part is still in austenitic phase, the stress-induced martensitic phase transformation is activated by the stress concentration at the hinge location subjected to bending. The martensitic volume fraction contour for the flexure during the crimping process is shown in figure \ref{fig:Stent_SDV}.  The global structural response of the SMA flexure for a material point P (see figure \ref{fig:Stent_stress}) is provided in terms of a 3-D stress-temperature-displacement curve in figure.\ref{fig:stent_path}, in which the red curve indicates the actual response while the blue curve is the projection of blue curve on the stress-radial reduction ratio plane. This well captured non-proportional stress evolution and martensitic phase transformation in this example demonstrates that the proposed model can be used as an efficient tool for the 3-D analysis and design of complex SMA-based structures.}

\section{Concluding remarks}\label{Conc}
Based on the SMA model proposed by Lagoudas and coworkers for small deformation analysis, a three-dimensional constitutive model for martensitic transformation in polycrystalline SMAs accounting for large deformation has been proposed in this work. Three important characteristics in SMA reponse are considered, i.e., the smooth transition during the phase transformation, the stress dependent transformation strain to account for the coexistence of oriented/self-accommodated martensitic variants, and a stress dependent critical driving force to consider the effect of applied stress levels on the size of hysteresis loop. The proposed model is formulated based on the finite deformation framework that utilizes logarithmic strain and rate such that it not only accounts for the large strains and rotations that an SMA component may undertake, but also resolves the artificial stress errors that are caused by using other non-integrable objective rates. The proposed model is able to predict the stress-induced and thermally induced phase transformations in SMAs under general three-dimensional thermomechanical loading. In particular, it was shown in the example of an SMA bar that the proposed model accounts for the geometry nonlinearity induced by large strains, so that it corrects the spurious material softening in the results from its infinitesimal counterpart. In the numerical examples of an SMA beam and an SMA torque tube, it was demonstrated that the proposed model captures the large rotations that SMA-based components may undertake. By comparing the predicted cyclic response to the results obtained through the Abaqus nonlinear solver, the proposed model demonstrated that it effectively resolves the artificial stress errors. In the end, a 3-D solid flexible structure experiencing local, non-proportional stress and strain evolution was analyzed by the proposed model, which shows the proposed model can be used as an efficient tool for the 3-D analysis and design of complex SMA-based structures. The detailed formulation of the proposed model and its implementation procedures make it readily used by other researchers. The model can be further extended to incorporate additional nonlinear phenomena exhibited by SMAs, such as transformation-induced plasticity, viscoplasticity, and damage evolution.

\section*{Acknowledgments}\label{Ack}
The authors would like to acknowledge the financial support provided by the Qatar National Research Fund (QNRF) under grant number: NPRP 7-032-2-016, and the NASA University Leadership Initiative (ULI) project under the grant number: NNX17AJ96A. The conclusions in
this work are solely made by the authors and do not necessarily represent the perspectives of QNRF and NASA.

\section*{Data availability}
The user-defined material subroutine implemented in commercial finite element software Abaqus for the proposed finite strain SMA model is available to download from the GitHub: https://github.com/Aero-tomato/SMA-UMAT.

\bibliographystyle{ieeetr} 
\bibliography{myarticle}

\begin{thebibliography}{10}

\bibitem{lagoudas2008}
Lagoudas D.~C, {\em Shape memory alloys: modeling and engineering
  applications}.
\newblock Springer Science \& Business Media, 2008.

\bibitem{peraza2014origami}
Peraza-Hernandez E.~A, Hartl D.~J, Malak~Jr R.~J, and Lagoudas D.~C,
  ``Origami-inspired active structures: a synthesis and review,'' {\em Smart
  Materials and Structures}, vol.~23, no.~9, p.~094001, 2014.

\bibitem{hartl2007aerospace}
Hartl D.~J and Lagoudas D.~C, ``Aerospace applications of shape memory
  alloys,'' {\em Proceedings of the Institution of Mechanical Engineers, Part
  G: Journal of Aerospace Engineering}, vol.~221, no.~4, pp.~535--552, 2007.

\bibitem{lazzara2019}
Lazzara D.~S, Magee T, Shen H, and Mabe J.~H, ``Off-design sonic boom
  performance for low-boom aircraft,'' in {\em AIAA Scitech 2019 Forum},
  p.~0606, 2019.

\bibitem{xu2019}
Xu L, Solomou A, Baxevanis T, and Lagoudas D.~C, ``A three-dimensional
  constitutive modeling for shape memory alloys considering two-way shape
  memory effect and transformation-induced plasticity,'' in {\em AIAA Scitech
  2019 Forum}, p.~1195, 2019.

\bibitem{raniecki1992}
Raniecki B, Lexcellent C, and Tanaka K, ``Thermodynamic models of pseudoelastic
  behaviour of shape memory alloys,'' {\em Archiv of Mechanics, Archiwum
  Mechaniki Stosowanej}, vol.~44, pp.~261--284, 1992.

\bibitem{boyd1996}
Boyd J.~G and Lagoudas D.~C, ``A thermodynamical constitutive model for shape
  memory materials. part i. the monolithic shape memory alloy,'' {\em
  International Journal of Plasticity}, vol.~12, no.~6, pp.~805--842, 1996.

\bibitem{birman1997}
Birman V, ``Review of mechanics of shape memory alloy structures,'' {\em
  Applied Mechanics Reviews}, vol.~50, no.~11, pp.~629--645, 1997.

\bibitem{raniecki1998}
Raniecki B and Lexcellent C, ``Thermodynamics of isotropic pseudoelasticity in
  shape memory alloys,'' {\em European Journal of Mechanics-A/Solids}, vol.~17,
  no.~2, pp.~185--205, 1998.

\bibitem{patoor1996}
Patoor E, Eberhardt A, and Berveiller M, ``Micromechanical modelling of
  superelasticity in shape memory alloys,'' {\em Le Journal de Physique IV},
  vol.~6, no.~C1, pp.~C1--277, 1996.

\bibitem{levitas1998}
Levitas V.~I, ``Thermomechanical theory of martensitic phase transformations in
  inelastic materials,'' {\em International Journal of Solids and Structures},
  vol.~35, no.~9, pp.~889--940, 1998.

\bibitem{levitas2002}
Levitas V.~I and Preston D.~L, ``Three-dimensional landau theory for
  multivariant stress-induced martensitic phase transformations. i. austenite
  martensite,'' {\em Physical review B}, vol.~66, no.~13, p.~134206, 2002.

\bibitem{patoor2006}
Patoor E, Lagoudas D.~C, Entchev P.~B, Brinson L.~C, and Gao X, ``Shape memory
  alloys, part i: General properties and modeling of single crystals,'' {\em
  Mechanics of materials}, vol.~38, no.~5, pp.~391--429, 2006.

\bibitem{zaki2007}
Zaki W and Moumni Z, ``A three-dimensional model of the thermomechanical
  behavior of shape memory alloys,'' {\em Journal of the Mechanics and Physics
  of Solids}, vol.~55, no.~11, pp.~2455--2490, 2007.

\bibitem{saint2009}
Saint-Sulpice L, Chirani S.~A, and Calloch S, ``A 3d super-elastic model for
  shape memory alloys taking into account progressive strain under cyclic
  loadings,'' {\em Mechanics of materials}, vol.~41, no.~1, pp.~12--26, 2009.

\bibitem{hackl2008}
Hackl K and Heinen R, ``A micromechanical model for pretextured polycrystalline
  shape-memory alloys including elastic anisotropy,'' {\em Continuum Mechanics
  and Thermodynamics}, vol.~19, no.~8, pp.~499--510, 2008.

\bibitem{chemisky2011}
Chemisky Y, Duval A, Patoor E, and Zineb T.~B, ``Constitutive model for shape
  memory alloys including phase transformation, martensitic reorientation and
  twins accommodation,'' {\em Mechanics of Materials}, vol.~43, no.~7,
  pp.~361--376, 2011.

\bibitem{sedlak2012}
Sedlak P, Frost M, Bene{\v{s}}ov{\'a} B, Zineb T.~B, and {\v{S}}ittner P,
  ``Thermomechanical model for niti-based shape memory alloys including r-phase
  and material anisotropy under multi-axial loadings,'' {\em International
  Journal of Plasticity}, vol.~39, pp.~132--151, 2012.

\bibitem{Zineb2016}
Cisse C, Zaki W, and Zineb T.~B, ``A review of constitutive models and modeling
  techniques for shape memory alloys,'' {\em International Journal of
  Plasticity}, vol.~76, pp.~244--284, 2016.

\bibitem{chen2002}
Chen L.-Q, ``Phase-field models for microstructure evolution,'' {\em Annual
  review of materials research}, vol.~32, no.~1, pp.~113--140, 2002.

\bibitem{steinbach1999}
Steinbach I and Pezzolla F, ``A generalized field method for multiphase
  transformations using interface fields,'' {\em Physica D: Nonlinear
  Phenomena}, vol.~134, no.~4, pp.~385--393, 1999.

\bibitem{steinbach2006}
Steinbach I and Apel M, ``Multi phase field model for solid state
  transformation with elastic strain,'' {\em Physica D: Nonlinear Phenomena},
  vol.~217, no.~2, pp.~153--160, 2006.

\bibitem{mamivand2013}
Mamivand M, Zaeem M.~A, and El~Kadiri H, ``A review on phase field modeling of
  martensitic phase transformation,'' {\em Computational Materials Science},
  vol.~77, pp.~304--311, 2013.

\bibitem{zhong2014}
Zhong Y and Zhu T, ``Phase-field modeling of martensitic microstructure in niti
  shape memory alloys,'' {\em Acta Materialia}, vol.~75, pp.~337--347, 2014.

\bibitem{tham2001}
Thamburaja P and Anand L, ``Polycrystalline shape-memory materials: effect of
  crystallographic texture,'' {\em Journal of the Mechanics and Physics of
  Solids}, vol.~49, no.~4, pp.~709--737, 2001.

\bibitem{Wang2008}
Wang X, Xu B, and Yue Z, ``Micromechanical modelling of the effect of plastic
  deformation on the mechanical behaviour in pseudoelastic shape memory
  alloys,'' {\em International Journal of Plasticity}, vol.~24, no.~8, pp.~1307
  -- 1332, 2008.

\bibitem{yu2013}
Yu C, Kang G, Kan Q, and Song D, ``A micromechanical constitutive model based
  on crystal plasticity for thermo-mechanical cyclic deformation of niti shape
  memory alloys,'' {\em International Journal of Plasticity}, vol.~44,
  pp.~161--191, 2013.

\bibitem{auricchio1997}
Auricchio F and Taylor R.~L, ``Shape-memory alloys: modelling and numerical
  simulations of the finite-strain superelastic behavior,'' {\em Computer
  methods in applied mechanics and engineering}, vol.~143, no.~1, pp.~175--194,
  1997.

\bibitem{lagoudas2012}
Lagoudas D, Hartl D, Chemisky Y, Machado L, and Popov P, ``Constitutive model
  for the numerical analysis of phase transformation in polycrystalline shape
  memory alloys,'' {\em International Journal of Plasticity}, vol.~32,
  pp.~155--183, 2012.

\bibitem{brinson1993finite}
Brinson L and Lammering R, ``Finite element analysis of the behavior of shape
  memory alloys and their applications,'' {\em International Journal of Solids
  and Structures}, vol.~30, no.~23, pp.~3261--3280, 1993.

\bibitem{brinson1993one}
Brinson L.~C, ``One-dimensional constitutive behavior of shape memory alloys:
  thermomechanical derivation with non-constant material functions and
  redefined martensite internal variable,'' {\em Journal of intelligent
  material systems and structures}, vol.~4, no.~2, pp.~229--242, 1993.

\bibitem{lexcellent1996general}
Leclercq S and Lexcellent C, ``A general macroscopic description of the
  thermomechanical behavior of shape memory alloys,'' {\em Journal of the
  Mechanics and Physics of Solids}, vol.~44, no.~6, pp.~953--980, 1996.

\bibitem{lexcellent2013shape}
Lexcellent C, {\em Shape-memory alloys handbook}.
\newblock John Wiley \& Sons, 2013.

\bibitem{reese2008}
Reese S and Christ D, ``Finite deformation pseudo-elasticity of shape memory
  alloys--constitutive modelling and finite element implementation,'' {\em
  International Journal of Plasticity}, vol.~24, no.~3, pp.~455--482, 2008.

\bibitem{shaw2000}
Shaw J.~A, ``Simulations of localized thermo-mechanical behavior in a niti
  shape memory alloy,'' {\em International Journal of Plasticity}, vol.~16,
  no.~5, pp.~541--562, 2000.

\bibitem{jani2014}
Jani J.~M, Leary M, Subic A, and Gibson M.~A, ``A review of shape memory alloy
  research, applications and opportunities,'' {\em Materials \& Design},
  vol.~56, pp.~1078--1113, 2014.

\bibitem{bh2016}
Haghgouyan B, Shafaghi N, Ayd{\i}ner C.~C, and Anlas G, ``Experimental and
  computational investigation of the effect of phase transformation on fracture
  parameters of an sma,'' {\em Smart Materials and Structures}, vol.~25, no.~7,
  p.~075010, 2016.

\bibitem{bh2019}
Haghgouyan B, Hayrettin C, Baxevanis T, Karaman I, and Lagoudas D.~C,
  ``Fracture toughness of niti--towards establishing standard test methods for
  phase transforming materials,'' {\em Acta Materialia}, vol.~162,
  pp.~226--238, 2019.

\bibitem{wheeler2015}
Wheeler R, Saunders R, Pickett K, Eckert C, Stroud H, Fink T, Gakhar K, Boyd J,
  and Lagoudas D, ``Design of a reconfigurable sma-based solar array deployment
  mechanism,'' in {\em ASME 2015 Conference on Smart Materials, Adaptive
  Structures and Intelligent Systems}, pp.~V001T02A010--V001T02A010, American
  Society of Mechanical Engineers, 2015.

\bibitem{ziolkowski2007}
Ziolkowski A, ``Three-dimensional phenomenological thermodynamic model of
  pseudoelasticity of shape memory alloys at finite strains,'' {\em Continuum
  Mechanics and Thermodynamics}, vol.~19, no.~6, pp.~379--398, 2007.

\bibitem{evangelista2010}
Evangelista V, Marfia S, and Sacco E, ``A 3d sma constitutive model in the
  framework of finite strain,'' {\em International Journal for Numerical
  Methods in Engineering}, vol.~81, no.~6, pp.~761--785, 2010.

\bibitem{wang2017sms}
Wang J, Moumni Z, Zhang W, Xu Y, and Zaki W, ``A 3d finite-strain-based
  constitutive model for shape memory alloys accounting for thermomechanical
  coupling and martensite reorientation,'' {\em Smart Materials and
  Structures}, vol.~26, no.~6, p.~065006, 2017.

\bibitem{wang2017ijes}
Wang J, Moumni Z, Zhang W, and Zaki W, ``A thermomechanically coupled finite
  deformation constitutive model for shape memory alloys based on hencky
  strain,'' {\em International Journal of Engineering Science}, vol.~117,
  pp.~51--77, 2017.

\bibitem{stupkiewicz2013}
Stupkiewicz S and Petryk H, ``A robust model of pseudoelasticity in shape
  memory alloys,'' {\em International Journal for Numerical Methods in
  Engineering}, vol.~93, no.~7, pp.~747--769, 2013.

\bibitem{damanpack2017}
Damanpack A, Bodaghi M, and Liao W, ``A finite-strain constitutive model for
  anisotropic shape memory alloys,'' {\em Mechanics of Materials}, vol.~112,
  pp.~129--142, 2017.

\bibitem{xiao2006}
Xiao H, Bruhns O, and Meyers A, ``Elastoplasticity beyond small deformations,''
  {\em Acta Mechanica}, vol.~182, no.~1-2, pp.~31--111, 2006.

\bibitem{xiao1997}
Xiao H, Bruhns I.~O, and Meyers I.~A, ``Logarithmic strain, logarithmic spin
  and logarithmic rate,'' {\em Acta Mechanica}, vol.~124, no.~1-4, pp.~89--105,
  1997.

\bibitem{xiao1997hypo}
Xiao H, Bruhns O, and Meyers A, ``Hypo-elasticity model based upon the
  logarithmic stress rate,'' {\em Journal of Elasticity}, vol.~47, no.~1,
  pp.~51--68, 1997.

\bibitem{bruhns1999self}
Bruhns O, Xiao H, and Meyers A, ``Self-consistent eulerian rate type
  elasto-plasticity models based upon the logarithmic stress rate,'' {\em
  International Journal of Plasticity}, vol.~15, no.~5, pp.~479--520, 1999.

\bibitem{bruhns2001large}
Bruhns O, Xiao H, and Meyers A, ``Large simple shear and torsion problems in
  kinematic hardening elasto-plasticity with logarithmic rate,'' {\em
  International journal of solids and structures}, vol.~38, no.~48,
  pp.~8701--8722, 2001.

\bibitem{bruhns2001self}
Bruhns O, Xiao H, and Meyers A, ``A self-consistent eulerian rate type model
  for finite deformation elastoplasticity with isotropic damage,'' {\em
  International Journal of Solids and Structures}, vol.~38, no.~4,
  pp.~657--683, 2001.

\bibitem{meyers2003elastic}
Meyers A, Xiao H, and Bruhns O, ``Elastic stress ratcheting and corotational
  stress rates,'' {\em Tech. Mech}, vol.~23, pp.~92--102, 2003.

\bibitem{meyers2006choice}
Meyers A, Xiao H, and Bruhns O, ``Choice of objective rate in single parameter
  hypoelastic deformation cycles,'' {\em Computers \& structures}, vol.~84,
  no.~17, pp.~1134--1140, 2006.

\bibitem{zhu2014}
Zhu Y, Kang G, Kan Q, and Bruhns O.~T, ``Logarithmic stress rate based
  constitutive model for cyclic loading in finite plasticity,'' {\em
  International Journal of Plasticity}, vol.~54, pp.~34--55, 2014.

\bibitem{zhu2016}
Zhu Y, Kang G, Kan Q, Bruhns O.~T, and Liu Y, ``Thermo-mechanically coupled
  cyclic elasto-viscoplastic constitutive model of metals: theory and
  application,'' {\em International Journal of Plasticity}, vol.~79,
  pp.~111--152, 2016.

\bibitem{muller2006}
M{\"u}ller C and Bruhns O, ``A thermodynamic finite-strain model for
  pseudoelastic shape memory alloys,'' {\em International Journal of
  Plasticity}, vol.~22, no.~9, pp.~1658--1682, 2006.

\bibitem{teeriaho2013}
Teeriaho J.~P, ``An extension of a shape memory alloy model for large
  deformations based on an exactly integrable eulerian rate formulation with
  changing elastic properties,'' {\em International Journal of Plasticity},
  vol.~43, pp.~153--176, 2013.

\bibitem{xiao2014explicit}
Xiao H, ``An explicit, straightforward approach to modeling sma pseudoelastic
  hysteresis,'' {\em International Journal of Plasticity}, vol.~53,
  pp.~228--240, 2014.

\bibitem{yu2015}
Yu C, Kang G, Kan Q, and Zhu Y, ``Rate-dependent cyclic deformation of
  super-elastic niti shape memory alloy: thermo-mechanical coupled and physical
  mechanism-based constitutive model,'' {\em International Journal of
  Plasticity}, vol.~72, pp.~60--90, 2015.

\bibitem{xu2017}
Xu L, Baxevanis T, and Lagoudas D.~C, ``A finite strain constitutive model for
  martensitic transformation in shape memory alloys based on logarithmic
  strain,'' in {\em 25th AIAA/AHS Adaptive Structures Conference}, p.~0731,
  2017.

\bibitem{Xu2017trip}
Xu L, Baxevanis T, and Lagoudas D, ``A finite strain constitutive model
  considering transformation induced plasticity for shape memory alloys under
  cyclic loading,'' in {\em 8th ECCOMAS Thematic Conference on Smart Structures
  and Materials}, pp.~1645--1477, 2017.

\bibitem{xu2018}
Xu L, Baxevanis T, and Lagoudas D, ``A three-dimensional constitutive model for
  polycrystalline shape memory alloys under large strains combined with large
  rotations,'' in {\em ASME 2018 Conference on Smart Materials, Adaptive
  Structures and Intelligent Systems}, pp.~V002T02A007--V002T02A007, American
  Society of Mechanical Engineers, 2018.

\bibitem{simo2006}
Simo J.~C and Hughes T.~J, {\em Computational inelasticity}, vol.~7.
\newblock Springer Science \& Business Media, 2006.

\bibitem{khan1995continuum}
Khan A.~S and Huang S, {\em Continuum theory of plasticity}.
\newblock John Wiley \& Sons, 1995.

\bibitem{xiao1998objective}
Xiao H, Bruhns O, and Meyers A, ``Objective corotational rates and unified
  work-conjugacy relation between eulerian and lagrangean strain and stress
  measures,'' {\em Archives of Mechanics}, vol.~50, no.~6, pp.~1015--1045,
  1998.

\bibitem{qidwai2000}
Qidwai M and Lagoudas D, ``On thermomechanics and transformation surfaces of
  polycrystalline niti shape memory alloy material,'' {\em International
  Journal of Plasticity}, vol.~16, no.~10, pp.~1309--1343, 2000.

\bibitem{hill1948}
Hill R, ``A variational principle of maximum plastic work in classical
  plasticity,'' {\em The Quarterly Journal of Mechanics and Applied
  Mathematics}, vol.~1, no.~1, pp.~18--28, 1948.

\bibitem{miehe1996}
MIEHE C, ``Exponential map algorithm for stress updates in anisotropic
  multiplicative elastoplasticity for single crystals,'' {\em International
  Journal for Numerical Methods in Engineering}, vol.~39, no.~19,
  pp.~3367--3390, 1996.

\bibitem{abaqus2014}
Abaqus, ``6.14 documentation,'' {\em Dassault Systemes Simulia Corporation},
  vol.~651, 2014.

\bibitem{saunders2014}
Saunders R, Hartl D, Herrington J, Hodge L, and Mabe J, ``Optimization of a
  composite morphing wing with shape memory alloy torsional actuators,'' in
  {\em ASME 2014 Conference on Smart Materials, Adaptive Structures and
  Intelligent Systems}, pp.~V002T02A014--V002T02A014, American Society of
  Mechanical Engineers, 2014.

\bibitem{mabe2014}
Mabe J, Brown J, and Calkins F, ``Flight test of a shape memory alloy actuated
  adaptive trailing edge flap, part 1,'' in {\em Proceedings of SMST 2014 the
  International Conference on Shape Memory and Superelastic Technologies},
  2014.

\bibitem{calkins2016}
Calkins F and Mabe J, ``Flight test of a shape memory alloy actuated adaptive
  trailing edge flap,'' in {\em ASME 2016 Conference on Smart Materials,
  Adaptive Structures and Intelligent Systems}, pp.~V001T04A007--V001T04A007,
  American Society of Mechanical Engineers, 2016.

\bibitem{icardi2009}
Icardi U and Ferrero L, ``Preliminary study of an adaptive wing with shape
  memory alloy torsion actuators,'' {\em Materials \& Design}, vol.~30, no.~10,
  pp.~4200--4210, 2009.

\bibitem{jay2018Stent}
Jayendiran R, Nour B, and Ruimi A, ``Fluid-structure interaction (fsi) analysis
  of stent-graft for aortic endovascular aneurysm repair (evar): Material and
  structural considerations,'' {\em Journal of the mechanical behavior of
  biomedical materials}, vol.~87, pp.~95--110, 2018.

\bibitem{stoeckel2004}
Stoeckel D, Pelton A, and Duerig T, ``Self-expanding nitinol stents: material
  and design considerations,'' {\em European radiology}, vol.~14, no.~2,
  pp.~292--301, 2004.

\bibitem{kapoor2017}
Kapoor D, ``Nitinol for medical applications: A brief introduction to the
  properties and processing of nickel titanium shape memory alloys and their
  use in stents,'' {\em Johnson Matthey Technology Review}, vol.~61, no.~1,
  pp.~66--76, 2017.

\bibitem{liu2008}
Liu X, Wang Y, Yang D, and Qi M, ``The effect of ageing treatment on
  shape-setting and superelasticity of a nitinol stent,'' {\em Materials
  Characterization}, vol.~59, no.~4, pp.~402--406, 2008.

\bibitem{arghavani2011}
Arghavani J, Auricchio F, and Naghdabadi R, ``A finite strain kinematic
  hardening constitutive model based on hencky strain: general framework,
  solution algorithm and application to shape memory alloys,'' {\em
  International Journal of Plasticity}, vol.~27, no.~6, pp.~940--961, 2011.

\bibitem{hartl2010}
Hartl D.~J, Chatzigeorgiou G, and Lagoudas D.~C, ``Three-dimensional modeling
  and numerical analysis of rate-dependent irrecoverable deformation in shape
  memory alloys,'' {\em International Journal of Plasticity}, vol.~26, no.~10,
  pp.~1485--1507, 2010.

\bibitem{jay2018Dacron}
Jayendiran R, Nour B, and Ruimi A, ``Computational fluid--structure interaction
  analysis of blood flow on patient-specific reconstructed aortic anatomy and
  aneurysm treatment with dacron graft,'' {\em Journal of Fluids and
  Structures}, vol.~81, pp.~693--711, 2018.

\bibitem{Johnson2015}
Medical J.~M, ``An overview of nitinol: superelastic and shape memory,'' {\em
  Medical Design Briefs}, 2015.

\bibitem{xiao1999acta}
Xiao H, Bruhns O, and Meyers A, ``Existence and uniqueness of the
  integrable-exactly hypoelastic equation and its significance to finite
  inelasticity,'' {\em Acta Mechanica}, vol.~138, no.~1, pp.~31--50, 1999.

\end{thebibliography}

%%%%%%%%%%%%%%%%%%%%%%%%%%%%%%%%%%%%%%%%%%%%%%%%%%%%%%%%%%%%%%%%%%%%%%%%%
%%%%%%%%%%%%%%%%%%%%%%%%%%%%%%%%%%%%%%%%%%%%%%%%%%%%%%%%%%%%%%%%%%%%%%%%%
%%%%%%%%%%%%%%%%%%%%%%%%%%%%%%%%%%%%%%%%%%%%%%%%%%%%%%%%%%%%%%%%%%%%%%%%%

\newpage
\appendix
\section{Artificial stress residuals due to the non-integrable objective rates}\label{sec:appendix}
In this appendix section, BVP is investigated to study the artificial stress residuals caused by using other non-integrable objective rates. As it is discussed in the introduction, the rate form hypoelastic constitutive theory has been criticized for its inconsistent choices on objective rates \cite{simo2006}, this includes the well-known objective rates (\emph{e.g.}, Zaremba-Jaumann rate, Green-Naghdi rate, Truesdell rate, \emph{etc}). In other words, the rate form hypoelastic constitutive equation fails to be integrated to deliver an algebraic hyperelastic constitutive equation via the so-called objective rates, because of which spurious phenomena (\emph{e.g.}, shear stress oscillation, artificial stress residuals, \emph{etc}.) are often observed in a simple elastic deformation \cite{xiao2006}.  It was not until recently such self-inconsistent issues about hypoelastic constitutive models has been resolved by the logarithmic rate proposed by \cite{xiao1997,xiao1997hypo,xiao2006,bruhns1999self,bruhns2001large,bruhns2001self,meyers2003elastic,meyers2006choice}. As their work proved that the logarithmic rate of the logarithmic strain $\mathbf{h}$ of its Eulerian type is equivalent to the strain rate $\mathbf{D}$, by which a grade-zero hypoelastic model can be exactly integrated into an finite strain elastic model based on logarithmic strain. 

\subsection{The cyclic response of an elastic square}\label{sec:cyc_sqaure}
\begin{figure}[H]
	\centering
	\centering
	\includegraphics[width=0.5\textwidth]{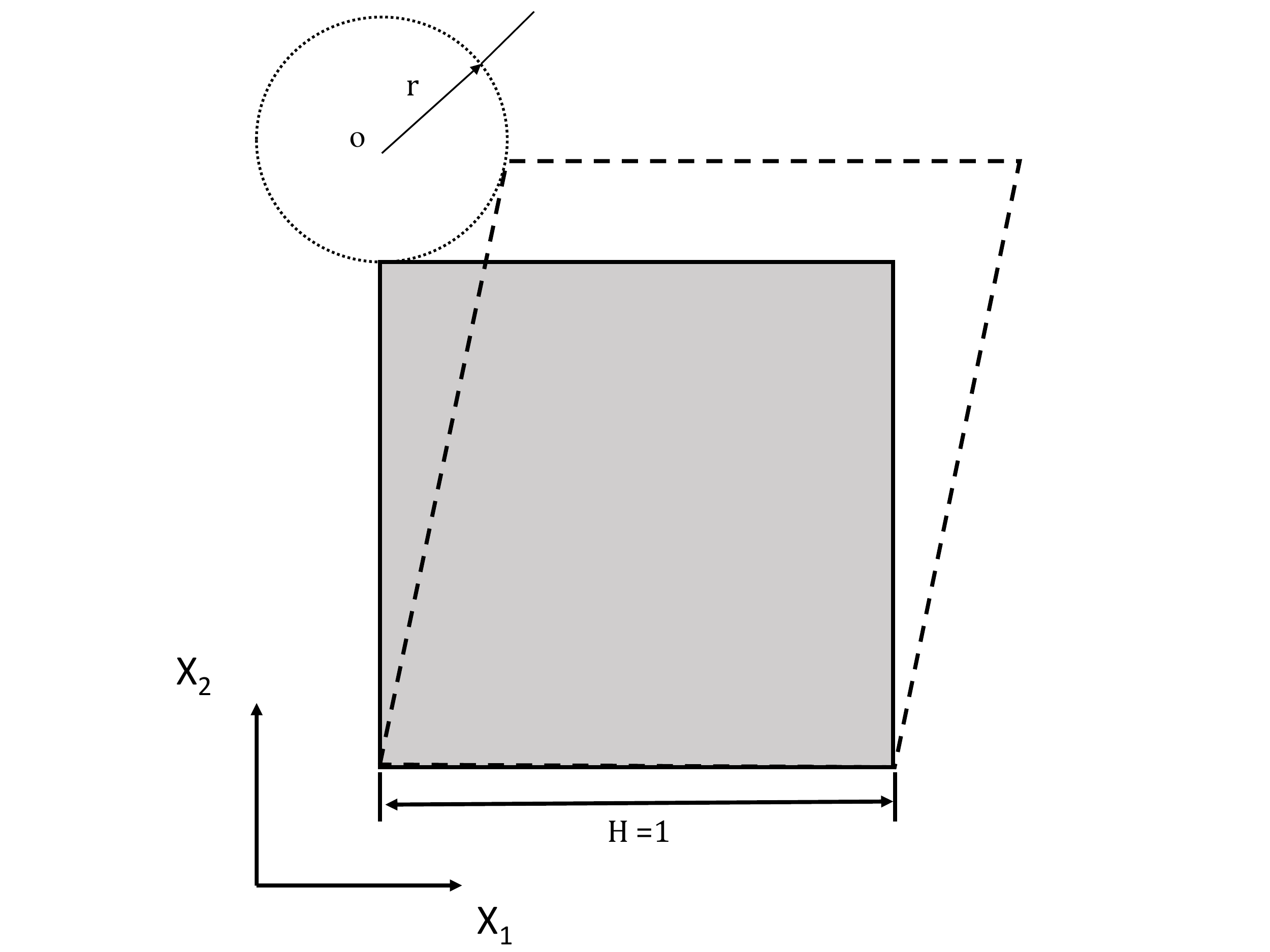}
	\caption{The schematic of a simple elastic square under closed path cyclic loading.}
	\label{fig:ElasticCube}
\end{figure}

\begin{figure}[H]
	\centering\vspace{-0.5cm}
	\includegraphics[width=0.8\textwidth]{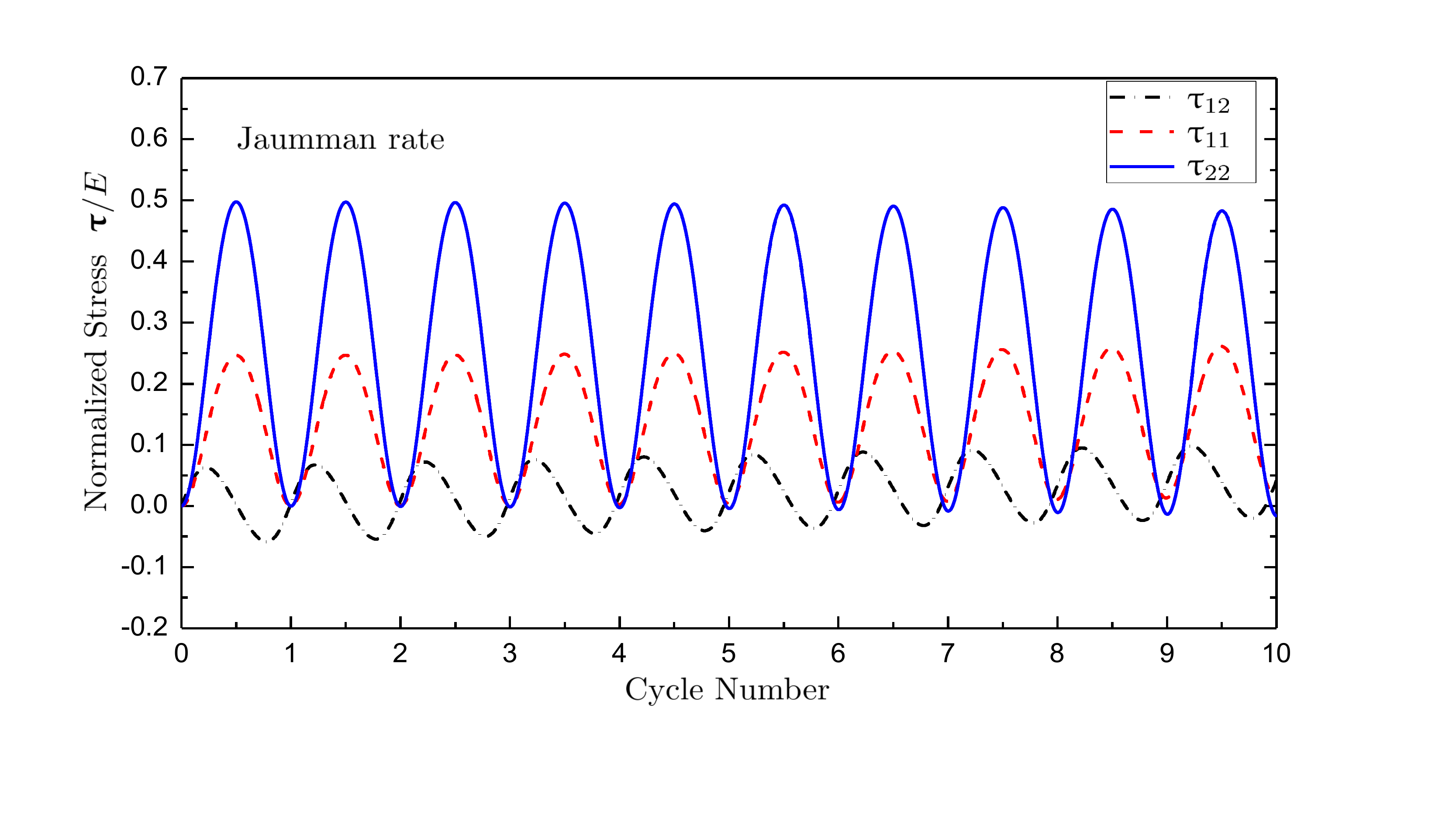}\vspace{-1cm}
	\caption{Kirchhoff stress components predicted by hypoelastic equation using Jaumman rate under 10 loading cycles}
	\label{fig:Jaumman}
\end{figure}

\begin{figure}[H]
	\centering\vspace{-0.5cm}
	\includegraphics[width=0.8\textwidth]{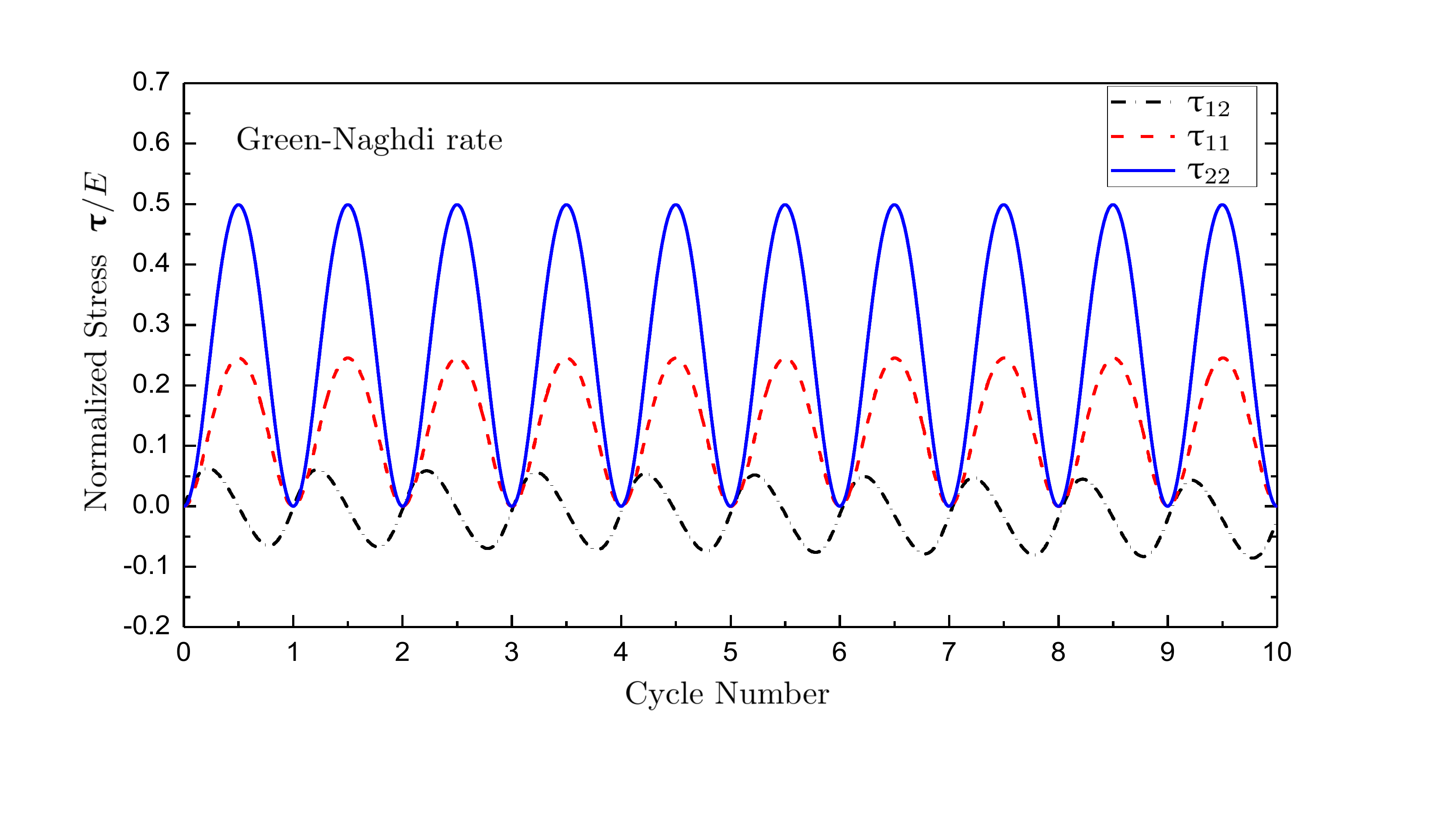}\vspace{-1cm}
	\caption{Kirchhoff stress components predicted by hypoelastic equation using Green-Naghdi rate under 10 loading cycles}
	\label{fig:Green}
\end{figure}

\begin{figure}[H]
	\centering\vspace{-0.5cm}
	\includegraphics[width=0.8\textwidth]{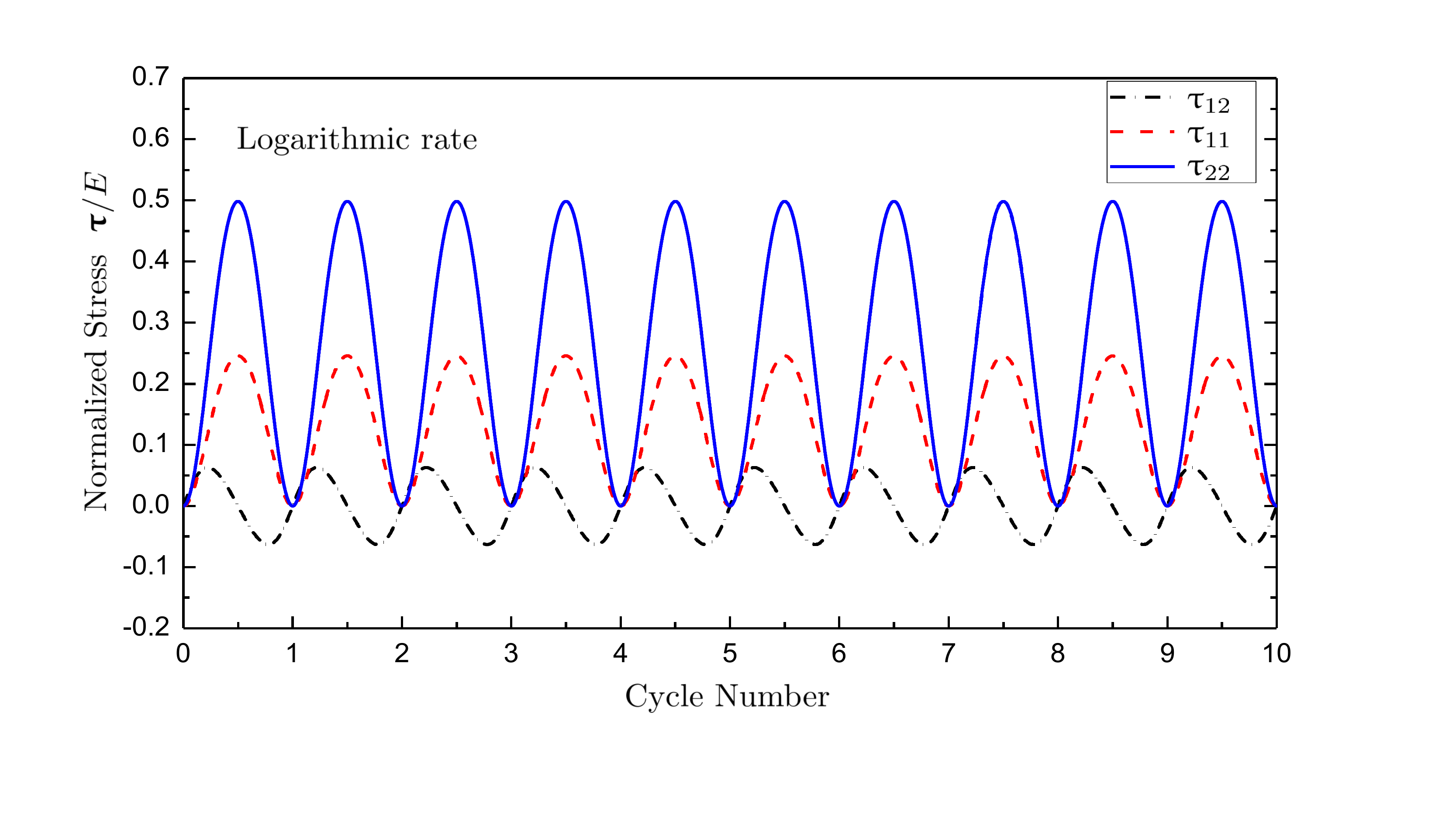}\vspace{-1cm}
	\caption{Kirchhoff stress components predicted by hypoelastic equation using Logarithmic rate under 10 loading cycles}
	\label{fig:Logarithmic}
\end{figure}

Refer to figure \ref{fig:ElasticCube} for the BVP schematics, a two-dimensional elastic square with length $H$ is under a closed path cyclic loading. The upper line of the square is subjected to a displacement control circular deformation, the deformation over geometry ratio is as $r/H=0.2$ to induce large deformation strain. The stress components are examined by the hypoelastic equation (\ref{eq:Cube_Con_Hypo}) for 10 loading cycles. In equation (\ref{eq:Cube_Con_Hypo}), $ \mathcal{\bm C}$ is the stiffness tensor and an circle over $\bm{\uptau}$ means the different objective rates adopted.
\begin{equation}\label{eq:Cube_Con_Hypo}
\mathring{\bm \uptau}  = \mathcal{\bm C}:{\mathbf{D}} 
\end{equation}
Based on the results from \cite{xiao1997}, hypoelastic equation (\ref{eq:Cube_Con_Hypo}) can be self-consistently integrated to a hyperelastic constitutive equation (\ref{eq:Cube_Con_hyper}) based on logarithmic strain through the logarithmic corotational integration \cite{khan1995continuum}.
\begin{equation}\label{eq:Cube_Con_hyper}
{\bm \uptau}  = \mathcal{\bm C}:{\mathbf{h}} 
\end{equation}
Kirchhoff stress components are obtained by equation (\ref{eq:Cube_Con_Hypo}) with three different objective rates, \emph{i.e.} Jaumman rate, Green-Naghdi rate and Logarithmic rate. The predicted results are presented in figure \ref{fig:Jaumman}, figure \ref{fig:Green} and figure \ref{fig:Logarithmic}. The stress results are normalized by the material Young's modulus $E$.


\begin{table}[h]
	\centering
	\caption{Stress residuals obtained by hypoelastic equation (\ref{eq:Cube_Con_Hypo}) with different objective rates for the elastic square}
	\label{tab:E_ReStress}
	\begin{tabular}{@{}lccccc@{}}
		\toprule
		&Normalized stress    & Logarithmic rate  & Jaumman rate  &Green-Naghdi rate &\\ \midrule
		&${\uptau}_{11}/E$ &-1.71e-5 & 0.0164 &2.17 e-6 &\\ 
		&${\uptau}_{12}/E$ &2.17e-6  & 0.0411 &-0.023&\\ 
		&${\uptau}_{22}/E$ &1.71e-5  &-0.0164 &1.76e-5&\\ 
		\bottomrule
	\end{tabular}
\end{table}
The stress residuals are examined at the end of the loading cycle and summarized in table \ref{tab:E_ReStress}. First, the stress components in all the three cases showed periodic oscillation. Since the material is confined to behave elastically, the deformation should be indissipative to anticipate that all the stress components should return to zero value in the end. However, the predicted stress components in the cases of Jaumman rate and Green-Naghdi rate showed artificial stress residuals are introduced. Refer to figure \ref{fig:Jaumman} for the case of Jaumman rate, stress residuals $\uptau_{11}$ is 0.0164, $\uptau_{22}$ is -0.0164 and $\uptau_{12}$ is 0.0411 after the 10 loading cycles. In the case of Green-Naghdi rate, although there are inconsiderable stress residuals for $\uptau_{11}$ and $\uptau_{22}$ components, the shear residuals $\uptau_{12}$ is -0.023. In contrast, all the stress residuals are almost negligible in the case of Logarithmic rate, which demonstrates that the hypoelastic constitutive equation utilizing logarithmic rate can be self-consistently integrated to deliver a hyperelastic equation based on the logarithmic strain. Interested readers are encouraged to further read \cite{meyers2003elastic,xiao1999acta}.
%

%%%%%%%%%%%%%%%%%%%%%%%%%%%%%%%%%%%%%%%%%%%%%%%%%%%%%%%
%%%%%%%%%%%%%%%%%%%%%%%%%%%%%%%%%%%%%%%%%%%%%%%%%%%%%%%
%%%%%%%%%%%%%%%%%%%%%%%%%%%%%%%%%%%%%%%%%%%%%%%%%%%%%%%
\section{Calibration of the material parameters}\label{sec:calibration}
In this section, the material parameters utilized in the proposed model are identified from a set of one-dimensional experimental data. Note that the strain measure used here should be in the true (or logarithmic) scale rather than the engineering (or infinitesimal) scale. Material parameters used in the proposed model can be categorized into three groups, \emph{i.e.} the key material parameters, smooth hardening parameters and intermediate parameters. First, the material constants such as elastic modulus $E_A, E_M$, Poisson's ratios $\nu_A$ and $\nu_B$, the thermal expansion tensors $\bm{\alpha}_A$ and $\bm{\alpha}_M$, stress influenced coefficients $C_A$ and $C_M$ from the phase diagram (or called clausius clapeyron coefficient), critical phase transformation temperatures $ A_s, A_f, M_s, M_f$ at stress free state are determined. Secondly, the hardening parameters describing the smooth transition feature are discussed. Finally, the intermediate parameters are derived based on the aforementioned two parameter groups. All the material parameters used in this model are summarized in table \ref{tab:MaterialProperty_bar}.

Because the data is provided in one-dimensional case, all tensorial variables of the proposed model have to be reduced into 1-D scalar value. For example, the stress tensor is reduced as $\bm{\uptau} \rightarrow \uptau_{11}=\uptau$;  logarithmic strain tensor is reduced as $ \mathbf{h} \rightarrow h_{11}=h$, \emph{etc}. Constitutive equation (\ref{eq:h_Cons_f}) can be rewritten as one dimensional form as follows,
\begin{equation}\label{eq:Reduced_stress}
\uptau=E [h- \alpha(T-T_0)- h^{tr}]
\end{equation}   
where the effective elastic modulus $E$ is calculated by using the rule of mixture as follows,
\begin{equation}\label{eq:YoungModuli}
E = [1/E^{A}+\xi(1/E^{M}-1/E^{A})]
\end{equation}
The evolution equation (\ref{eq:Trans_Evol}) is also reduced in one-dimensional form as,
\begin{equation}\label{eq:Reduced_trans.}
\begin{aligned}
\Lambda=\Lambda_{11}=
\begin{cases}   
H^{cur}(\sigma) ~\text{sgn}({\uptau})     \qquad         &; \dot{\xi}>0,
\vspace{4pt} \\ \ \qquad \dfrac{h^{t-r}}{\xi^r}  &; \dot{\xi}<0, 
\end{cases}\\
\end{aligned}
\end{equation}   
the thermodynamic driving force $\pi$ in one-dimensional case can thus be obtained,
\begin{equation}\label{eq:Reduced_pi}
\begin{aligned}
\pi=\uptau\Lambda+ \dfrac{1}{2} \Delta\mathcal{S}\uptau^2+\uptau{\Delta}{\alpha}(T-T_0)+ \rho_0\Delta s_0 T-\rho_0\Delta c \big[ T-T_0-T\ln(\dfrac{T}{T_0}) \big]
- \rho\Delta u_0 - \frac{\partial f}{\partial \xi}
\end{aligned}
\end{equation} 
the transformation function in one-dimensional form can be calculated based on equation (\ref{eq:Reduced_pi}). Considering the phase difference for the thermal expansion $\Delta \alpha$ and specific heat $\Delta c$ are small enough to be ignored, the following  transformation functions for the forward case and the reverse case can be obtained respectively,    
\begin{equation}\label{eq:Reduced_Trans_Fun_fwd}
\begin{aligned}
\normalfont{\Phi}_{\textit{fwd}}(\uptau,T,\xi)=\big [ \uptau\Lambda+ \dfrac{1}{2} \Delta\mathcal{S}\uptau^2 + \rho_0\Delta s_0 T - \rho_0\Delta u_0 - \frac{\partial f}{\partial \xi} \big ] - Y = 0
\end{aligned}
\end{equation}
\begin{equation}\label{eq:Reduced_Trans_Fun_rev}
\begin{aligned}
\normalfont{\Phi}_{\textit{rev}}(\uptau,T,\xi)=- \big [ \uptau\Lambda+ \dfrac{1}{2} \Delta\mathcal{S}\uptau^2 + \rho_0\Delta s_0 T - \rho_0\Delta u_0 - \frac{\partial f}{\partial \xi} \big ] - Y = 0
\end{aligned}
\end{equation}
As described in the first paragraph of this section, there are three sets of material parameters that need to be identified. First, let's consider the material constants $(E_A, E_M, \nu_A, \nu_M, \alpha_A, \alpha_M )$. Elastic modulus  $ E_A, E_M $ can be determined through a pseudoelastic stress and strain curve by calculating the slopes at martensitic phase and austenite phase. Poisson’s ratio is attained using a widely accepted value of $ \nu_A=\nu_M=0.33$ found in \cite{lagoudas2012}. The thermal expansion coefficient are usually considered as $\alpha_A=\alpha_M$, which can be calibrated through an isobaric actuation experiment. The maximum transformation strain $H^{max}$ can be determined from the pseudoelastic experimental and the value of parameter $k_t$ are chosen to best fit the $H^{cur}$ curve. The stress influence coefficients and the critical phase transformation temperatures $(C_A, C_M, M_s, M_f, A_s, A_f)$ can be calibrated through the phase diagram. Second, the material parameters related to the smooth hardening features are discussed. The Coefficients $n_1, n_2, n_3, n_4$ without specific physical meanings are determined to best match the smoothness in corners of material response. Lastly, there are seven intermediate material parameters $(\rho_0\Delta s_0, \rho_0\Delta u_0, a_1, a_2,a_3, Y_0 ~\text{and} ~D)$ that need to be calculated to complete the model. Determination of such intermediate parameters requires a set of seven algebraic equations. The needed four equations come from transformation constraints as the Kuhn-Tucker condition Equation.\ref{eq:Kuhn-Tucker} (i.e. $\normalfont{\Phi}_{rev}(\uptau,T,\xi)=0$). The fifth equation comes from the continuity of Gibbs free energy at the end of the forward transformation $(\xi=1)$. The needed five algebraic equations are summarized as follows, 
\begin{enumerate}[1.]
	\item Start of the forward transformation at zero stress $(\uptau=0; T=M_s; \xi=0)$.\\
	$\normalfont{\Phi}_{\textit{fwd}}(0,M_s,0)= \rho_0\Delta s_0 M_s - \rho_0\Delta u_0 - a_3 - Y_0 =0 $
	\item Finish of the forward transformation at zero stress $(\uptau=0; T=M_f; \xi=1)$.\\
	$\normalfont{\Phi}_{\textit{fwd}}(0,M_f,1)=\rho_0\Delta s_0 M_f - \rho_0\Delta u_0 -a_1 - a_3 - Y_0 = 0 $
	\item Start of reverse transformation under zero stress $(\uptau=0; T=A_s; \xi=1)$.\\
	$\normalfont{\Phi}_{\textit{rev}}(0,A_s,1)=- \rho_0\Delta s_0 M_s + \rho_0\Delta u_0 + a_2 - a_3 - Y_0 =0 $
	\item Finish of the reverse transformation under zero stress $(\uptau=0; T=A_f; \xi=0)$.\\
	$\normalfont{\Phi}_{\textit{rev}}(0,A_f,0)= - \rho_0\Delta s_0 A_f + \rho_0\Delta u_0 - a_3 - Y_0 =0 $
	\item The gibbs free energy continuity at the end of forward transformation $(\xi=1)$. \\
	$f(\xi=1)|_{\dot{\xi}\ge 0}$ =  $f(\xi=1)|_{\dot{\xi}\le 0}$ 		
\end{enumerate}
The above five algebraic equations yield the following expression for the five out of seven intermediate model parameters,
\begin{equation}\label{eq:Reduced_5_Parameters}
\begin{aligned}
&a_1=\rho_0 \Delta s_0 (M_f-M_s); \quad a_2=\rho_0 \Delta s_0 (A_s-A_f)\\
&a_3 = \cfrac{1}{4}~a_2(1+\cfrac{1}{n_3+1})-\cfrac{1}{4}~a_1(1+\cfrac{1}{n_1+1})\\
&\rho_0 \Delta u_0 =\dfrac{1}{2}\rho \Delta s_0 (M_s+A_f)\\
&Y_0=\dfrac{1}{2} \rho_0 \Delta s_0 (M_s-A_f)-a_3
\end{aligned}
\end{equation}
Another two equations are derived from the Kuhn-Tucker condition in order to complete the calculation. For a one-dimensional uniaxial experiment, the Kuhn-Tucker condition (\ref{eq:Kuhn-Tucker}) requires equation (\ref{eq:Reduced_D_Phi}) to hold true at any specific stress level $\uptau^{*}$,
\begin{equation}\label{eq:Reduced_D_Phi}
\begin{aligned}
\text{d}\Phi = \partial_{\uptau}\Phi~\text{d}\uptau + \partial_{T}\Phi ~\text{d}T +\partial_{\xi}\Phi ~\text{d}\xi  =0 
\end{aligned}
\end{equation}
Evaluate $\text{d}\Phi$ at the start point of the forward phase transformation (i.e. $\xi=0$), and at the finish point of the forward phase transformation (i.e. $\xi=1$), the incremental part of martensitic volume fraction should be zero (i.e. $\text{d}\xi=0$) in both of the aforementioned cases. Therefore,  the relationships between the stress temperature coefficients $C_M$, $C_A$ and the stress temperature slopes $\frac{\text{d}\uptau}{\text{d} T}$ can be obtained.\\ 
For the forward transformation case, $\dot{\xi}>0$,
\begin{equation}\label{eq:Reduced_dsdt_fwd}
C_M= \dfrac{\text{d}\uptau}{\text{d} T} \Big| _{\uptau^{*},\dot\xi>0}= \frac{ - \rho \Delta s_0}{ \Lambda + \uptau:\partial_{\uptau}\Lambda + \Delta\mathcal{S}\uptau  -  \partial_{\uptau} Y}\Big| _{\uptau^{*}} 
\end{equation}
For the reverse transformation case, $\dot{\xi}<0$
\begin{equation}\label{eq:Reduced_dsdt_rev}
C_A= \dfrac{\text{d}\uptau}{\text{d} T} \Big| _{\uptau^{*},\dot\xi<0}= \frac{ - \rho \Delta s_0}{ \Lambda + \uptau:\partial_{\uptau}\Lambda + \Delta\mathcal{S}\uptau  +  \partial_{\uptau} Y}\Big| _{\uptau^{*}}  
\end{equation}
Using the equations (\ref{eq:Reduced_dsdt_fwd}) and (\ref{eq:Reduced_dsdt_rev}), the rest two intermediate material parameters $\rho_0\Delta s_0$ and $D$ can thus be expressed as follows,
\begin{equation}\label{eq:Reduced_D}
\begin{aligned}
D = \dfrac{ (C_M-C_A)\big[ H^\textit{cur}+ \uptau\partial_{\uptau}H^\textit{cur}+ \uptau\Delta\mathcal{S}\big]}
{(C_M+C_A) ( H^\textit{cur} + \uptau\partial_{\uptau}H^\textit{cur})}
\end{aligned}
\end{equation}

\begin{equation}\label{eq:Reduced_rds0}
\begin{aligned}
\rho_0 \Delta s_0 = -\dfrac{ 2 C_M C_A \big[ H^\textit{cur}+ \uptau\partial_{\uptau}H^\textit{cur}+ \uptau\Delta\mathcal{S}\big]}
{C_M+C_A} 
\end{aligned}
\end{equation}

\section{Model validation}\label{sec:validation}
\begin{figure}[h]
	\centering
	\includegraphics[width=0.65\textwidth]{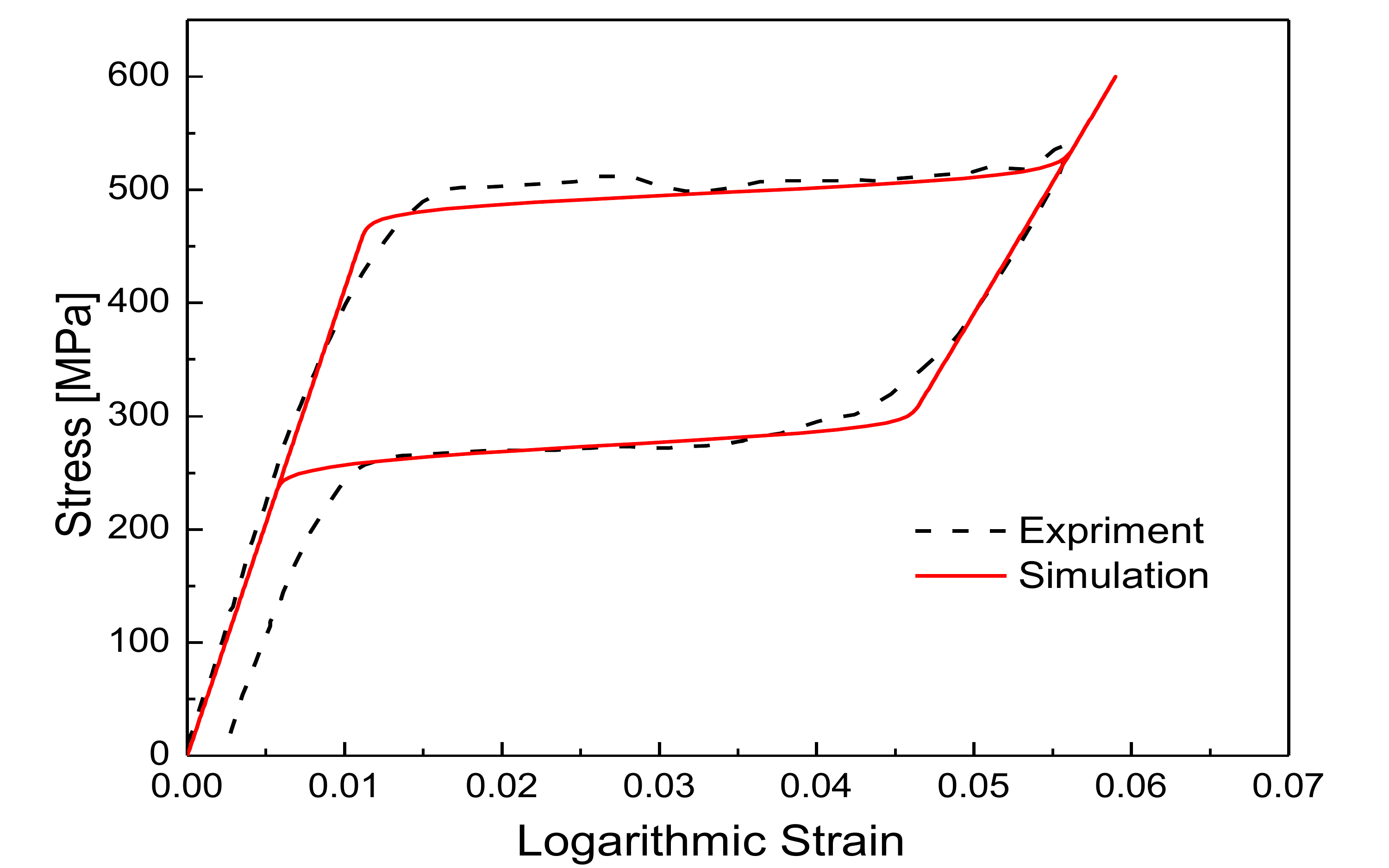}%\vspace{-0.2cm}
	\caption{ {The validation of the proposed model against the experimental data through a uniaxial stress-strain response of NiTi SMA, the dashed black line indicates the experimental data and the solid red line is the prediction by the proposed model.}}
	\label{fig:Model_Calibration}
\end{figure}

{
For the validation of proposed model against experimental results, a uniaxial pseudoelastic tensile test is performed on a NiTi SMA \cite{lagoudas2008}. In the experiment, a NiTi SMA strip is loaded at a constant temperature of $320 $ K, which is larger than the austenistic finish temperature $A_f$. The strip is subjected to a traction up to $600 $ MPa then unloaded to $0$ MPa. The material parameters listed in table \ref{tab:model_calibration} are calibrated based on the experimental result. Figure \ref{fig:Model_Calibration} shows the comparison between the experimental results and the simulations from the proposed model, it clearly demonstrates that the proposed model predicts the stress-strain response of the NiTi strip quite well, including the phase transformation starting and finishing points, and the size of the hysteresis loop. As the experimentally tested specimen is an untrained SMA, there is a amount of transformation-induced plastic strain remained at the end of the loading. This phenomenon can be captured by extending the current model with the consideration of the transformation-induced plasticity.}
 
\begin{table}[h] 
	\centering
	\caption{ {Calibrated material parameters of a NiTi SMA used for the model validation.}}\vspace{-0.2cm}
	\renewcommand{\arraystretch}{1}
	\begin{tabular}{c|lr|ll} \toprule
		Type                         &Parameter                        & Value                                   &Parameter            & Value  \\                                       \midrule
		&$E_A$                            & 41   [GPa]                              & $C_A$                & 5.5  [MPa/K]\\
		&$E_M$                            & 22   [GPa]                              & $C_M$               & 5.5  [MPa/K]\\
		Key material parameters       &$\nu_A=\nu_M$                          & 0.33~~~~~~~~                              & $M_s$                & 237  [K]\\
		12                      &$\alpha_A=\alpha_M$    &    1.0$\times$10$^{-5}$ [K$^{-1}$]                           & $M_f$                &217.5  [K]\\
		& $ H^\textit{max}$                        & 3.35\%          & $A_s$                 &254  [K]\\
		& $k_t$                      &    N/A   & $A_f $                 & 282  [K]\\                                   \midrule
		
		Smooth hardening parameters 		&  $n_1$         &   0.15                &  $n_3$             & 0.25 \\
		4							   &  $n_2$         &   0.17                &  $n_4$        	 & 0.15 \\                                   
		\bottomrule
	\end{tabular}
	\label{tab:model_calibration}
\end{table}

\vspace{0.2cm}

%%%%%%%%%%%%%%%%%%%%%%%%%%%%%%%%%%%%%%%%%%%%%%%%%%%%%%%
%%%%%%%%%%%%%%%%%%%%%%%%%%%%%%%%%%%%%%%%%%%%%%%%%%%%%%%
%%%%%%%%%%%%%%%%%%%%%%%%%%%%%%%%%%%%%%%%%%%%%%%%%%%%%%%
\section{Supplementary calculation for consistent tangent stiffness and thermal matrix}\label{sec:cons_append}
The consistent tangent stiffness and thermal matrix are derived in section \ref{sec:Jacobian}. 
In order to determine the explicit values for $\mathcal{L}$ and ${\Theta}$ during the implementation of the proposed model, the explicit expressions of the following terms $\partial_{\bm\uptau}\Phi$, $\partial_{\xi}\Phi$, $\partial_{T}\Phi$ used in equation (\ref{eq:Jacobian_exp}) are needed. First, the partial derivative of transformation function $\Phi$ with respect to stress $\bm{\uptau}$ can be obtained through differentiating equation (\ref{eq:Transfor_Fun}) by $\bm{\uptau}$. Utilize the expression for $\pi$ in equation (\ref{eq:Driving_Force}), it obtains,
\begin{equation}\label{eq:dpds}
\normalfont{\partial_{\bm\uptau}\Phi}=\begin{cases}~~\partial_{\bm\uptau}\pi - \partial_{\bm\uptau}Y, \; \dot{\xi}>0, \vspace{5pt} \\ -\partial_{\bm\uptau}\pi - \partial_{\bm\uptau}Y, \; \dot{\xi}<0 \end{cases}\\
\end{equation}
where the partial derivative of the thermodynamic driving force $\pi$ with respect to stress $\bm{\uptau}$ is,
\begin{equation}\label{eq:dpds}
\normalfont{\partial_{\bm\uptau}\pi}=\bm\Lambda+ (\partial_{\bm{\uptau}}\bm\Lambda)\bm\uptau +  {\Delta}\mathcal{S}\bm\uptau+{\Delta}\bm{\alpha}(T-T_0)
\end{equation}
and the partial derivative of critical driving force value $Y$ with respect to stress $\bm{\uptau}$ is,
\begin{equation}
\partial_{\bm{\uptau}} Y= D \Big[\bm{\Lambda}+(\partial_{\bm{\uptau}}\bm\Lambda)\bm\uptau \Big]
\end{equation}
based on the expression for the transformation direction tensor in equation (\ref{eq:direction}), the partial derivative of $\bm{\Lambda}$ with respect to stress $\bm{\uptau}$ are provided for the forward and reverse transformation cases as follows,  
\begin{equation}
\partial_{\bm{\uptau}}\bm{\Lambda} = \begin{cases}~~\dfrac{3}{2} \partial_{\bm{\uptau}} H^{cur}\otimes\dfrac{\bm{\uptau}^{'}}{\bar{\bm\uptau}} + \dfrac{3}{2}H^{cur} \partial_{\bm{\uptau}} \Big( \dfrac{\bm{\uptau}^{'}}{\bar{\bm\uptau}}\Big) , \; ~~~\dot{\xi}>0, \vspace{3pt} \\~~0, \; ~~~~\dot{\xi}<0 \end{cases}
\end{equation} 
where the partial derivative of the term $\Big( \dfrac{\bm{\uptau}^{'}}{\bar{\bm\uptau}} \Big)$ with respect to stress $\bm{\uptau}$ is provided in the following equation, in which $\mathbf{I}$ is the forth order identity tensor and $\mathbf{1}$ is the second order identity tensor. It can be observed that $\partial_{\bm{\uptau}}\bm{\Lambda}$ only has value for the forward transformation case while it has value zero for the reverse transformation case.  
\begin{equation}
\partial_{\bm{\uptau}}\Big( \dfrac{\bm{\uptau}^{'}}{\bar{\bm\uptau}} \Big) = \frac{1}{\bar{\bm\uptau}} \Big( \mathbf{I} - \frac{1}{3}\mathbf{1}\otimes\mathbf{1} -\frac{2}{3}\frac{\bm{\uptau}^{'}}{\bar{\bm\uptau}} \otimes \frac{\bm{\uptau}^{'}}{\bar{\bm\uptau}} \Big) 
\end{equation}
to calculate the partial derivative of the current maximum transformation strain $H^{cur}$ with respect to stress $\bm{\uptau}$, the following result can be obtained based on equation (\ref{eq:Hcur}),
\begin{equation}
\partial_{\bm{\uptau}} H^{cur} = \frac{3}{2} H^{max} k_t \frac{\bm\uptau^{'}}{\bar{\bm\uptau}} 
\end{equation}

Follow the similar procedure to obtain $\partial_{\bm\uptau}\Phi$, the partial derivative of the transformation function $\Phi$ with respect to martensitic volume fraction $\xi$, and the partial derivative of the transformation function $\Phi$ with respect to temperature $T$ can be calculated as follows,
\begin{equation}
\partial_{\xi} \Phi = \begin{cases}
~~\dfrac{1}{2} a_1 \Big[ n_1 \xi^{n_1-1} + n_2 (1-\xi)^{n_2-1} \Big],  ~~~~\dot{\xi}>0, \\[6pt]
-\dfrac{1}{2} a_2 \Big[ n_3 \xi^{n_3-1} + n_4 (1-\xi)^{n_4-1} \Big],  ~~~~\dot{\xi}<0 \end{cases}\\
\end{equation}

\begin{equation}
\partial_{T} \Phi = \begin{cases}
~~~\bm{\uptau}:\Delta\bm{\alpha} + \rho_0\Delta c \ln(\frac{T}{T_0}) + \rho_0\Delta s_0 ,  ~~~~\dot{\xi}>0, \\[4pt]
-\Big[ \bm{\uptau}:\Delta\bm{\alpha} + \rho_0\Delta c \ln(\frac{T}{T_0}) + \rho_0\Delta s_0 \Big] ,  ~~\dot{\xi}<0 \end{cases}\\
\end{equation}

%%%%%%%%%%%%%%%%%%%%%%%%%%%%%%%%%%%%%%%%%%%%%%%%%%%%%%%
%%%%%%%%%%%%%%%%%%%%%%%%%%%%%%%%%%%%%%%%%%%%%%%%%%%%%%%
%%%%%%%%%%%%%%%%%%%%%%%%%%%%%%%%%%%%%%%%%%%%%%%%%%%%%%%

\end{document}